\documentclass[12pt]{article}
\pdfoutput=1  
\usepackage{amsmath}
\usepackage{graphicx}
\usepackage{subfig}
\usepackage{slashed}

\def\LSP{\tilde{\chi}^0_1}

\def\chipm{\tilde\chi_1^\pm}

\def\neu2{\tilde \chi_2^0}
\def\neu1{\tilde \chi_1^0}

\def\Re{{\cal R \mskip-4mu \lower.1ex \hbox{\it e}\,}}
\def\Im{{\cal I \mskip-5mu \lower.1ex \hbox{\it m}\,}}
\def\ie{{\it i.e.}}
\def\eg{{\it e.g.}}

\def\etal{{\it et al.}}

\def\sub#1{_{\lower.25ex\hbox{$\scriptstyle#1$}}}
\def\tev{\,{\rm TeV}}
\def\gev{\,{\rm GeV}}
\def\mev{\,{\rm MeV}}
\def\ev{\,{\rm eV}}
\def\kpc{\,{\rm kpc}}
\def\dof{\,{\rm dof}}
\def\to{\rightarrow}

\def\subw{_{\rm w}}
\def\mh{\ifmmode m\sbl H \else $m\sbl H$\fi}
\def\mch{\ifmmode m_{H^\pm} \else $m_{H^\pm}$\fi}
\def\mt{\ifmmode m_t\else $m_t$\fi}
\def\mc{\ifmmode m_c\else $m_c$\fi}
\def\mz{\ifmmode M_Z\else $M_Z$\fi}
\def\mw{\ifmmode M_W\else $M_W$\fi}
\def\mws{\ifmmode M_W^2 \else $M_W^2$\fi}
\def\mhs{\ifmmode m_H^2 \else $m_H^2$\fi}
\def\mzs{\ifmmode M_Z^2 \else $M_Z^2$\fi}
\def\mts{\ifmmode m_t^2 \else $m_t^2$\fi}
\def\mcs{\ifmmode m_c^2 \else $m_c^2$\fi}
\def\mchs{\ifmmode m_{H^\pm}^2 \else $m_{H^\pm}^2$\fi}
\def\ztwo{\ifmmode Z_2\else $Z_2$\fi}
\def\zone{\ifmmode Z_1\else $Z_1$\fi}
\def\mtwo{\ifmmode M_2\else $M_2$\fi}
\def\mone{\ifmmode M_1\else $M_1$\fi}
\def\tb{\ifmmode \tan\beta \else $\tan\beta$\fi}
\def\xw{\ifmmode x\subw\else $x\subw$\fi}
\def\ch{\ifmmode H^\pm \else $H^\pm$\fi}
\def\lum{\ifmmode {\cal L}\else ${\cal L}$\fi}
\def\inpb{\ifmmode {\rm pb}^{-1}\else ${\rm pb}^{-1}$\fi}
\def\infb{\ifmmode {\rm fb}^{-1}\else ${\rm fb}^{-1}$\fi}
\def\epem{\ifmmode e^+e^-\else $e^+e^-$\fi}
\def\ppb{\ifmmode \bar pp\else $\bar pp$\fi}
\def\pbp{\ifmmode ~^(\bar p^)p\else $~^(\bar p^)p$\fi}
\def\bsg{\ifmmode B\to X_s\gamma\else $B\to X_s\gamma$\fi}
\def\bsll{\ifmmode B\to X_s\ell^+\ell^-\else $B\to X_s\ell^+\ell^-$\fi}
\def\bstt{\ifmmode B\to X_s\tau^+\tau^-\else $B\to  X_s\tau^+\tau^-$\fi}

\newskip\zatskip \zatskip=0pt plus0pt minus0pt
\def\matth{\mathsurround=0pt}
\def\lsim{\mathrel{\mathpalette\atversim<}}
\def\gsim{\mathrel{\mathpalette\atversim>}}
\def\atversim#1#2{\lower0.7ex\vbox{\baselineskip\zatskip\lineskip\zatskip
  \lineskiplimit
  0pt\ialign{$\matth#1\hfil##\hfil$\crcr#2\crcr\sim\crcr}}}

\def\sigv{\ifmmode \langle\sigma v\rangle\else $\langle\sigma v\rangle$\fi}
\def\tsigv{\ifmmode \langle\sigma v\rangle R^{2}\else $\langle\sigma v\rangle
  R^{2}$\fi}
\def\Dxx{\ifmmode D_{xx} \else $D_{xx}$\fi}
\def\Dpp{\ifmmode D_{pp} \else $D_{pp}$\fi}
\def\ddp{\ifmmode \frac{\partial}{\partial p} \else $\frac{\partial}{\partial p}$\fi}
\def\alle{\ifmmode (e^{+}+e^{-}) \else $(e^{+}+e^{-})$ \fi}
\def\pamr{\ifmmode e^{+}/(e^{+}+e^{-}) \else $e^{+}/(e^{+}+e^{-})$ \fi}
\def\pbarp{\ifmmode \bar{p}/p \else $\bar{p}/p$ \fi}
\def\gams{$\gamma$'s\hspace{0.2cm}}
\def\gamsn{$\gamma$'s}
\def\boverc{\ifmmode B/C \else $B/C$ \fi}
\def\chisq{\ifmmode \chi^2 \else $\chi^2$ \fi}
\def\rchisq{ \ifmmode \chi^2/\dof \else $\chi^2/\mathrm{dof}$ \fi}

\def\beq{\begin{equation}}
\def\eeq{\end{equation}}
\def\bit{\begin{itemize}}
\def\eit{\end{itemize}}

\begin{document} \begin{titlepage}
\rightline{\vbox{\halign{&#\hfil\cr
&SLAC-PUB-14211\cr
&Bonn-TH-2010-05\cr
&ANL-HEP-PR-10-40\cr
&NUHEP-TH/10-10\cr
}}}
\vspace{0.5in}
\begin{center}

{\Large\bf
Cosmic Ray Anomalies from the MSSM?\footnote{Work supported by the Department of
Energy, Contract DE-AC02-76SF00515}}
\medskip

{\normalsize
R.C. Cotta$^{(a)}$, J.A. Conley$^{(b)}$, J.S. Gainer$^{(c,d)}$,
J.L. Hewett$^{(a)}$ and T.G. Rizzo$^{(a)}$} \\
\vskip .3cm
$^{(a)}$ SLAC National Accelerator Laboratory, 2575 Sand Hill Rd., Menlo
Park, CA 94025 USA \\
$^{(b)}$Physikalisches InstitutUniversit\"at Bonn, Bonn, Germany\\
$^{(c)}$High Energy Physics Division, Argonne National Laboratory, Argonne, IL 60439 USA\\
$^{(d)}$Department of Physics and Astronomy, Northwestern University,
Evanston, IL 60208 USA
\vskip .3cm
\end{center}

\begin{abstract}

The recent positron excess in cosmic rays (CR) observed by the PAMELA
satellite 
may be a signal for dark matter (DM) annihilation. When these
measurements are 
combined with those from FERMI on the total ($e^++e^-$) flux and 
from PAMELA itself on 
the $\bar p/p$ ratio, these and other results are difficult to
reconcile with 
traditional models of DM, including the conventional mSUGRA
version of Supersymmetry 
even if boosts as large as $10^{3-4}$ are allowed. In this paper, we
combine the 
results of a previously obtained scan over a more general
19-parameter subspace 
of the MSSM with a corresponding scan over astrophysical parameters that
describe the propagation of CR.  We then ascertain 
whether or not a good fit to this CR data can be obtained with
relatively small 
boost factors while simultaneously satisfying the additional
constraints arising 
from gamma ray data. We find that a specific subclass of MSSM models
where the LSP 
is mostly pure bino and annihilates almost exclusively into
$\tau$ pairs 
comes very close to satisfying these requirements. 
The lightest $\tilde \tau$ 
in this set of models is found to be relatively close in mass to 
the LSP and is in some cases the nLSP.
These models lead to a significant improvement in the overall fit to the
data by an amount 
$\Delta \chi^2 \sim 1/$dof in comparison to the best fit
without Supersymmetry while employing boosts $\sim 100$.
The implications of these
models for future experiments are discussed. 
\end{abstract}


\renewcommand{\thefootnote}{\arabic{footnote}} \end{titlepage}

  \section{Introduction}
  \label{sec:intro}

Evidence for physics beyond the Standard Model (SM) has been uncovered
with the cosmological observations that $\sim 25\%$ of our universe is
comprised of Dark Matter (DM).  Numerous astrophysical observations
point to the existence of cold DM that cannot be accommodated in the 
SM \cite{silk}.
This cold DM
could take many forms, for example a weakly interacting massive
particle
(WIMP), axion, dark photon, or something more exotic are all
compatible
with the data \cite{Bergstrom:2010zz}.  
The WIMP-like DM scenario has received the most consideration as
WMAP measurements \cite{Komatsu:2008hk} of the relic density point to the
TeV-scale 
for weak-strength
DM annihilation and WIMPs occur naturally in many theories which go
beyond the
SM.  As a result, much attention is focused on various means of
detecting WIMPs,
through direct detection experiments deep underground, direct
production at
colliders, or indirect detection in cosmic ray (CR) spectra.

In fact, an anomaly has recently been reported in CR data.  The
PAMELA
collaboration observes \cite{Adriani:2008zr} 
a steep rise in the positron fraction (the
ratio of positrons over
the sum of electrons plus positrons) at energies between 10 and 100
GeV.  The Fermi Gamma
Ray Space Telescope (FERMI) reports \cite{Abdo:2009zk} a slightly stiffer
electron
spectrum than expected,
but with no dramatic features.  These are countered by observations of
the
anti-proton fraction (ratio of $\bar p$ to $p$) by PAMELA \cite{Adriani:2008zq} 
and various gamma ray
measurements by FERMI \cite{Abdo:2010ex,Abdo:2010nc,Abdo:2009mr,tporter:sympdiff}
which reveal no unexpected characteristics.

Much attention has been given to explaining these results in theories
with
new physics, with Supersymmetry (SUSY), arguably the most
widely-studied theory of physics
beyond the SM \cite{SUSYrev}, receiving the most consideration. 
In these models the DM particle is identified
with the Lightest Supersymmetric particle (LSP) which is typically 
assumed to be the lightest neutralino.
However, it was demonstrated early on \cite{toobig} 
that the most-studied Supersymmetric models, such as minimal
Supergravity
(mSUGRA), coupled with a thermal cosmology, have difficulty
accommodating this data.  In this case, separate issues arise
concerning
the reproduction of the sharp rise
in the positron fraction, while simultaneously not affecting the
anti-proton or gamma
ray spectra.  In addition, the typical Supersymmetric thermal DM
annihilation 
cross section is found
to be much too small, by a factor of order $10^{3-4}$, to explain the
PAMELA
positron spectrum excess.  This requires the introduction of a large
boost factor, $B$, which
multiples the annihilation cross section, hence increasing the rate.
It has
been hypothesized that such a boost factor could arise from clumpiness
\cite{Lavalle:2006vb} in the DM
distribution, from resonant \cite{resDM} 
or Sommerfeld enhancement \cite{summf} of the
annihilation cross section, from
non-minimal SUSY contributions \cite{Bai:2009ka}\cite{Bajc:2010qj},
or from a non-thermal cosmology \cite{gordy}.
Meanwhile, once such
a boost factor is incorporated it can create problems with the SUSY
contributions to both the
$\bar p/p$ and gamma ray fluxes, rendering them too large
\cite{toobig}.

In this work, we study a broader and more comprehensive region of the
parameter space than is usually studied
in the Minimal Supersymmetric Standard Model (MSSM) with the
expectation of obtaining a variety of contributions to cosmic rays
that differ from those found in mSUGRA.  We employ
the large
set of models ($\sim$ 68.5k) generated in the 19-dimensional
phenomenological MSSM
(pMSSM) parameter space by Berger \etal\ \cite{Berger:2008cq}.
These models contain no
theoretical prejudice
concerning the origin of Supersymmetry breaking or interactions at the
GUT scale.
As noted in this previous work, the characteristics of these pMSSM
models can be
quite different from those of mSUGRA.  In this paper we examine the
predictions of these models
for the various CR spectra in the hope of obtaining a good fit to the
data
without the need to incorporate such very large boost factors.

To this end, we couple this scan of the pMSSM parameter space with an
exploration of the
uncertainties inherent in the calculation of the astrophysical
contributions to cosmic rays.
There are numerous steps in the computation of the propagation of
cosmic rays
across the galaxy, each fraught with their own approximations and
uncertainties.  We account for
these by performing an extensive scan over the astrophysical
parameters entering
the propagation calculation and find a set of $\sim$ 6k
phenomenologically
viable CR `models' that are consistent with the CR data set (except
for the high energy bins in the PAMELA positron flux) and are
subject to a number
of additional constraints.  We then
compute both the standard astrophysical background spectra and the
SUSY 
DM annihilation contributions
using the same propagation parameters, for each combination of the 
pMSSM and astrophysics
models.  Performing a global $\chi^2$ fit to the ($e^++e^-$), 
$e^+/(e^++e^-)$ and ($\bar p/p$) data, we find a region of
pMSSM
parameter space that yields a good fit to the positron and anti-proton
fractions as
well at the ($e^++e^-$)  data (\eg, $\chi^2\sim 1.55$/dof) with boost
factors in the range $\sim 100-200$, far lower than those typically
required in
conventional mSUGRA.
The presence of the SUSY DM component is found to significantly
increase
the quality of the
fit to this data, by $\sim$ 1 unit of $\chi^2/$dof, in comparison to
the null hypothesis
of no new physics with only conventional
astrophysical sources. We next perform a consistency check, employing
data from both the
mid-latitude diffuse gamma ray spectrum as well as from the total
$\gamma$ flux
from faint dwarf galaxies for a small subset of 10 models which are
consistently found to provide the
best overall fit.
In these models we find that the pMSSM spectrum and couplings are such
that the LSPs tend
to annihilate almost exclusively into tau pairs. This mostly occurs
because: ($i$) the LSPs are
found to be dominantly bino-like, preventing annihilations into
$W,Z$-bosons, ($ii$) the staus are light, ($iii$)
while the sbottoms are reasonably heavy thus suppressing
annihilation into $b\bar{b}$
and avoiding significant contributions to the antiproton flux.

The rest of this paper proceeds as follows:  the next section
describes the generation
and properties of the pMSSM model sample that was originally
carried out in \cite{Berger:2008cq}.  We discuss our CR results
in Section 3,
first describing the CR propagation uncertainties and our scan of the
astrophysics parameter space, next, giving the results that are
obtained from astrophysical contributions alone, and, finally,
including 
the SUSY contributions.  In Section 4 we
compute the SUSY
contributions from WIMP annihilation in dwarf galaxies as well as
there contribution to the mid-latitude diffuse gamma ray flux, while
Section
5 contains
a description of the best-fit pMSSM models.  Our conclusions are given
in Section 6.

  \section{Generation and Properties of pMSSM Model Set}
  \label{sec:pMSSM}

We study a large set of Supersymmetric models that broadly sample the
parameter space, do not incorporate a specific SUSY-breaking mechanism
and
are free of theoretical assumptions at the GUT scale.  We do so in the
expectation
that a more extensive sampling of the MSSM parameter space will be
better able to identify models which
accommodate the cosmic ray data.  We use the MSSM database generated
in
the
work of Berger \etal\cite{Berger:2008cq}.  In this section, we briefly
describe the
procedure employed in constructing
these models and summarize the properties of these models that are
most relevant to DM phenomenology.
Throughout this work, we refer to a point in the Supersymmetric
parameter space as a model.

Berger \etal\ examined the CP-conserving MSSM with minimal flavor
violation
\cite{D'Ambrosio:2002ex}. The first two generations of sfermions were
taken to be
degenerate as motivated by constraints from flavor physics.  A thermal
cosmology was assumed with the lightest neutralino, $\tilde\chi_1^0$,
being the
lightest Supersymmetric particle (LSP) and stable.  Given these
assumptions, one is left with the 19 soft-SUSY breaking parameters of
the phenomenological MSSM (pMSSM). This set of real weak-scale SUSY
Lagrangian
parameters is given by the gaugino masses ($M_{1,2,3}$), 10 squared
masses of the sfermions ($m_{\tilde Q_{1,3}}\,, m_{\tilde u_{1,3}}\,,
m_{\tilde d_{1,3}}\,, m_{\tilde L_{1,3}}\,, m_{\tilde e_{1,3}}$, where
the subscripts
$1,3$ denote the first two and the third generations, respectively),
the Higgsino
mixing parameter $\mu$, the ratio of the Higgs vevs $\tan\beta$, the
mass
of the pseudoscalar Higgs boson $m_A$, and the third-generation $A$
terms
$A_{b,t,\tau}$.  A scan of $10^7$ points in the 19-dimensional
parameter
space was performed assuming flat priors for the ranges
\begin{eqnarray}
100\gev & \leq & m_{\tilde f}\leq 1\tev\,, \nonumber\\
 50\gev & \leq & |M_{1,2},\mu|\leq 1\tev\,,\nonumber\\
 100\gev & \leq & M_3\leq 1\tev\,,\nonumber\\
 |A_{b,t,\tau}| & \leq 1 & \tev\,, \\
1 & \leq & \tan\beta\leq 50\,,\nonumber\\
43.5\gev & \leq & M_A\leq 1\tev\,.  \nonumber
\end{eqnarray}
The lower bounds were chosen for general
consistency with collider data and the upper bounds were taken to
ensure
large sparticle production cross sections at the LHC.  A smaller
number of points
were scanned using logarithmic priors to ensure that the results
displayed little
dependency on the choice of priors.  These points
were
then subjected to a set of theoretical and experimental constraints
summarized
below.  The number of points scanned and analyzed was
limited by CPU availability.

The theoretical requirements imposed on models were that
their SUSY spectra contain no tachyons, that there be no color (or
charge) 
breaking minima in the scalar potential, that
electroweak symmetry breaking be consistent and that the Higgs
potential
be bounded from below.  Consistency with
the electroweak precision measurements $\Delta\rho$, the invisible
width of
the $Z$-boson, and the anomalous magnetic moment of the muon was
required.
Constraints from
the flavor sector were applied and agreement with the rates or limits
for
$b\to s\gamma$, $B_s\to\mu\mu$, $B\to\tau\nu$ and meson mixing was
imposed.
Limits from collider searches for sparticles and SUSY Higgs bosons
were
enforced.  The charged sparticle and SUSY Higgs bounds from LEPII
contain
many caveats and dependencies on the details of the individual model
spectrum;
these were carefully incorporated into the analysis.  The Tevatron
searches for squarks
and
gluinos were generalized to render them model independent and
multi-jet + MET events
were generated for each pMSSM model and compared to the data.  In
addition, bounds
from  the stable charged particle searches at LEPII and the Tevatron
were applied.
Two astrophysical constraints were imposed on the long-lived relic
$\tilde\chi_1^0$.
Agreement with the WMAP 5-year measurement of the relic density was
required
such that $\Omega h^2|_{\rm LSP}\leq 0.121$ ($1\sigma$ above the
central
value).  In not employing a lower
bound
on $\Omega h^2|_{\rm LSP}$, the possibility was left open that dark
matter may have
multiple components within or outside of the pMSSM and
thermal relic framework.  Lastly, restrictions on spin independent and
spin
dependent cross sections from
direct DM detection searches were incorporated.
For further details and references on this set of constraints, see
\cite{Berger:2008cq}.

\begin{figure}[htbp]
  \centering
  \includegraphics[width=0.65\textwidth]{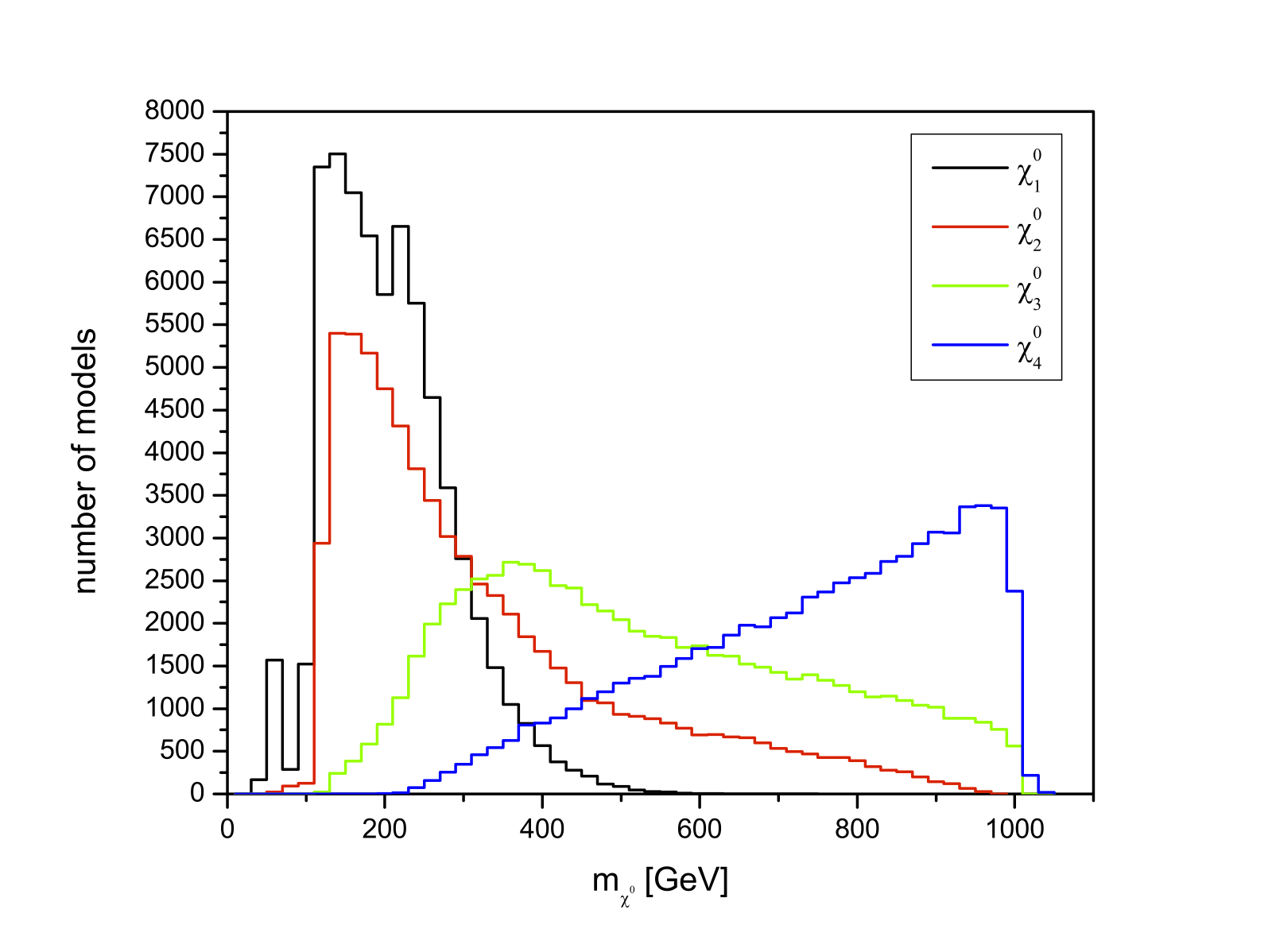}
  \caption{
    Distribution of neutralino masses for the pMSSM model sample.
    From \cite{Berger:2008cq}.}
  \label{lspmassdist}
\end{figure}

After this set of constraints was imposed, $\sim 68.5$k pMSSM models
from
the flat prior sample were found to
survive, satisfying all the restrictions.  Here, we discuss the
attributes of
these models which are most germane to the present analysis.  Figure
\ref{lspmassdist}
(reproduced here from \cite{Berger:2008cq})
presents a histogram of the masses of the four neutralino species for
this
model set.  We see that the LSP mass lies mostly in the range
$100-250$ 
GeV in these models.   Models with a mostly Higgsino or
Wino-like
LSP generally have a chargino with nearly the same mass as the LSP,
and as
sufficiently
light charginos would have been detected at LEP or the Tevatron, there
are
fewer models with an LSP of mass $m_{\LSP}\lsim 100$ GeV.
  The gauge
eigenstate
content of the LSPs in these models is described in Table
\ref{lsptable}.
We emphasize that the percentages given in this Table apply strictly
to this pMSSM model sample and are not related to the likelihood of
models being realized in Nature.
We note that
most $\tilde\chi_1^0$'s are relatively pure eigenstates, with models
where
the LSP is Higgsino or mostly Higgsino being the most common
case.  Within mSUGRA, the LSP is generally close to being a pure Bino,
suggesting
that this pMSSM model set has substantially different properties that
will
affect DM annihilation rates.

\begin{table}
\centering
\begin{tabular}{|c|c|c|} \hline\hline
LSP Type & Definition & Fraction of Models  \\ \hline
Bino & $|Z_{11}|^2 > 0.95$ & 0.14 \\
Mostly Bino & $0.8 < |Z_{11}|^2 \leq 0.95$ & 0.03\\
Wino & $|Z_{12}|^2 > 0.95$ & 0.14 \\
Mostly Wino & $0.8 < |Z_{12}|^2 \leq 0.95$ & 0.09 \\
Higgsino & $|Z_{13}|^2 + |Z_{14}|^2 > 0.95$ & 0.32 \\
Mostly Higgsino & $0.8 < |Z_{13}|^2 + |Z_{14}|^2 \leq  0.95$ & 0.12 \\
All other models & & 0.15 \\ \hline\hline
\end{tabular}
\caption{\small The fraction of the pMSSM model set for which the LSP
  has
the specified gauge eigenstate content.  This is defined by the
  modulus
squared of the elements of the neutralino mixing matrix in the SLHA
  convention.
See \cite{SUSYrev} for details.}
\label{lsptable}
\end{table}

The identity of the next-to-lightest Supersymmetric particle (nLSP) is
quite varied
in this pMSSM model sample.  The lightest chargino is the nLSP in
roughly 78\%
of the models.  This is because a large fraction of models have a Wino
or Higgsino eigenstate LSP (as seen in Table \ref{lsptable})
and there is generally a small mass splitting between
a mostly Wino/Higgsino neutralino and the corresponding chargino.  The
second
lightest neutralino is the nLSP $\sim 6\%$ of the time.  These are
generally
models where the LSP is dominantly Higgsino.  While $\tilde\chi_1^\pm$
and
$\tilde\chi_2^0$ are the nLSP in the vast majority of cases, in the
remaining 16\%
of the model sample 10 other sparticles
are the nLSP with approximately equal probability.  These are the
$\tilde\tau_1\,,
\tilde e(\mu)_R\,, \tilde\nu_\tau\,, \tilde\nu_{e(\mu)}\,, \tilde
b_1\,,
\tilde t_1\,,
\tilde u_{R,L}\,, \tilde d_R$ and $\tilde g$.  Lastly, two
models have
the $\tilde d_L$ as the nLSP.
Scenarios in which these sparticles are the nLSP may lead to
uncharacteristic
DM annihilation rates in various channels (such as a large rate into
$\tau$ pairs),
as well as interesting signatures at the LHC \cite{uslhc}.  Many
models are
found to have small mass splittings between the LSP and nLSP, leading
to potentially
large DM annihilation rates.

\begin{figure}[htbp]
  \centering
  \includegraphics[width=0.65\textwidth]{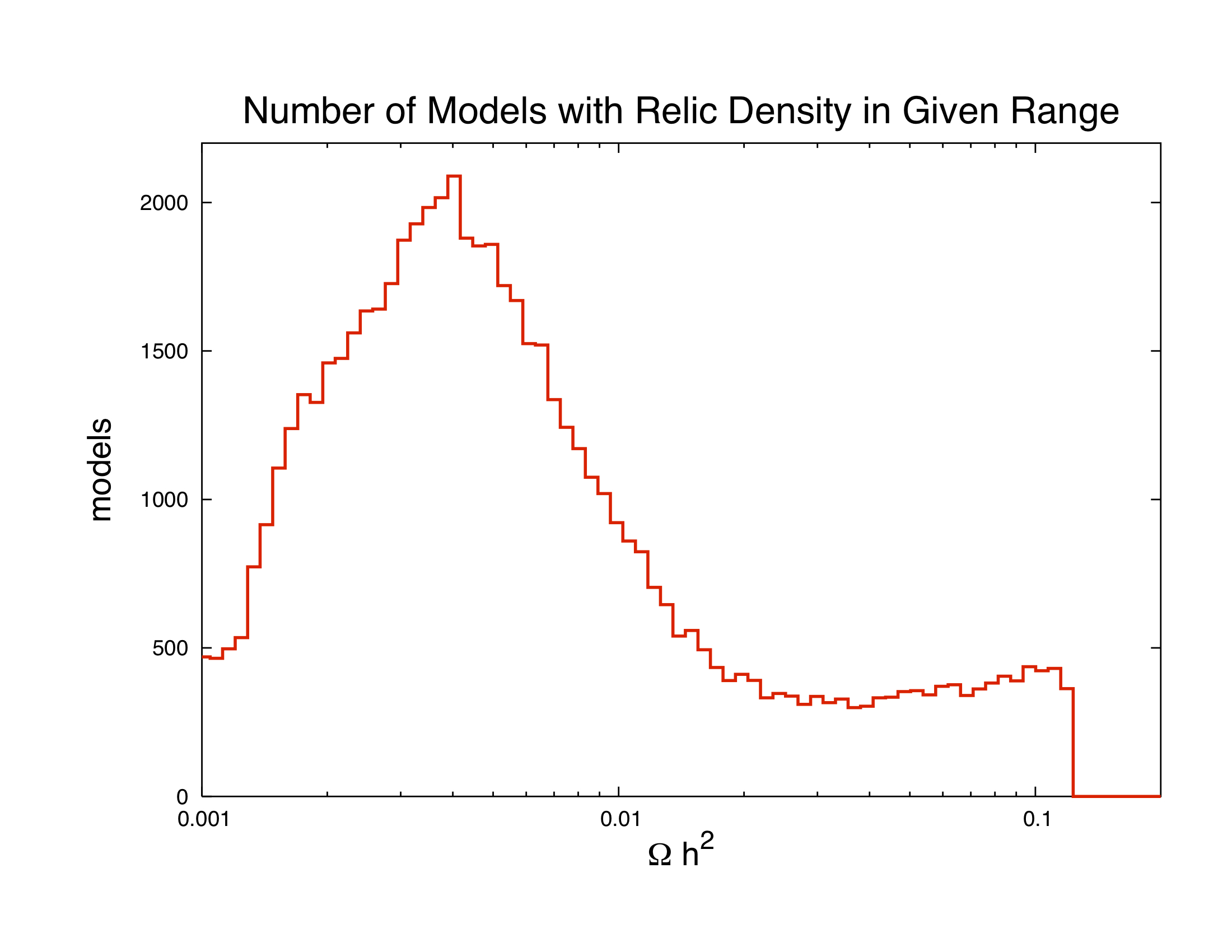}
  \caption{The distribution of predictions for the relic density in
  our pMSSM
    model sample.  From \cite{Berger:2008cq}.}
  \label{lsprelicden}
\end{figure}

\begin{figure}
    \centering
    \subfloat
    {\label{fig:svVmass}\includegraphics[width=0.48\textwidth]{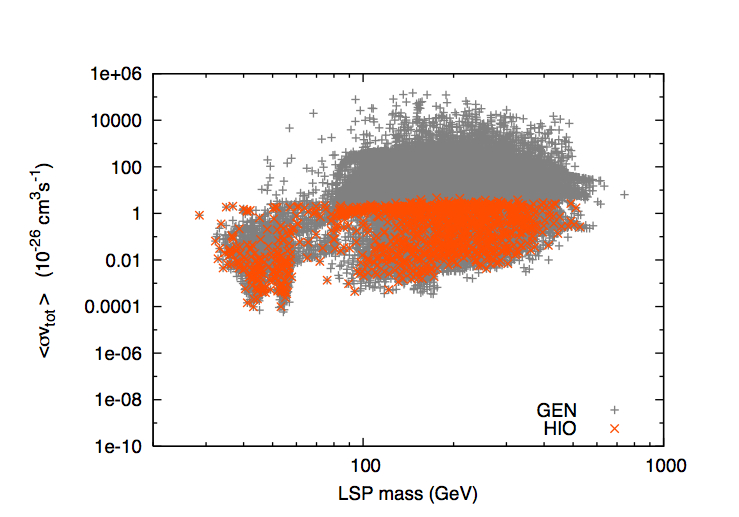}}
    \subfloat
    {\label{fig:svVmass-res}\includegraphics[width=0.48\textwidth]{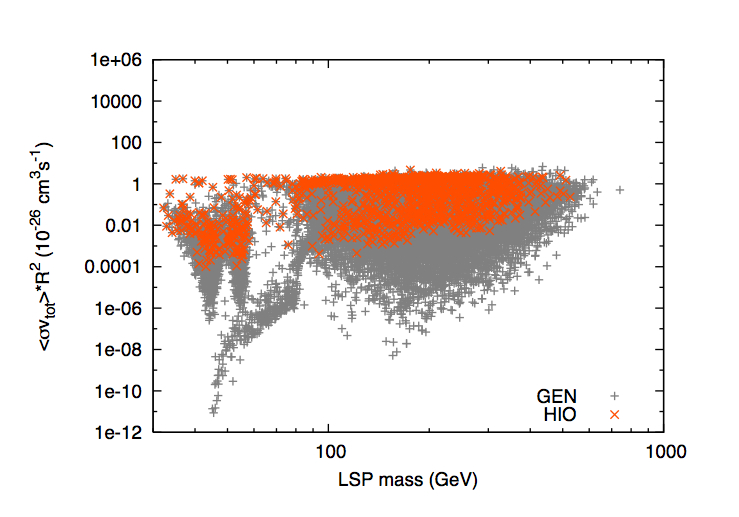}}
    \caption{The unscaled (left-hand-side) and scaled
    (right-hand-side) total annihilation
cross sections for our model set.   The points highlighted in orange
    correspond to the
high-$\Omega$ models.}
    \label{fig:svVmass-both}
  \end{figure}

As stated above, it was not required that the LSP account for all of
the dark
matter. Rather the only constraint was that the LSP relic density not be
so large as
to be inconsistent with WMAP.  Figure
\ref{lsprelicden} (reproduced here from \cite{Berger:2008cq}) displays
the distribution
for the LSP relic density in our pMSSM model set.  Note
that this
distribution is peaked at rather small values of $\Omega h^2|_{LSP}$.
In
particular, the mean value for this quantity is 
$\Omega h^2|_{LSP}\sim 0.012$.  There are
1240 models that produce a relic density in the range $0.10\leq\Omega
h^2|_{LSP}
\leq 0.121$, thus saturating the WMAP measurement; we will refer to
these
hereafter as the
high-$\Omega$ model set.  For many of the observables that we compute
here, we will
need to scale by the ratio
\begin{equation}
    R=\frac{\rho}{\rho_0}=\frac{\Omega h^2|_{\LSP}}{\Omega
    h^2|_{\tiny WMAP}}\,.
    \label{relicrescale}
\end{equation}
This gives the fraction of the local relic density as determined by
WMAP that
arises from Supersymmetric DM.  In our numerical computations, we take
the
central value $\Omega h^2|_{\tiny WMAP}=0.1143$.  The importance of
including
this scale factor is demonstrated in Fig. \ref{fig:svVmass-both},
which
shows both the scaled and unscaled total annihilation cross sections
for our
pMSSM model set.   The points highlighted in orange correspond to the
high-$\Omega$ models.  We see from the figure that this scale factor
is
necessary for consistency with WMAP, which bounds the total
annihilation
cross section.  The quantity $\langle\sigma v\rangle R^2$
is more indicative of the annihilation signals that result from a
given model.

\begin{figure}[hbtp]
    \centering
    \includegraphics[width=0.65\textwidth]{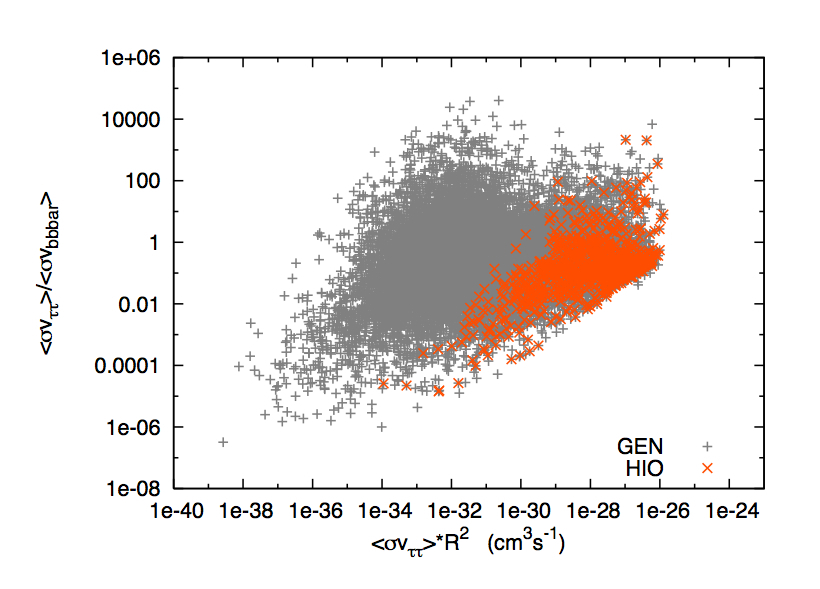}
    \caption{ We display the relation between the annihilation rates
      $\LSP\LSP\rightarrow\tau^{+}\tau^{-}$ and
      $\LSP\LSP\rightarrow b\bar{b}$ for models in our pMSSM model
      set as $\langle\sigma v\rangle_{\tau^{+}\tau^{-}}/\langle\sigma
      v\rangle_{b\bar{b}}$ versus $\langle\sigma
      v\rangle_{\tau^{+}\tau^{-}}R^2$. Here we only show models for
      which $m_{\LSP}>100\gev$. The orange points indicate models
      for which $\Omega h^2|_{LSP}\geq0.1$.}
    \label{figs:allheavyonTauVB}
  \end{figure}

Of particular interest to us below, are the relative DM annihilation
rates
into $\tau$ pairs versus $b\bar b$ final states in our pMSSM model
set.
This
is presented in Figure \ref{figs:allheavyonTauVB} which shows the
ratio
of these two final state annihilation rates versus the scaled rate for
$\langle\sigma
v\rangle_{\tau^+\tau^-}$ for models with $m_{\tilde\chi_1^0}>100$
GeV.  The
points highlighted in orange correspond to the high-$\Omega$ models.
We
note that for the full model sample, the values of the ratio
$\langle\sigma v
\rangle_{\tau^{+}\tau^{-}}/\langle\sigma v\rangle_{b\bar{b}}$ range
over more
than ten orders of magnitude. We see that there are many models with a
high
annihilation rate into $\tau$ pairs at relatively large rates 
with high purity (\ie, the rate into
$b\bar b$
is significantly lower).  These models are perhaps the best candidates
for ``leptophilic'' DM contributions to CR spectra necessary to 
simultaneously explain the $\alle$ and $\pbarp$ fluxes
observed by PAMELA.  In the analysis that follows, we will repeatedly
find reasons to focus on this subset of models.

  \section{Cosmic-Rays}
  \label{sec:CRs}

  We now turn to a discussion of cosmic-ray (CR) signals from the
  annihilation of WIMPs in our pMSSM model set. We pay particular
  attention to the anomalous measurements of CR $e^{\pm}$ species (here 
  termed, ``CRE''), investigating the properties of models that do
  best to alleviate the tension between measurements of the \alle, \pamr
  and \pbarp flux spectra. 

  It is generally expected that the usual thermal cosmological evolution
  of the MSSM particle content should give a DM annihilation
  cross-section of $\tsigv \lsim 3\cdot10^{-26}$
  \cite{Jungman:1995df}, a number well shy of that required for an easy fit to the
  PAMELA measurement of the \pamr spectrum. This number, though necessarily somewhat
  approximate, has served to crystallize the lore that the CRE anomaly
  must be described by non-standard DM models (if by DM at all). In
  this work we test this belief by performing an extensive exploration
  of both the parameter space describing phenomenologically-viable
  supersymmetric models and the space of parameters used to model
  the propagation of cosmic-rays in our galaxy. We will look for
  regions in both spaces of parameters which provide a good
  fit to all relevant data. 

  We will follow the now typical practice of adding a so-called 
  boost factor, that is: $\tsigv \rightarrow B\cdot\tsigv$. While
  boost factors have been motivated by $\eg$, non-thermal 
  cosmological history \cite{gordy} or novel halo 
  distribution effects \cite{Lavalle:2006vb},
  here we will remain agnostic as to their origin and just quantify
  the boost that is necessary to best fit the data with respect
  to the usual scenario: a thermal cosmology and the canonical value
  for the local dark matter density ($0.3\gev$ cm$^{-3}$). We use the 
  same boost factor in adding the SUSY contributions to all 
  measurements, for example in the \pamr and in \pbarp 
fluxes\footnote{This is non-trivial because different quantities probe 
  different regions of the galaxy and so distinct effective boosts
  may be appropriate in the various cases.}. It is not difficult to find
  MSSM models for which boosts of $\mathcal{O}(10^{3}-10^{4})$ allow
  good fits to the data. However,
  it has been argued that effects in the halo distribution 
  alone can only be expected to account for a boost factor of $\sim
  2-3$, due to the uncertainty in the estimation of the local halo
  abundance \cite{Gates:1995dw}. A possible additional factor $\lsim
  10$ could arise from substructure within the Milky-Way halo 
  \cite{Diemand:2008in}, and, for signals from dwarf galaxies,
  there could possibly be a factor $\lsim 100$ \cite{Strigari:2007at}. In this
  work we will focus on models that best fit the data with $B<200$.

  Of course the size of $B$, and the effect of SUSY annihilations in
  general, can only be sensibly determined if one has a good estimate
  of the relevant astrophysical backgrounds. In this respect, the treatment of
  CRE spectra is particularly difficult due to the large uncertainty
  inherent in modeling their sources and propagation
  \cite{Delahaye:2007fr}\cite{Delahaye:2010ji}.
  In this section we discuss how we take account of these uncertainties
  by performing a large scan over some of the astrophysical parameters
  used to model cosmic-ray propagation, thereby creating a variety of 
  astrophysical background spectra that we regard as
  internally consistent and plausible with respect to relevant
  astrophysical measurements.  

  As mentioned in Section \ref{sec:pMSSM} the LSPs in our model
  set typically have $m_{\LSP}\lsim 500\gev$, well shy of what would be
  required to provide a DM explanation to the putative
  anomaly seen by FERMI-LAT in the \alle spectrum near 1 TeV. As has
  already been emphasized \cite{Grasso:2009ma}, the FERMI-LAT
  measurement of the \alle flux can
  be well described in a conventional diffusion model if the injection
  spectra of CREs coming from the average background of CR sources is
  stiffer than previously thought, though this interpretation only 
  exacerbates the apparent anomaly in the PAMELA observation of the
  \pamr spectrum.
  Here we search for a region of astrophysical and pMSSM parameter space such
  that the excess measured by PAMELA in the \pamr spectrum is alleviated by a sizeable
  contribution of positrons from pMSSM DM annihilation while \emph{at the same
  time} the FERMI-LAT measurement of the \alle flux is accommodated by the astrophysical
  background model (without including any particular astrophysical sources, 
  such as pulsars).

  At the end of this section we combine the two scans, over both
  astrophysical and pMSSM parameter spaces, and perform a simultaneous
  fit of the PAMELA \pamr and $(\bar{p}/p)$ flux measurements as well
  as the FERMI-LAT \alle flux measurement. We then discuss the
  properties of both the astrophysical models and the SUSY models that
  lead us to the best global fits with the smallest boost factors.
  
  \subsection{Consistent Modeling of the High-Energy Sky}

  The prevailing theory describing CR propagation throughout the
  galaxy is based on the diffusion model, for which a detailed
  description can be found in
  \cite{Strong:2007nh}\cite{Ginzburg:1990sk}.
  The diffusive paradigm follows naturally from the
  expectation that charged CR species follow tangled paths as they
  scatter resonantly off of the turbulent features in the galactic
  magnetic field whose size match their gyroradii; this explains the observed
  isotropy of CRs and the retention of CR species in the galaxy. 
  In its simplest guise one can solve for the local spectra of a
  single CR species using a model with relatively few parameters. The
  density of a CR species at galactic radius $r$ and
  with momentum $p$, i.e. $\psi(r,p,t)$, can be calculated using the
  CR transport equation \cite{Strong:2007nh}:

  \begin{eqnarray}
    {\frac{\partial \psi (\vec r,p,t) }{ \partial t}} 
    &= &
    q(\vec r, p, t)                                             
    + \vec\nabla \cdot ( \Dxx\vec\nabla\psi - \vec V_c\psi )   \nonumber
    \\
    &   +& {\frac{\partial}{\partial p}} p^2 \Dpp \ddp\, {\frac{1}{p^2}}\, \psi                  
    - {\frac{\partial}{\partial p}} \left[\dot{p} \psi
      - {\frac{p}{3}} \, (\vec\nabla \cdot \vec V_c )\psi\right]
    - {\frac{1}{\tau_f}}\psi - {\frac{1}{\tau_r}}\psi\ .
    \label{diffloss}
  \end{eqnarray}

  One solves this equation over a diffusion zone that is usually
  represented by a thick cylinder whose origin and axis coincide
  with that of our galactic disk and whose half-height is typically 
  in the range $L \sim 1-10\kpc$. The diffusive term has an 
  \emph{energy-dependent} diffusion coefficient,
  $\Dxx=\Dxx(\vec{r},p)$, that describes scattering off of the turbulent
  features of the galactic magnetic field, $\delta B(\vec{r},p)$ (the
  full field being thought of as the sum of smooth and turbulent
  components, $B(\vec{r},p)=B_{sm}(\vec{r},p)+\delta B(\vec{r},p)$). One 
  typically simplifies to a spatially uniform diffusion coefficient with
  free-escape boundary conditions (i.e. $\psi(r,p,t)=0$) at the edges
  of the diffusion zone. The other terms describe effects such as energy losses by
  radiation ($\dot{p}$),  diffusion in momentum space (so-called
  ``re-acceleration,'' \Dpp), interactions of single CR particles
  with the convective ``winds'' created by the bulk population of
  galactic CRs (parameterized by the convection velocity, $\vec{V}_c$),
  and the possible fragmentation and radioactive
  decay of unstable nuclei ($\tau_f$ and $\tau_r$), as well as the
  source term, $q(\vec r, p, t)$. We will describe
  some of these effects in more detail below. 

  One can often use the transport equation and simplify the
  problem enough to obtain explicit analytic solutions
  for individual CR spectra. However, a network of relations
  connect various aspects of the astrophysical environment and
  the spectra of CRs/\gamsn, making a more holistic analysis
  desirable. For example, primary protons\footnote{When referring to CR species the
  terms ``primary'' and ``secondary'' refer to species that have
  either been injected into the Inter-stellar Medium (ISM) via stellar
  expulsion events such as SNe, or to species that have been generated
  in inelastic spallation scatterings between primary species
  (typically protons) and the ISM.}, the most abundant CR
  species, source all of the other species of cosmic-rays (including $e^{\pm}$) by
  spallative interactions with dust and gas. CRE are strongly related
  to a vast spectrum of photons in the galaxy as they generate copious
  amounts of radio photons by synchrotron radiation and gamma-rays by
  Inverse Compton (IC) up-scattering of ambient photons in the
  microwave-to-optical range. The parameters in Equation
  (\ref{diffloss}) represent a simple phenomenological model that
  encodes the effects of these relations on a given CR
  species. However, arbitrary tuning of these parameters without
  consideration of the effect on related quantities may lead to
  meaningless results. Using CREs as an example, one may wish to
  change the term describing electron energy losses in order to see
  the effect on locally measured CRE spectra while forgetting the 
  consequences of this change for previously measured quantities such
  as diffuse gamma-ray (via IC scattering) or radio photon spectra
  (via synchrotron radiation). 

  The GALPROP numerical package is by far the most developed tool
  for consistent cosmic-ray analysis
  \cite{Strong:1998pw}\cite{Moskalenko:1997gh}. The code solves 
  a network of transport equations for Z$\geq$1 nuclei as well
  as for electrons and positrons while computing energy losses using
  realistic maps of galactic gas \cite{Strong:2004td} and of the
  far-infrared, optical and CMB photons that make up the Inter-Stellar
  Radiation Field (ISRF) \cite{Porter:2005qx}. One also has the ability
  to transfer tabulated Green's functions from GALPROP to DarkSUSY
  5.0.4 \cite{Gondolo:2004sc}, where the injected spectra of CREs from dark matter annihilations are
  calculated using all of the details of the SUSY model spectrum and
  couplings. In this way we can propagate the CRE coming from dark 
  matter annihilations in the same manner as those arising from standard 
  astrophysical sources so that signal and background may be combined
  in a consistent manner.   

  \subsection{Designing Astrophysical Models}
  
  Here we investigate astrophysical background uncertainties 
  by carrying out a scan over the parameters associated with cosmic
  ray propagation. Our scan is not ``without prejudice,'' in the sense that we do not scan all
  parameters over their full range of plausible values. In principle there are
  many combinations of parameters that do not allow for a simultaneous fit of
  CRE measurements, with or without a SUSY contribution. We instead
  focus on designing astrophysical models that are both plausible from the
  sense of astrophysical measurements and that have the potential for
  a good fit with a supersymmetric DM contribution and a low boost
  factor. In this Section we describe the process of generating such models.

  We use GALPROP av50.1p and in all cases we generate astrophysical
  models by starting with a conventional 
  model\footnote{ galdef\_50p\_599278, which can be found at \cite{GALPROP}.} 
  and changing only the parameters that are explicitly mentioned below. 

  The connection between locally-measured CR spectra and the nature of
  their inter-stellar propagation is confounded by the fact that
  charged particles are heavily affected by the currents and magnetic
  fields of the heliosphere. This so-called solar modulation
  effect is more important for low energy CRs (see $\eg$,
  \cite{Shikaze:2006je}) and its modeling is currently an active area
  of CR research. In this work, the
  inter-stellar spectra of all species are solar modulated 
  according to a spherical force field approximation
  \cite{Gleeson:1968zz} with potential $\phi=500$GV,
  though in all cases we fit only to data in the range 
  $E>10\gev$ to minimize the importance of uncertainties
  in modeling this effect. This method of computing modulation effects
  does 
  not depend on the sign of charged species, and so factors out 
  in ratios such as \pamr, but nevertheless has some effect
  on fitting absolute spectra ($\eg$, \alle ). We note that,
  while we minimize the effect of solar modulation on our fit to the positron
  fraction by simply leaving out the data bins below $10\gev$,
  previous works (for example, \cite{Bai:2009ka}) have attempted
  to model charge-sign-dependent solar modulation using the data
  presented in \cite{clemetal}. We make no attempt at
  charge-sign-dependent solar modulation in this analysis but note
  that recent work \cite{Tarle}, using data that is significantly more
  recent that that of \cite{clemetal}, has shown that the various positron
  fraction data sets, as well as the predictions of diffusion models
  such as GALPROP, can be brought into accord (at energies \emph{below}
  $10\gev$) in this way.

  The astrophysical parameters that we scan can be classified into related
  subsets: those describing nucleonic CR sources, those that model the diffusion
  of CRs, those describing electron CR sources and those that model electron energy
  losses. In each of these subsets one is constrained by associated
  measured quantities: the proton absolute flux and \pbarp spectrum are
  directly related to the nucleonic CR source distribution, \boverc
  and $^{10}$Be/$^{9}$Be flux spectra ratios are very sensitive to changes in
  the parameters describing diffusion, and the \alle, \pamr and diffuse
  mid-latitude gamma-ray spectra are all strongly affected by changes in
  the electron source and energy loss models. We will describe our
  treatment of each of these subsets of parameters in the subsections
  that follow. The phenomenological fact that Z$\geq$1 nuclei undergo negligible
  radiative energy losses leads to a decoupling of the hadronic quantities
  ($\eg$, \pbarp, \boverc and similar ratios) from the
  astrophysical parameters that affect only the CRE spectra ($\eg$,
  \alle, \pamr and similar quantities), so there is a logical sequence for
  scanning parameters using minimal CPU time. This approach proceeds
  as follows.

  \subsubsection{Nucleon Source and Turbulent Diffusion Parameters}

  As protons are by far the most numerous CR species, their
  absolute spectrum has been observed by many experiments, producing
  data with very little statistical error. We thus begin by finding
  the values of the nucleon source parameters that best fit this
  data. We use the GALPROP default spatial distribution
  for nucleon sources and tune the overall normalization and 
  spectral index ($\ie$, the exponent of the power-law describing
  the energy-dependence of the sources) of these sources to best fit
  measurements of the proton absolute spectrum from BESS/BESS-TeV 
  \cite{Shikaze:2006je}\cite{Haino:2004nq}, AMS-01
  \cite{Alcaraz:2000vp}, CAPRICE98 \cite{Boezio:2002ha} and ATIC 
  \cite{Panov:2006kf}. It is clear that the data favor a (locally-measured)
  spectral index of $\approx2.75\pm0.05$, though the ATIC points stiffen
  somewhat above $\sim1\tev$; including all four data sets, we find a
  best fit spectral index of $\gamma_n=2.73$ and normalization
  $3.96 \times 10^{-6} \:
  \gev^{-1} \: \mathrm{cm}^{-2} \: \mathrm{sr}^{-1} \:
  \mathrm{s}^{-1}$ @ $100\gev$. Once the normalization and spectral
  index are fit in this way, these values are held fixed for the rest
  of our analysis.

  The diffusion sector is described by five parameters: $D_{0xx}$,
  $\delta$, $L$, $V_a$ and $\partial V_c/\partial z$. Here, $D_{0xx}$ and $\delta$
  parameterize the energy-dependent diffusion coefficient
  as\footnote{To be more precise, we use $\Dxx                   
  (\vec{r},p) = D_{0xx}\mathcal{R}^{\delta}$, where $\mathcal{R}=p/Z$
  is the particle rigidity, and so $\mathcal{R}^{\delta} \approx
  (E/cZ)^{\delta}$ for all energies of interest here.}
  $\Dxx (\vec{r},E) = D_{0xx}E^{\delta}$ ($D_{0xx}$ is taken to be 
  spatially constant inside the diffusion zone) and $L$ is the 
  half-height of the diffusion 
  zone. $V_c$ describes the convection of single CRs caused by the bulk
  population of CRs and is taken to be directed vertically away from
  the galactic disk with constant gradient, $\partial V_c/\partial z$.
  $V_a$ is the Alfven velocity; it is related to $\Dpp$ from
  Eq. (\ref{diffloss}) via the approximate relation
  $\Dpp = p^2V_a^2/(9D_{xx})$, and so parameterizes the effect of
  diffusion in momentum space.

  We can gain some intuition for the effects of these parameters in
  the context of a simple 1D diffusion model. We can simplify
  Eq. (\ref{diffloss}) by specializing to stable nuclei ($\tau_r,
  \tau_f \rightarrow \infty$ and $\dot p \approx 0$, as nuclei undergo
  negligible radiative energy losses) and dropping the terms 
  involving $V_a$ and $V_c$, which have a small effect on spectra
  above $\approx 20\gev$
  \cite{Delahaye:2007fr}\cite{Delahaye:2010ji}.
  We are then left with only the spatial diffusion term. Taking the 
  source $q(z,E) = N\delta(z)E^{-\gamma}$, $\ie$, a source localized to
  the galactic plane having spectral index $\gamma$ and normalization
  $N$, Eq. (\ref{diffloss}) becomes:

  \begin{equation}
    {\frac{\partial^2 \psi (z,E) }{ \partial z^2}} = - \frac{N\delta(z)E^{-\gamma}}{D_{0xx}E^{\delta}}	  
    \label{1Ddiffonly} \: .
  \end{equation}  

  \noindent With free-escape boundary conditions, $\psi(\pm L,E) = 0$,
  this is easily solved. We may also obtain a 
  solution for secondary CRs (produced by spallation events 
  involving primary CRs) by substituting the primary spectra
  as a source for secondaries. At our location ($z=0$) we have

  \begin{eqnarray}
    \psi_p (0,E) &=& \frac{N}{2}\bigg(\frac{L}{D_{0xx}}\bigg)E^{-\gamma_{n} -\delta} ,\\
    \psi_s (0,E) &=& \frac{N\sigma}{2}\bigg(\frac{L}{D_{0xx}}\bigg)^2E^{-\gamma_{n} - 2\delta} ,
    \label{1Dspectra}
  \end{eqnarray}
  
  \noindent where the source spectral index, $\gamma=\gamma_{n}$, is
  universal in the GALPROP model for nuclei of all (Z,A). Secondary
  species are injected with index $\gamma_{n}+\delta$ (as they are formed by 
  proton spallation events) and $\sigma$ represents the
  emissivity for primaries moving through the disk. Although this result
  was obtained in a somewhat approximate
  fashion, it is very close to the numerical results
  obtained using GALPROP to propagate stable nuclei. From this we
  note that the ratio of secondary to primary spectra is expected to
  scale as $(secondary)/(primary)\sim (L/D_{0xx})E^{-\delta}$, and thus, for
  example, the ratio of the measurements of boron CRs (which are known to be almost
  entirely secondarily produced) to that for carbon CRs (which are known to be
  dominantly primary in origin) may constrain plausible values of
  $\delta$ and the ratio $(L/D_{0xx})$. In fact the \boverc flux ratio is quite
  sensitive to the choice of diffusion parameters and can tightly
  constrain candidate diffusion models, as it has been
  measured in many experiments with good statistics. We describe our scan in this space
  and the results of our fit to the \boverc flux data below, but first, it is
  worthwhile to make one more observation about this simplified 1D result.

  To consider the propagation of CRE species, which lose energy
  rapidly through radiative processes ($\eg$, synchrotron radiation and
  IC scattering), the 1D model above should also include the radiative
  energy loss term.  Despite this, we can get a feel for how to choose
  $\delta$ for our purposes here, by examining this diffusion-only result.
  To the extent to which the flux ratio $\pamr\sim e^{+}/e^{-}$ (CRE
  electrons far outnumbering CRE positrons), the background-only 
  positron ratio scales as $\pamr\sim
  E^{-\gamma_{n}+\gamma_{e}-\delta}$. Here, the primary electrons are
  injected with source spectral index $\gamma=\gamma_{e}$, while the
  positrons (being formed predominantly by proton CR spallation) are
  injected with spectral index $\gamma=\gamma_n+\delta$. To
  this background we would like to add a primary positron contribution
  from SUSY DM annihilation; this contribution will scale as
  $e^{+}_{susy}\sim E^{-\gamma_{susy}-\delta}$. Hence, from both the
  standpoint of attaining large $e^{+}$ fluxes from DM
  annihilations, and from the standpoint of generating \pamr flux
  backgrounds that will not require large boost factors when
  combined with the DM signal, we would like to choose $\delta$ to
  be as small as possible.

  As $\delta$ describes the spectrum of turbulent features in the 
  bulk magnetic field, there has been a large industry devoted to 
  theoretical and experimental considerations of what its value should be
  \cite{deltalow}\cite{deltahi}. Particularly well-motivated
  theoretical values include $\delta=0.33$ (Kolmogorov turbulence) and
  $\delta=0.5$ (Kraichnan turbulence).
  Experimentally, there are measurements of \boverc and various
  phenomenological propagation models that are used to constrain $\delta$ given
  the measured data. These works have found ranges of values
  for $\delta$ that provide consistency with \boverc measurements,
  some favoring lower values $\delta=0.3-0.5$ \cite{deltalow} 
  and others finding that larger values $\delta=0.5-0.8$ are preferred
  \cite{deltahi}. In this work we use only propagation models that 
  have $\delta=0.33$.

  One should also note that it is not entirely trivial to identify the
  ``$\delta$'' that is being measured by the \boverc data with the
  $\delta$ parameter that effectively describes the diffusion of CRE
  species. This is because CRE energy losses are dominantly radiative,
  with rates proportional to the Thomson cross-section $\sigma_T$, 
  whereas the corresponding radiative loss rate for CR nuclei is 
  proportional to $(Z^2/A)^2(m_e/m_p)^2\sigma_T$. Such losses are thus
  completely negligible for nuclei such as boron and carbon, which lose energy predominantly
  via interactions with matter, and so should be able to diffuse
  substantially throughout the galactic magnetic field from their
  point of creation. Conversely, high-energy ($\gsim 10\gev$) $e^{\pm}$ lose
  energy very efficiently through synchrotron radiation and IC
  scattering and so must come from a region within the local
  $\sim\kpc$ to be detected near our solar system \cite{Delahaye:2007fr}\cite{Delahaye:2010ji}.
  In such a picture the ``$\delta$'' appropriate to fitting \boverc
  effectively describes an average over the entire diffusion zone
  while the $\delta$ parameter appropriate for modeling CRE
  flux spectra near earth is that which describes diffusion in the
  local galactic neighborhood, a scale at which
  diffusion is likely highly complicated\footnote{For instance
    we are conjectured to live on the edge of the ``Local Bubble,''
    a morphologically complex object of size $\sim100$pc near which
    diffusion is likely varied and highly anisotropic \cite{Cox:1987ww}.}.

  We fix $\delta=0.33$ and scan over the remaining diffusion
  parameters, $D_{0xx}$, $L$, $V_a$ and $\partial V_c/\partial z$, to
  find configurations that best fit the \boverc data. Although there are many
  measurements of \boverc spectra at various energies (see
  $\eg$, the list presented in \cite{Simet:2009ne}), we choose to fit
  only the data sets from the CREAM \cite{Ahn:2008my},
  HEAO-3 \cite{Engelmann:1990zz} and ATIC experiments
  \cite{Panov:2007fe}. These sets are the most recent, cover the 
  largest range of energies and have the smallest quoted 
  errors\footnote{The PAMELA Collaboration have also recently made a
    measurement of \boverc (see $\eg$, \cite{picozza:tevpa}) but
    this is unpublished as yet and thus has not been included in
    our analysis.}. 

  As noted above, \boverc provides quite a sensitive constraint on the ratio
  $(L/D_{0xx})$, but cannot separately constrain either $L$ or
  $D_{0xx}$. This degeneracy can be broken by measuring the abundance
  of unstable radioactive nuclei relative to their stable isotopic
  counterparts, $\ie$, flux ratios such as $^{10}$Be/$^{9}$Be, which are referred
  to as ``radio-clocks.'' There are a relatively small number of
  measurements of such ratios, however. Rather than trying to fit
  $^{10}$Be/$^{9}$Be directly in this analysis, we note that
  the data seem to favor $L\gsim 4.0\kpc$ \cite{Strong:1998pw}
  \cite{Regis:2009md}, and choose diffusion zone heights of
  2.0, 3.0, 4.0, 7.0 and 10.0$\kpc$ (with 2.0$\kpc$ being somewhat
  disfavored and with 3.0-7.0$\kpc$ being the most commonly used 
  values \cite{Strong:2007nh}). For each value of $L$, we 
  scan over the remaining parameters, $D_{0xx}$, $V_a$ and 
  $\partial V_c/\partial z$, and compute a \chisq fit to the
  three data sets for each configuration (taking only the 
  bins which cover energies above $10\gev$ to avoid the effects of
  solar modulation), first using a 
  coarse grid and then zeroing in with a finer scan. The resulting 
  best fit configurations are listed in Table \ref{diffpartab}.

  \begin{table}
    \renewcommand{\arraystretch}{1.1}
    \centering
    \begin{tabular}{|c|c|c|c|c|c|c|} \hline\hline 
      $\delta$ &  & 0.33 & 0.33 & 0.33 & 0.33 & 0.33 \\ \hline
      L & \kpc & 2.0 & 3.0 & 4.0 & 7.0 & 10.0 \\ \hline
      $D_{0xx}$ & $\times 10^{28}$ cm$^{2}$/s (@ $\mathcal{R}=4$GV) & 2.83 & 4.20 & 5.40 & 8.25 & 9.97 \\ \hline
      $V_a$ & km/s & 33.67 & 34.33 & 33.67 & 32.83 & 32.00 \\ \hline
      $\partial V_c/\partial z$ & km/s/\kpc & 0.5 & 0.5 & 0.1 & 0.1 & 0.1 \\ \hline
      \hline
    \end{tabular}
    \caption{Best fit parameter configurations in the diffusion sector.}
    \label{diffpartab}
  \end{table}

  It should be noted that ignoring data below $10\gev$, as we do here,
  likely has an impact on the determination of best fit parameters, 
  especially those describing convection and re-acceleration. However, this
  should have little impact on our analysis as the effects of
  convection and re-acceleration are largely unimportant to
  CRE spectra above $10\gev$, at least for commonly accepted sizes of
  these effects \cite{Delahaye:2007fr}\cite{Delahaye:2010ji}.
  Our analysis is similar in this regard to that of \cite{Simet:2009ne},
  especially in the finding that $\partial V_c/\partial z\approx 0$.
  We note that one should refer to more serious attempts at fitting
  \boverc in the $\sim 1\gev$ range for applications that may be 
  particularly sensitive to these effects.

  \subsubsection{ Electron Source and Energy Loss Parameters }
  \label{sec:crepars}

  In the case of nucleonic CRs, where measurements of
  protons have been made with large statistics, and for which
  propagation modeling is relatively simple, the spectral
  index of the sources can be relatively unambiguously connected to the
  locally measured spectral index\footnote{As previously noted, in our 1D model for stable
  nuclei the nucleon source and diffusion terms are the only
  important terms in Eqn. (\ref{diffloss}). Taking $\delta=0.33$ for
  example, and with a locally measured spectral index of
  $\gamma_{local}=2.73$, the spectral index of the nucleon sources is then
  easily found to be $\gamma_{n}=2.40=2.73-0.33$.}. In the case of
  CRE, however, energy losses play an important role in propagation and there is
  much greater uncertainty in relating locally measured fluxes to the
  properties of the sources. We vary the nature of the electron sources
  in a manner similar to that for the nucleon sources, scanning only the overall
  normalization and spectral index. The normalization of the primary
  electron sources is treated somewhat differently than in the nucleon
  case in that we allow an \emph{a posteriori} adjustment of the
  normalization of the source to best-fit the data, as we will
  describe below. We study the nature of CRE energy losses by scanning
  over the overall rates of synchrotron and IC radiative losses by
  tuning the galactic magnetic field and inter-stellar photon
  densities, respectively. We vary the nature of IC energy losses,
  $\ie$, whether the IC scatterings are dominantly in the Thomson
  regime or dominantly in the Klein-Nishina regime, by changing the
  average energy of inter-stellar photons (in practice, the relative
  amounts of far-infrared versus optical photons). In this section we
  give a detailed description of this process.

  Since primary electrons dominate over other (secondarily produced)
  CRE species we expect that the fluxes behave as $\alle\approx e^{-}$
  and $\pamr\approx e^{+}/e^{-}$, at least at energies $\gsim 10
  \gev$. Hence, the choice of primary electron injection index affects
  each quantity in its own way: decreasing $\gamma_e$ will stiffen
  the combined $\alle$ spectrum while \emph{simultaneously softening} the
  energy dependence of the positron fraction. As has been noted previously 
  \cite{Delahaye:2007fr}, the
  uncertainty inherent in $\gamma_e$ can drastically affect the
  visibility of a potential DM annihilation signal in the positron 
  fraction. The FERMI-LAT measurement of the local $\alle$ spectrum
  is consistent with a power-law of spectral index
  $\gamma_{local}\approx3.05$. As was shown in \cite{Grasso:2009ma},
  this result can be interpreted in the context
  of a simple diffusion model with a relatively stiff 
  $\gamma_e=2.42$ (``Model 1'' of \cite{Grasso:2009ma}). In such
  a scenario, however, the expected astrophysical contribution to the 
  positron fraction is even softer than what was previously predicted,
  diving away from the PAMELA $\pamr$ flux data, and further 
  exacerbating the anomaly.  

  Here we scan CRE source spectral indices in the range: 
  $2.42 \leq \gamma_e \leq 2.60$.
  This range is chosen because we want astrophysical backgrounds 
  that will require relatively low boost factors when including a DM
  signal. We will rely on the scan of parameters in the radiative
  loss sector to make up for the softer spectrum arising from this
  range of $\gamma_e$ values to obtain the measured
  dependence, $\alle\approx E^{-3.05}$.

  We also vary the overall normalization of the CRE source, $N_e$.
  However, we do not do this from the outset ($\ie$, in the galdef 
  input files used in GALPROP) as we do for the other parameters, but
  rather during the fitting process and in the calculation of the
  $\chi^2$ fits. This is possible as the overall normalization factors out of the 
  numerical process in GALPROP. Thus $N_e$ can be treated in the same
  manner as the SUSY boost factor: $N_e$ is allowed to float in 
  determining the best-fit $\chi^2$ and we subtract an additional
  degree of freedom in calculating the reduced $\chi^2$.
  
  Let us now discuss CRE energy losses in more detail. At the energies
  with which we are concerned ($\sim 10-100\gev$) CRE losses
  dominantly occur through IC up-scattering of ambient photons and 
  through the synchrotron radiation induced in transiting the
  large scale galactic magnetic field. We can write a general
  expression for such losses as 
  \cite{Moderksi:2005jw}:

  \begin{equation}
    \dot{E}(E) = \dot{E}_{syn}\Big\{ 1 + \frac{(u_0/u_B)}{u_0}\int
    f_{KN}(E,\epsilon_0)u_{\epsilon_0}d\epsilon_0\Big\} \: ,
    \label{radlosses}
  \end{equation}

  \noindent where the first term describes losses 
  via synchrotron radiation, with

  \begin{equation}
    \dot{E}_{syn}=\bigg(\frac{4\sigma_T}{3m_e^2c^3}\bigg) \cdot u_B\cdot E^2 \: ,
    \label{synchloss} 
  \end{equation}

  \noindent where $E=\gamma m_{e}c^2$ and $u_B=B^2/8\pi$. The second
  term describes IC energy losses, with $u_{\epsilon_0}$ being the distribution
  of ambient photons, $\epsilon_0$ is the ambient photon energy scaled
  as $\epsilon_0=h\nu_0/m_ec^2$ and $u_0=\int u_{\epsilon_0}d\epsilon_0$.
  This term would have the Thomson-like energy dependence present
  in the synchrotron term ($\ie$, $\dot{E}_{syn} \sim E^2$), if not
  for the energy dependence of the kernel $f_{KN}(E,\epsilon_0)$. An
  analytical solution is known for $f_{KN}(E,\epsilon_0)$ 
  \cite{Jones:1968zza}, but for our purposes here we merely summarize
  some key features: (i) that $f_{KN}(E,\epsilon_0)<1$, (ii) that
  $f_{KN}(E,\epsilon_0)\approx 1$ in the Thomson regime and (iii) that
  $f_{KN}(E,\epsilon_0)$ falls off significantly as it transitions to 

  \begin{eqnarray}
    f_{KN}(E,\epsilon_0)
    &\simeq&
    \frac{9}{2}\cdot\frac{1}{(4\epsilon_{0}\gamma)^2}\Big\{ln(4\epsilon_{0}\gamma)-11/6\Big\}
    \label{knlike}
  \end{eqnarray}

  \noindent in the Klein-Nishina (KN) regime (where
  $4\epsilon_{0}\gamma\gg 1$).

  In the idealized case of a monochromatic photon ($\epsilon_0$) bath
  and a low-enough magnetic field, this transition of
  $f_{KN}(E,\epsilon_0)$ would become manifest in the CRE spectra as
  a ``step'' feature \cite{Schlickeiser:2009qq}. For CREs with energy
  high enough (w.r.t. photons of energy $\epsilon_0$) to be in the KN
  regime, the spectral index stiffens relative to the Thomson result.
  As energies increase, the overall energy loss rate goes from being dominantly due to IC
  scattering to being dominantly due to synchrotron radiation (whose
  contribution always grows as $\dot{E}_{syn} \sim E^2$), and the CRE
  spectral index returns to a Thomson-like slope (with normalization 
  larger by a factor $u_{\epsilon_0}/u_B$). The transition between these
  two regimes occurs at $\epsilon_0\approx(1/4\gamma)$, corresponding 
  to photons of energy $\approx 1\ev$ ($\approx 10\ev$) for electrons 
  of energy $\approx 100\gev$ ($\approx 10\gev$), respectively. 

  Of course the Milky Way ISRF is more
  complicated than this monochromatic idealization, being composed of starlight
  photons from an array of different types of stars, of lower energy
  photons reprocessed by warm dust and gas and also the CMB. Energies
  typical for starlight photons in our galaxy are $\approx 1-5\ev$, while
  photons from hot matter and the CMB are less energetic,
  $\approx 0.1\ev$. In this case the shape of CRE spectra at a given energy
  reflects the combination of energy losses off of various components of the
  ISRF (as well as of the bulk magnetic field) and the issue of
  whether IC losses are dominantly Thomson-like or dominantly KN-like
  essentially becomes a question of the relative intensity of $0.1\ev$
  photons to $1.0\ev$ photons in our local few \kpc. 

  This point has been studied in \cite{Stawarz:2009ig}, where the
  authors discuss the impact of tuning the relative amounts of $1\ev$
  and $0.1\ev$ photons in the context of an analytical model. They note
  that by using parameter choices that could plausibly describe our astrophysical
  environment, one may generically expect a ``spectral pileup'' (a
  feature that ends up looking rather more like a ``bump'' than a
  ``step'') in the \alle spectrum in the neighborhood of a few hundred
  \gev. Their work demonstrated that one could explain any small
  excess that may exist in the FERMI-LAT \alle measurement around
  $\sim 1\tev$ while noting that the relatively large and sharp 
  excess apparent in the ATIC \alle measurement could not be described
  in this way. They also demonstrated that the 
  putative anomaly represented in the PAMELA \pamr data cannot be 
  simultaneously explained without tuning astrophysical parameters
  (such as the energy density of starlight and matter density in the
  ISM) away from their estimated values by several orders of magnitude.

  Despite the fact that tuning the ISRF does not plausibly
  explain the anomalous positron fraction on its own, one certainly
  sees that the uncertainty in the description of the ISRF can have a
  large effect on CRE spectra. We thus expect that a tuning of
  these parameters can be helpful in our present aims. In varying
  more parameters than just $\gamma_e$ we will be afforded more options
  with which to fit the FERMI-LAT \alle data, each of
  which necessitates a particular propagation of the signal and
  background positrons that also enters the calculation of the \pamr
  spectrum. 

  To implement this customization we use the ``ISRF-Factors'' that are
  available as parameters in the GALPROP galdef input files. The
  ISRF is implemented in GALPROP as a multidimensional array split
  into three components that represent maps of optical (starlight),
  FIR (reprocessed) and CMB photons as generated in the work
  \cite{Porter:2005qx}. These ISRF-Factors, referred to as
  $\mathcal{F}_{op}$, $\mathcal{F}_{FIR}$ and $\mathcal{F}_{CMB}$
  here, are simply global normalizations of the three
  components\footnote{It should be noted that in the public GALPROP
  v50.1p source code these factors are applied to the ISRF components
  only after the CRE spectra have been calculated. This is 
  inconsistent as changes in the ISRF via these input variables
  should have an impact on the overall rate and nature of CRE energy
  losses via IC scattering. We use a modified version of the code
  that accounts for this.}. Here, we scan the sum
  ($\mathcal{F}_{op}+\mathcal{F}_{FIR}$) to vary the overall IC loss
  rate and we do not vary $\mathcal{F}_{CMB}$ as there is little
  uncertainty in the description of the CMB. We scan the
  normalization of the bulk magnetic field, $\mathcal{F}_{B}$, to
  vary the overall synchrotron loss rate. We also vary the ratio
  ($\mathcal{F}_{op}/\mathcal{F}_{FIR}$) to study the
  Thomson/KN nature of IC losses. We make no serious attempt to
  estimate the uncertainties in these quantities but simply limit
  ourselves to values within an order of magnitude of the GALPROP 
  default set. The exact values used in this scan are given in 
  Table \ref{losspartab}.

  \begin{table}
    \centering
    \renewcommand{\arraystretch}{1.1}
    \begin{tabular}{|c|c|} \hline\hline
      $\gamma_e$ & \textbf{2.42}, 2.45, 2.48, 2.51, 2.54, 2.57, 2.60 \\ \hline
      $\mathcal{F}_{B} \cdot B(r=0,z=0)$ ($\mu$G) & 0.2, 0.4, 0.6, 0.8, 1.0, 2.0, 3.0, 4.0, \textbf{5.0}  \\ \hline
      $\mathcal{F}_{op}+\mathcal{F}_{FIR}$ & 0.5, 1.0, \textbf{2.0} \\ \hline
      $\mathcal{F}_{op}/\mathcal{F}_{FIR}$ & 0.1, 0.25, 0.5,
      \textbf{1.0}, 2.0, 4.0, 10.0 \\ \hline
      \hline
    \end{tabular}
    \caption{ Parameter scan ranges in the CRE source and
    radiative energy loss sector. Parameters in bold type are default
    (i.e. as in galdef\_50p\_599278, except in the case of $\gamma_e$
    where $\gamma_e=2.42$ is default in the sense that it matches the benchmark
    ``Model 1'' of \cite{Grasso:2009ma}). }
    \label{losspartab}
  \end{table}

  Finally, we note that tuning the parameters that describe
  CRE sources and propagation will also affect the observed
  gamma-ray and radio photon spectra. The measured diffuse gamma-ray
  spectrum has many components: the IC and
  bremsstrahlung scattering of the galactic CREs, the decay of
  hadronically-produced $\pi^0$ mesons, known ($\ie$, resolved)
  energetic point sources, an isotropic population of \gams from 
  extragalactic processes (such as unresolved energetic point-sources)
  and, in practice, contamination due to detector mis-identification 
  of charged particles. Each of these contributions comes with its own
  uncertainties, whether from modeling the physical process involved
  or from the measurements of relevant data, that make the combination
  of components a non-trivial exercise.
  
  In what follows we will show the effects of tuning these parameters
  on the spectrum of diffuse gamma-rays at mid-latitudes\footnote{In galactic coordinates this is
  $0^{\circ}\leq l < 360^{\circ}$ and $10^{\circ}\leq |b| \leq
  20^{\circ}$, where the galactic longitude $l$ is measured in the
  plane of the galactic disk from the sun-galactic-center axis and the
  galactic latitude $b$ is the angle measured in a plane perpendicular to that
  of the disk, containing the sun-galactic-center axis, with the sun
  at its vertex.}. We note, however, that this calculation is
  approximate as the modeling of the various components that make the diffuse
  mid-latitude gamma-ray spectrum is still very much an active area of
  research. In constructing our custom astrophysical models we started
  with a model (galdef\_50p\_599278) that was designed to fit
  the EGRET measurement \cite{Hunger:1997we} of the diffuse mid-latitude gammas 
  (see $\eg$, \cite{Ptuskin:2005ax}), which is now known to be in tension with the
  more recent FERMI-LAT measurements \cite{Abdo:2009mr}\cite{tporter:sympdiff}.
  In addition, the work \cite{Kamae:2006bf} has also presented a more accurate parameterization of the gamma
  spectra arising from the decays of spallatively-produced $\pi^0$'s,
  which is not yet incorporated into the latest public version of
  GALPROP. An as-yet unreleased version of GALPROP is expected to
  include this and other data and may also be accommodated by a new
  conventional astrophysical model. A recent analysis, \cite{Cumberbatch:2010ii}, discusses
  the importance of uncertainties in the calculation of the diffuse
  mid-latitude gamma-ray spectrum in the context of constraining
  contributions from the annihilation of SUSY DM.

  \subsubsection{ Astrophysical Scan Results }
  
  Before discussing our global fits that include the SUSY DM 
  contributions, we can get a sense of what we have already 
  accomplished by tuning the astrophysical parameters alone. Performing
  all of the steps described above leads to a set of 6615 astrophysical
  models. From these we take a smaller subset (524 models) to be the
  ones that we eventually combine with each of our 
  $\approx 69$k models from the pMSSM, so that the combined analysis
  could be done with a reasonable amount of CPU time. 

\begin{figure}[hbtp]
    \centering
    \includegraphics[width=0.70\textwidth]{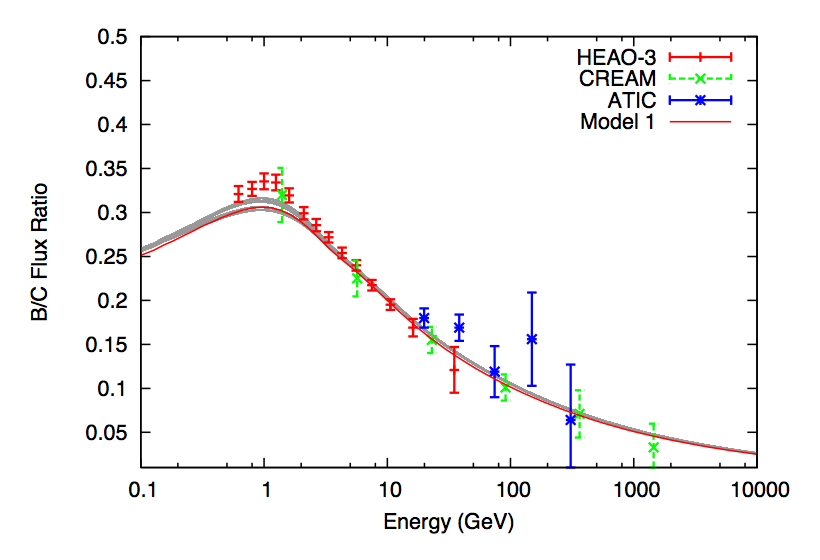}
    \caption{\boverc spectra for all 524
    astrophysical models with no SUSY contributions (grey curves).
    Also shown is the \boverc spectrum for the benchmark model
    (red curve).  Data shown are from the CREAM \cite{Ahn:2008my},
      HEAO-3 \cite{Engelmann:1990zz} and ATIC experiments
    \cite{Panov:2007fe}. 
    }
    \label{figs:bgboverc}
  \end{figure}

\begin{figure}[hbtp]
    \centering
    \includegraphics[width=0.70\textwidth]{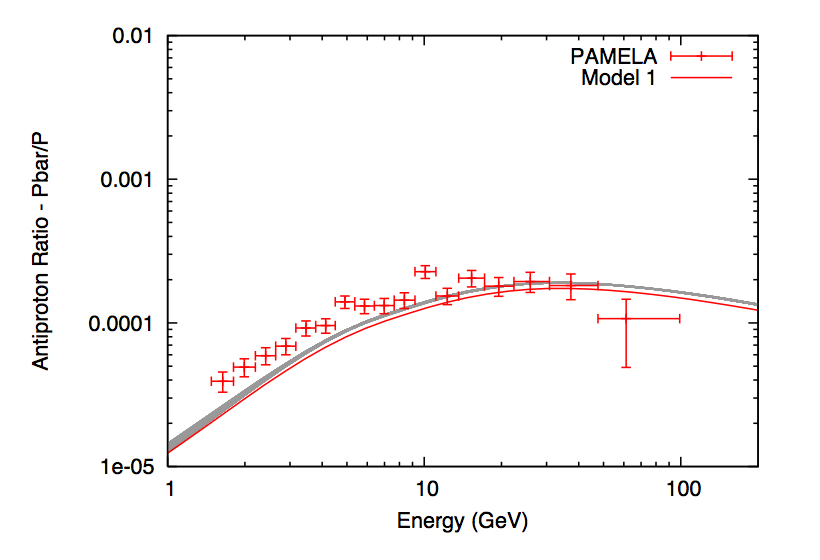}
    \caption{\pbarp spectra for all 524 astrophysical 
      models with no SUSY contributions (grey curves). Also
      shown is the \pbarp spectrum for the benchmark model (red
      curve), along with the PAMELA \pbarp data \cite{Adriani:2008zq}.
    }
    \label{figs:bgpbarp}
  \end{figure}

\begin{figure}[hbtp]
    \centering
    \includegraphics[width=0.70\textwidth]{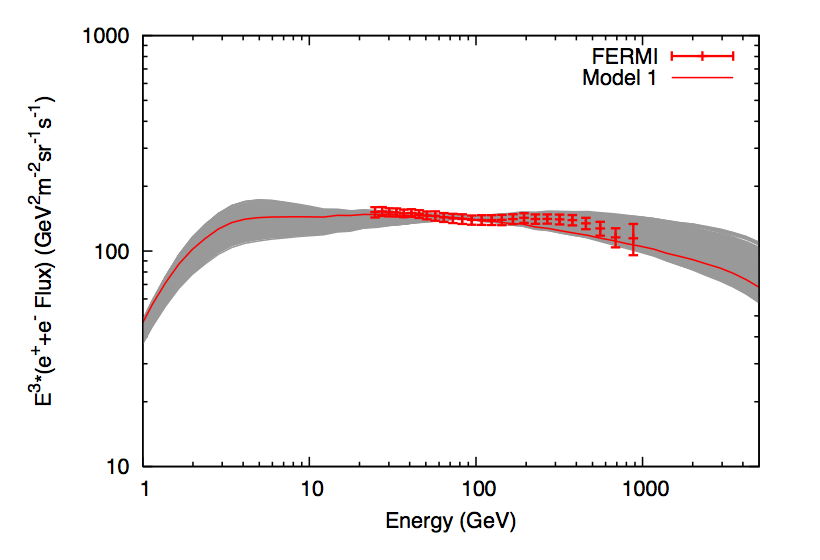}
    \caption{\alle spectra for all 524 astrophysical
      models with no SUSY contributions (grey curves). Also
      shown is the \alle spectrum for the benchmark model (red
      curve). The FERMI \alle data \cite{Abdo:2009zk} is also shown.
    }
    \label{figs:bgalle}
  \end{figure}

\begin{figure}[hbtp]
    \centering
    \includegraphics[width=0.70\textwidth]{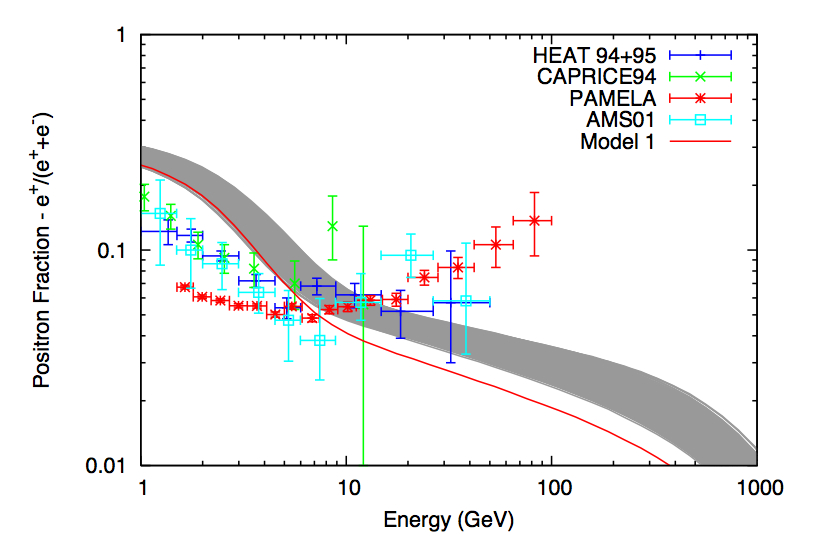}
    \caption{\pamr spectra for all 524 astrophysical
      models with no SUSY contributions (grey curves). Also
      shown is the \pamr spectrum for the benchmark model (red
      curve). Although we only fit to data from PAMELA
      \cite{Adriani:2008zr} we include also data from
      HEAT\cite{DuVernois:2001bb}, AMS01\cite{Aguilar:2002ad}, and
      CAPRICE94\cite{CAPRICE}, for comparison.
    }
    \label{figs:bgpep}
  \end{figure}

  Figures \ref{figs:bgboverc}-\ref{figs:bgpep} show the results for each
  of these models in comparison to the \boverc\nolinebreak,\linebreak \alle, \pamr and \pbarp
  data, respectively. In each figure we include as a benchmark the astrophysical
  model, ``Model 1'' of \cite{Grasso:2009ma}. One can see that there
  is far less variation in the hadronic quantities than in the
  leptonic ones, as leptonic propagation parameters are far less 
  constrained. In fact, we find that all of our 524 astrophysical
  models provide an excellent fit to the \boverc and \pbarp data, that
  is as good, if not better, than the benchmark model. Our models also
  afford a good description of the \alle data in the energy range
  measured by the FERMI-LAT. In the case of the positron fraction, our
  models have a harder spectrum at high energies compared to the
  benchmark case and thus have the potential to provide a better fit
  to the data once the supersymmetric contributions are included.

\begin{figure}[hbtp]
    \centering
    \includegraphics[width=0.70\textwidth]{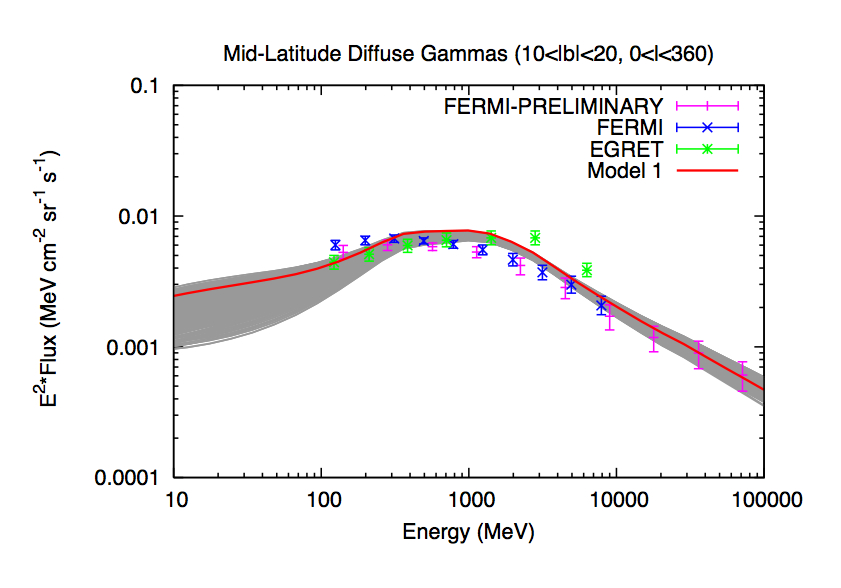}
    \caption{ Mid-latitude diffuse gamma spectra for all 524
    astrophysical models with no SUSY contributions (grey curves).
    Also shown is the spectrum for the benchmark model (red curve). We
    display for comparison the EGRET \cite{Hunger:1997we} and FERMI 
    \cite{Abdo:2009mr} data, as well as preliminary FERMI data
    presented in \cite{tporter:sympdiff}.
    }
    \label{figs:bggams}
  \end{figure}

  In Figure \ref{figs:bggams} we show the spectra of diffuse
  gamma-rays at mid-latitudes that result from these models along with
  the spectrum calculated with the benchmark astrophysical model. We
  emphasize that this is an approximate calculation, containing many
  uncertainties, but is nonetheless suitable
  for comparing the results from our custom astrophysical models
  with those of the benchmark. Due to these uncertainties, we do
  not include this data in our 
  calculation of the global $\chisq$ fit. Each curve includes IC, 
  bremsstrahlung and $\pi^0$-decay
  contributions. One can see that our custom models exhibit more 
  variation at low energies than they do at high energies. This is
  again because leptonic quantities exhibit more variation than
  hadronic quantities: the low-energy portion of the spectrum
  is dominated by gammas produced in IC up-scattering of ISRF photons
  by galactic CREs while the high-energy portion of the spectra is 
  dominated by secondarily-produced $\pi^0$-decay. 

\begin{figure}[hbtp]
    \centering
    \includegraphics[width=0.70\textwidth]{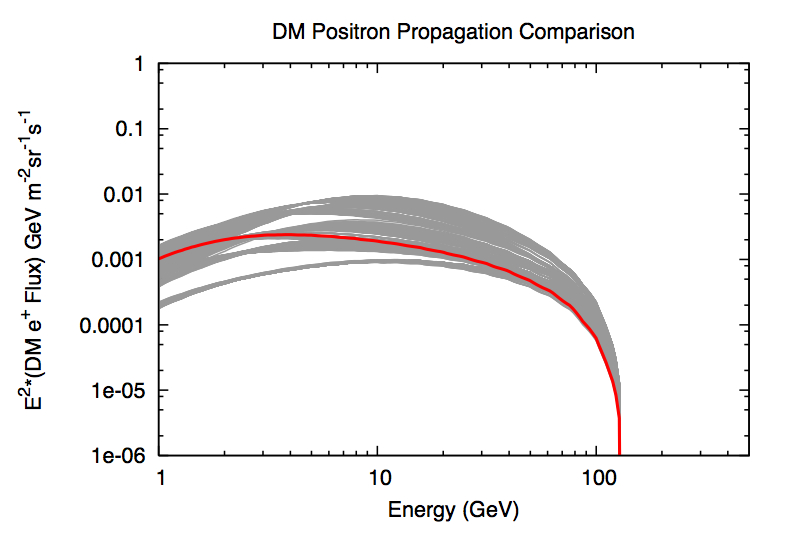}
    \caption{ The propagated DM positron flux (scaled by $E^2$) from a representative
    pMSSM model for each of our 524 astrophysical models (grey
    curves). Also shown is the corresponding DM positron signal from this pMSSM
    model propagated in the benchmark ``Model 1'' (red curve). The pMSSM
    model has an LSP with $m_{\LSP}=128\gev$, which annihilates almost
    purely to $\tau$ pairs.
    }
    \label{figs:proppep}
  \end{figure}

\begin{figure}[hbtp]
    \centering
    \includegraphics[width=0.70\textwidth]{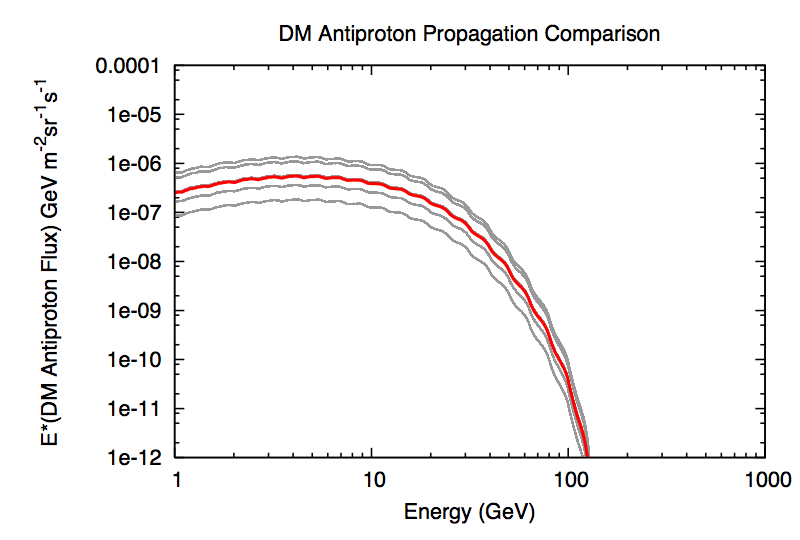}
    \caption{ The propagated DM antiproton flux (scaled by $E^2$) from a representative
    pMSSM model for each of our 524 astrophysical models (grey
    curves). Also shown is the corresponding DM antiproton signal from this pMSSM
    model propagated in the benchmark ``Model 1'' (red curve). The
    pMSSM model has an LSP with $m_{\LSP}=128\gev$, which annihilates almost purely to
    $\tau$ pairs.
    }
    \label{figs:propppb}
  \end{figure}

  As emphasized above, we propagate our pMSSM DM signal contributions
  using Green's functions appropriate to each set of parameters
  employed in the computation of the astrophysical background spectra.
  Figures \ref{figs:proppep}-\ref{figs:propppb} demonstrate how the DM
  signal predictions change as we scan the astrophysical parameter
  space. Taking one representative SUSY model (a model with
  $m_{\LSP}=128\gev$ that annihilates almost purely to 
  $\tau$ pairs) we display both the positron and antiproton spectra
  from the DM halo annihilation in each of the 524 astrophysical
  models. We employ the DarkSUSY default NFW profile
  \cite{NFW} to calculate all of the signals.
  The choice of halo profile is found to have little impact on 
  the positron signals, but may have a sizeable impact on the 
  antiproton signals (as commonly used profiles differ largely 
  only within $\approx 1\kpc$ of the galactic center). Nevertheless,
  the NFW cusp profile gives results that lie in between other 
  commonly chosen profiles and should be a reasonable standard 
  for our purposes.

\begin{figure}[hbtp]
    \centering
    \includegraphics[width=0.75\textwidth]{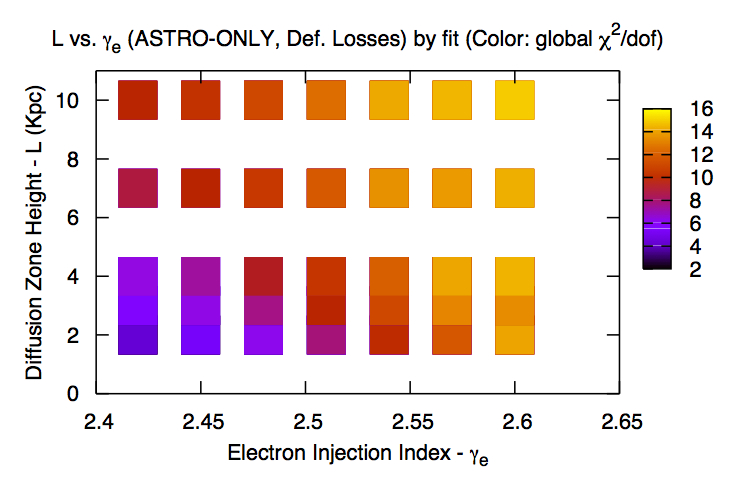}
    \caption{ The $\chisq$ value of the fit (without SUSY
      contributions) in the L-$\gamma_e$ plane, allowing only the
      default IC energy loss parameters. The stiffest injection index 
      ($\gamma_e=2.42$) and smallest diffusion zone ($L=2.0\kpc$)
      are heavily favored over the other configurations. 
    }
    \label{figs:height-gammae-deflosses}
  \end{figure}

\begin{figure}[hbtp]
    \centering
    \includegraphics[width=0.75\textwidth]{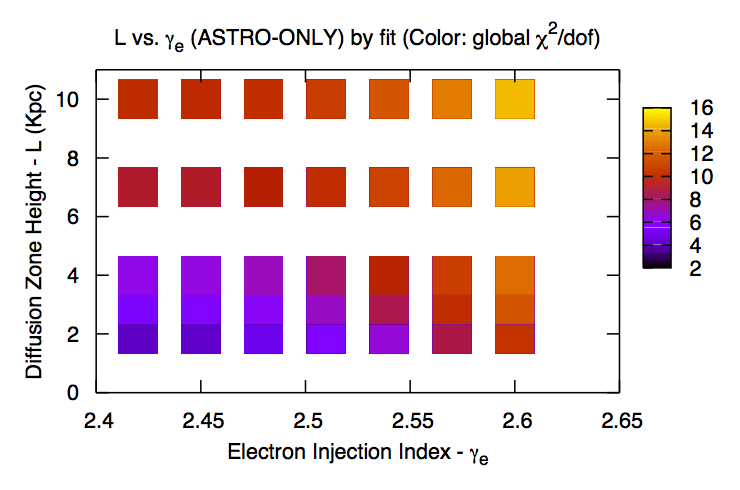}
    \caption{The $\chisq$ value of the fit (without SUSY  
      contributions) in the L-$\gamma_e$ plane, with all combinations
      of IC energy loss parameters. The result is similar to that 
      obtained with only default parameters but the best fit region is
      significantly broadened, allowing for good fits with
      larger values of $\gamma_e$.
    }
    \label{figs:height-gammae}
  \end{figure}

  While the benchmark ``Model 1'' illustrated that the FERMI-LAT \alle measurement
  could be accommodated simply by choosing a relatively stiff
  $\gamma_e$, our models show that the \alle data can also
  be accommodated in diffusion models for which the (background-only)
  \pamr ratio comes significantly closer to the recent 
  (anomalously large) PAMELA measurement. This occurs due to allowing
  for non-default electron energy loss parameters. We illustrate
  this point in Figures \ref{figs:height-gammae-deflosses}-\ref{figs:height-gammae}.
  Here, we color-code according to the value of resulting \chisq for
  the fit in the Diffusion Zone Height versus $\gamma_e$ plane,
  first taking only the subset of astrophysical models with default
  IC energy loss parameters (Fig. \ref{figs:height-gammae-deflosses})
  and then showing the results for all 524 models 
  (Fig. \ref{figs:height-gammae}). Although the best fits
  in each panel are found by taking a value for $\gamma_e$
  which is as low as possible, one can see that allowing 
  non-standard IC loss parameters widens the region of the best
  fit significantly. It is also apparent in Figures 
  \ref{figs:height-gammae-deflosses}-\ref{figs:height-gammae}
  that smaller diffusion zones yield markedly better fits to the data than
  do larger diffusion zones. We emphasize that these are \emph{astrophysics-only
  fits}; in the next section we will show that the astrophysical
  models that give the best fits with SUSY contributions included
  have a somewhat softer $\gamma_e\approx 2.51-2.55$.

  \begin{figure}[hbtp]
    \centering
    \includegraphics[width=0.65\textwidth]{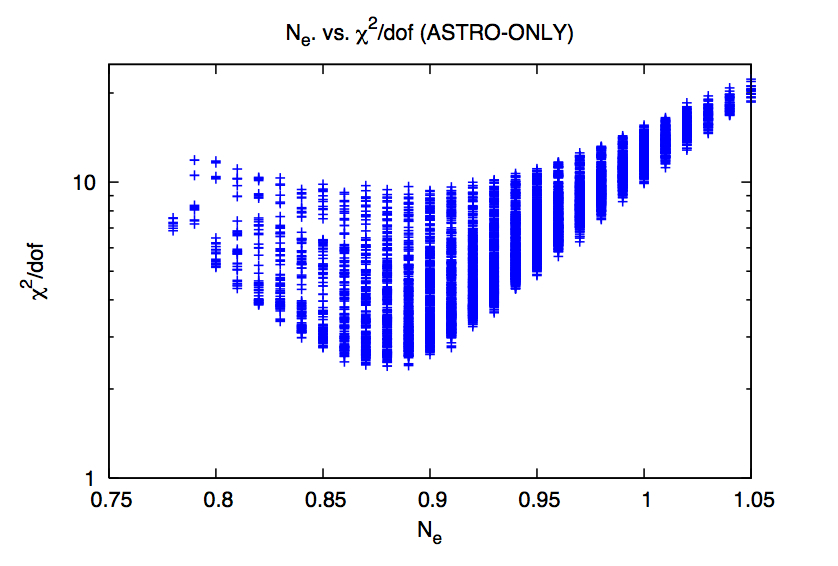}
    \caption{The relation between the global \rchisq values and
    best-fit $N_e$ values for all 524 astrophysical models with no
    SUSY contributions. The $N_e$ values on the horizontal axis should
    be multiplied by $0.4 \times 10^{-6}$ to obtain the flux of primary
    electrons of energy $34.5\gev$ in units of
    $\gev^{-1} \: \mathrm{cm}^{-2} \: \mathrm{sr}^{-1} \:
    \mathrm{s}^{-1}$. The horizontal gaps between points reflect the
    resolution of our scan over $N_e$.
    }
    \label{figs:chitot-enorm}
  \end{figure}

  \begin{figure}[hbtp]
    \centering
    \includegraphics[width=0.75\textwidth]{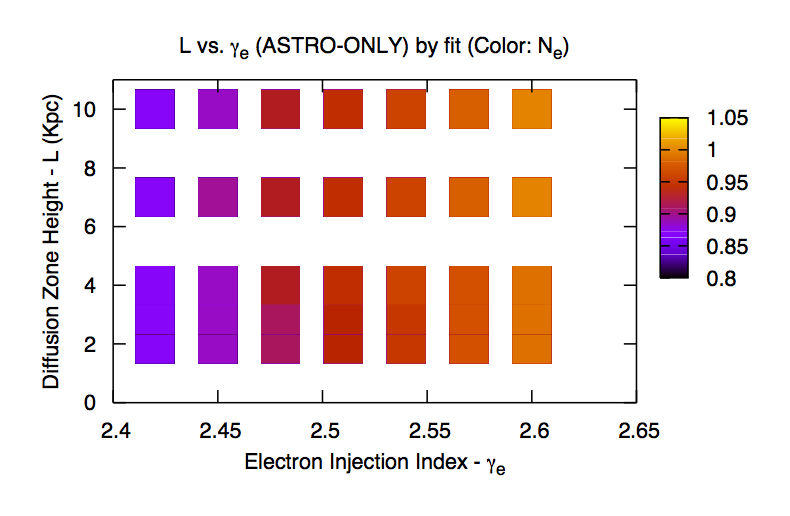}
    \caption{
      The value of the best-fit $N_e$ (without SUSY
      contributions) in the L-$\gamma_e$ plane. The $N_e$ values
      quoted here should be multiplied by $0.4 \times 10^{-6}$ to obtain
      the flux of primary electrons of energy $34.5\gev$ in units
      of $\gev^{-1} \: \mathrm{cm}^{-2} \: \mathrm{sr}^{-1} \:
      \mathrm{s}^{-1}$. We note that the color gradient in this figure
      is similar to that in Figs. 
      \ref{figs:height-gammae-deflosses}-\ref{figs:height-gammae},
      which were color-coded by the value of $\rchisq$.
    }
    \label{figs:height-gammae-enorm}
  \end{figure}

  Another important factor in determining the global best fit is
  $N_{e}$ the normalization of primary electrons. This normalization
  is mostly set by the need to fit the FERMI-LAT \alle measurement, as
  this data set has relatively many bins with relatively small errors
  as compared to the PAMELA \pamr data. The primary electron normalization
  obviously has a large impact on the resulting positron fraction
  curve, and hence a large impact on the portion of the fit due to the
  PAMELA \pamr data, especially from the low-energy data
  bins\footnote{Specifically: the lowest-energy bins \emph{above 10
      GeV}, i.e. the fifth and sixth highest-energy bins from the 
    entire data set, \cite{Adriani:2008zr}.}. Even while varying 
  other parameters, as we do here, a relatively small window of 
  $N_{e}$ values yield a good fit to both data sets. This is
  illustrated in Figure \ref{figs:chitot-enorm}, where we show the global
  \rchisq versus $N_e$. In figure \ref{figs:height-gammae-enorm} we
  color-code according to the primary electron
  normalization factor $N_e$ (rather than the global \chisq)
  in the Diffusion Zone Height versus $\gamma_e$ plane. We observe
  that the $N_e$ values corresponding to good fits to both the 
  FERMI-LAT \alle and the PAMELA \pamr data are found in the low
  $\gamma_e$ and small diffusion zone height corner of the plane.

  We again emphasize that these results include \emph{only} standard
  astrophysical contributions and that, with SUSY DM contributions
  included the primary electron normalization is allowed to float
  along with the boost-factor in order to determine the best fit. We
  now turn to a discussion of the results of the combined scan. 

  \subsection{Results From Combined SUSY/Astrophysical Scans}
  \label{sec:results}
  
  In this section we describe the results of a global \chisq to the CR
  data set for each combination of models in our dual scan over the
  astrophysical and pMSSM parameter spaces. The fit includes
  all ($\geq 10\gev$) data bins from the FERMI-LAT \alle
  measurement as well as the PAMELA \pamr and \pbarp measurements (26, 6 and
  6 bins, respectively). Errors for the FERMI-LAT data set are derived
  from \cite{Abdo:2009zk} by adding statistical and systematic errors
  in quadrature while those for the PAMELA data sets are purely 
  statistical, as quoted by the collaboration in \cite{Adriani:2008zr} and
  \cite{Adriani:2008zq}.   The contributions to individual
  experimental bins from the theory spectra obtained from GALPROP/DarkSUSY
  are determined by numerically integrating the curves over the
  relevant bin. We allow the values of the boost factor, $B$, and
  of the normalization of primary electrons, $N_e$, to float in order to
  obtain the best fit. This procedure leaves us with 36 degrees of
  freedom in calculating \rchisq.

\begin{figure}[hbtp]
    \centering
    \includegraphics[width=0.75\textwidth]{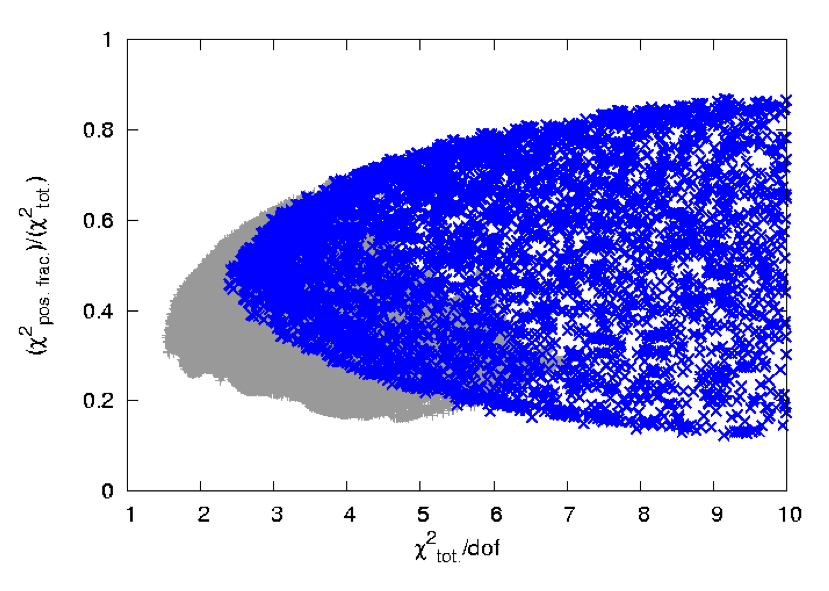}
    \caption{Fraction of the full \rchisq coming from the PAMELA
    \pamr data, $\chisq_{\mathrm{pos. frac.}}/\chisq_{\mathrm{tot}}$, versus the global
    \rchisq. The astrophysical background-only fits (blue) reach a minimum
    $\rchisq\approx 2.39$ while the fits which include a SUSY DM
    contribution (grey) reach a minimum $\rchisq\approx 1.55$. These
    fits were performed for each combination of our $\sim$69k pMSSM models
    and 524 astrophysical background models, though all of the grey
    points displayed here correspond to combinations which find their
    best fits for $B<200$.
    }
    \label{figs:susyvsnosusypep}
  \end{figure}

  We first examine the effect of a supersymmetric DM
  contribution to the global fit. Figure \ref{figs:susyvsnosusypep}
  shows the fraction of \rchisq due to the fit to PAMELA \pamr
  data alone as a function of the \rchisq value for the global fit. In
  this figure, the grey points correspond to fits including both
  supersymmetric and astrophysical contributions for the combinations
  of our $\sim$69k pMSSM models and 524 astrophysical models for which
  best-fits occur with boost factors $B<200$. The blue points
  represent the results that can be attained with astrophysical
  contributions alone. It is clear from the figure that the addition
  of supersymmetric DM can improve the fit to the data over the pure
  astrophysical case. The $\rchisq$ fits including SUSY DM reach a
  minimum of $\rchisq=1.55$, whereas the astrophysics-only fits obtain
  values reaching only as low as $\rchisq \approx 2.39$. Figure 
  \ref{figs:susyvsnosusypep} also suggests that such a reduction of
  \rchisq typically arises from an improvement in the fit to the
  PAMELA positron fraction data. 

  \begin{figure}[hbtp]
    \centering
    \includegraphics[width=0.75\textwidth]{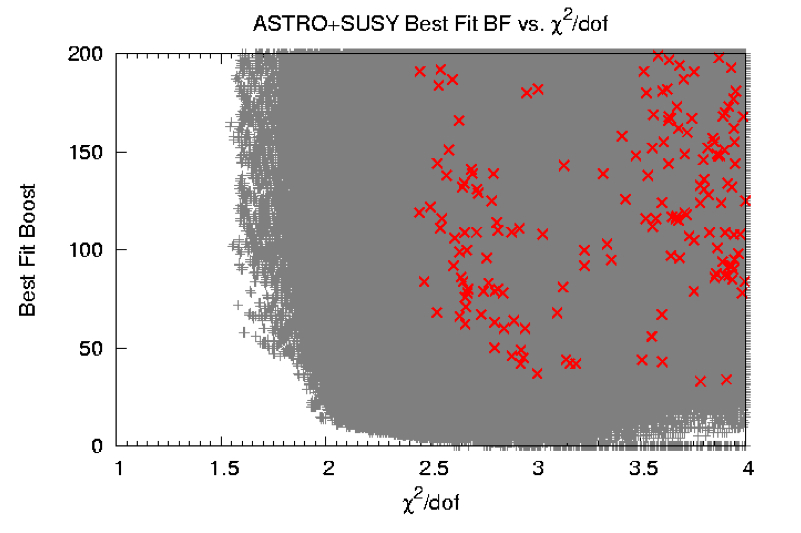}
    \caption{Best-fit boost factor versus the global \rchisq for all
      combinations in the astrophysics and pMSSM dual scan that
      attain their best fit with $B<200$ (grey points). Also shown are
      the results from some of the fits obtained by combining our pMSSM scan with the
      benchmark astrophysical ``Model 1'' (red points) discussed in
      the text. Note that most of the results from the fits obtained
      with the benchmark model are not found in the ranges displayed here.}
    \label{figs:globalscatter}
  \end{figure}

  We would also like to know what values of the boost factor are required
  in order to obtain the best fit to the data. In Figure \ref{figs:globalscatter} we
  display the best-fit boost factor versus the reduced \chisq for all
  combinations of pMSSM models and CR astrophysics models
  that attain their best fit when $B\leq 200$. The red points in this
  figure represent the results from combining our $\sim$69k pMSSM 
  models with the \emph{benchmark}
  astrophysical model. We observe that the combination of our $\sim$69k
  pMSSM models with the astrophysical models in our 524 custom model set
  yields much improved fits with much lower boost factors compared to the
  benchmark astrophysical case (in fact most of the $\sim$69k benchmark cases
  do not attain their best fits for values of \rchisq and $B$ that
  fall within the ranges displayed in
  Fig. \ref{figs:globalscatter}). Although this figure shows
  only the model combinations that have a best fit with $B<200$,
  we allowed for boosts as high as 5000 in our numerical
  calculations. It is amusing to note that, although the case with the
  overall best fit occurs for $B\approx 3000$ with $\rchisq=1.544$,
  this is hardly better than the global fits that can be obtained with
  $B<200$, where $\rchisq=1.545$ is obtained for several astrophysical/SUSY model
  combinations with boosts in the range $B=100-160$. 

  We should also remark on the perplexing feature of Figure 
  \ref{figs:globalscatter} that many model combinations seem to
  require little or no boost to attain their best fit to the data. We
  find that in these cases the astrophysical background models have
  relatively stiff electron injection indices ($\gamma_e=2.42-2.45$),
  low radiative loss rates
  ($\mathcal{F}_{op}+\mathcal{F}_{FIR}=0.5-1$)
  and relatively KN-like losses
  ($\mathcal{F}_{op}/\mathcal{F}_{FIR}=4-10$). These features combine
  to give an \alle spectrum that is very stiff, relative to other
  astrophysical models in our set. For these models we find that the 
  astrophysical contribution to the positron fraction is
  very large in the low energy bins, leaving little room to add a flux of
  positrons from SUSY DM annihilation without greatly exceeding the
  (most tightly-constraining) positron fraction data near $10-20\gev$.
  In the end, as is evident in Figure \ref{figs:globalscatter}, 
  these ``very-low-boost'' combinations attain global fits of at
  best $\rchisq\approx2.0$, significantly worse than the best 
  cases, which have $\rchisq\approx1.55$. 
  
  An interesting result of this analysis is that the cases with the
  best fit to the data set with low boost factors from among all the
  combinations of astrophysical and pMSSM models are dominated by both
  a small number of particular astrophysical models and by a small
  number of particular pMSSM models. This results in a few parameter sets
  that best describe the data, giving global fits that most
  frequently satisfy $\rchisq<1.70$ with best-fit boost factors
  $B<200$. These astrophysical models have relatively small diffusion
  zones (half-heights $L=2.0-3.0\kpc$) so 
  that, as discussed previously, $N_e$ takes a value appropriate for
  fitting both the \alle and \pamr data simultaneously. These models 
  have relatively soft injection spectra, $\gamma_e=2.51-2.54$, which 
  are compensated for by IC energy losses that are more KN-like than in 
  the default case (i.e. with
  $\mathcal{F}_{op}/\mathcal{F}_{FIR}=4-10$). In addition, smaller
  values of the magnetic field normalization 
  $\mathcal{F}_{B} \cdot B(0,0)\lsim 1.0 \mu G$ give better fits than
  do larger values, though this dependence is the weakest among the 
  parameters being scanned here. 

\begin{figure}[hbtp]
    \centering
    \includegraphics[width=0.75\textwidth]{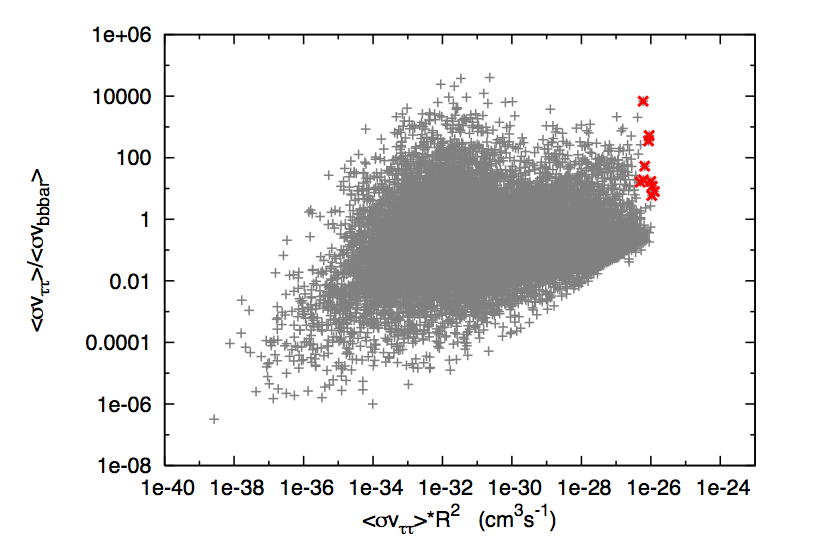}
    \caption{ Comparison of the $\tau^{\pm}$ and $b\bar{b}$ annihilation
        rates for the ten pMSSM models that most frequently
        (over the set of 524 astrophysical models) attain the best fits
        with $B<200$. 
    }
    \label{figs:goodmods}
  \end{figure}

  To gauge which of our $\sim$69k pMSSM models provide the ``best''
  fits to the astrophysical data, we consider which models with
  $m_{\LSP}\geq 100\gev$ most frequently (over the set of
  524 astrophysical models) satisfy $\rchisq<1.7$ with best fit boost
  factors $B<200$. The 10 models that most frequently satisfy these
  criteria are highlighted in red in Figure \ref{figs:goodmods}. These
  models are displayed in the plane comparing the ratio of DM 
  annihilation cross sections for $\tau$ pair and $b\bar{b}$ final
  states versus the scaled $\tau$ pair cross section. It is clear
  that this set of models has a significantly larger annihilation
  rate to $\tau^{+}\tau^{-}$ than to $b\bar{b}$. 

  These models also have relic densities of order 
  $\Omega h^2 \approx \Omega h^2|_{\tiny{WMAP}}$ and LSP masses that 
  are large enough ($\gsim100\gev$) to make a significant impact on 
  the CR positron fraction. We find that these models realize
  $\Omega h^2 \approx \Omega h^2|_{\tiny{WMAP}}$ via just the right 
  amount of co-annihilation with $\tilde{\tau}_1$ with the mass splitting: 
  $(m_{\tilde{\tau}}-m_{\LSP})\sim 15-20\gev$. This alone suggests that
  these models have sizeable $\tau^{\pm}$ annihilation rates
  (via t-channel exchange of the light $\tilde{\tau}$), but with the
  very general mass spectra available in the pMSSM. We find that these
  models also annihilate with high purity into $\tau^{\pm}$. In these
  models, the hadronic t-channel squark exchange is usually mass suppressed. Such
  high purity in the $\tau$ channel allows for a good fit with lower
  boost to the PAMELA positron fraction data without significantly 
  affecting the fit to the (non-anomalous) PAMELA antiproton data. The
  best fit boost factors for these models are in the range $B\sim
  70-190$. 

  We postpone a more detailed discussion of this set of 10 
  pMSSM models until Section \ref{sec:10models}. We now turn to a
  discussion of the search for DM contributions to gamma-ray 
  observations of dwarf satellite galaxies, where we will find that
  annihilation dominantly to $\tau$-pairs is an interesting and
  exceptional case.

  \section{Gamma-ray signals from WIMP annihilation in dwarf galaxies}
  \label{sec:gammas}

  Though ``dark,'' SUSY dark matter can, via its annihilation, 
  produce a rich gamma-ray spectrum. The detection of monochromatic
  \gams would represent a nearly undeniable detection
  of DM annihilation through loop-level channels with
  photons in the final state. Comparable fluxes of continuum gammas can
  also arise from radiation off of charged final and virtual
  states, as well as from the products of hadronization,
  fragmentation and decay ($\eg$, $\pi^{0}$'s). Such continuum spectra are 
  highly model-dependent but may also be distinguished from
  astrophysical background sources by using characteristic features,
  such as a kinematic edge at $m_{\LSP}$ or a discernable non-power-law
  shape that can help to determine their DM origin upon observation.

  Strategies for detecting \gams from DM annihilation are
  varied in their aims. If intuition from numerical simulations is
  correct, the inner-most part of the galaxy should house an
  extremely over-dense, and therefore luminous, DM cusp (along
  with an uncertain astrophysical background). Line 
  signals, however, are so distinct that searches can include
  diffuse \gams from the entire Milky Way Halo to increase their
  sensitivity (and possibly overcome the loop-suppression). Another possibility is to 
  look where there is essentially no astrophysical background so 
  that any observation of \gams may be hypothesized to have come
  only from DM.
  
  Dwarf galaxies may represent our best chance at carrying out this
  last type of search. Dwarfs are relatively small systems that 
  accompany the much larger Milky Way; they harbor small collections
  of stars, relatively small amounts of gas and have weak magnetic 
  fields. Their existence is anticipated from the hierarchical nature 
  of structure formation and therefore dwarfs are expected to be highly DM-dominated
  (with mass-to-light ratios $M/L\sim100-1000(M_{\odot}/L_{\odot})$,
  as determined from the kinematics of their stellar populations). 
  Dwarfs are not expected to contain any internal astrophysical
  sources of \gams and though the Milky Way's own diffuse \gams
  present a sizable foreground, this can be subtracted reliably for
  dwarfs at high enough galactic latitudes. This leaves the very low 
  background levels necessary for such low-statistics DM searches.
  
  The full gamma-ray signal originating from a DM distribution
  $\rho(\vec{r})$ factorizes into a piece dependent only on the DM
  distribution (or what is often called the ``DM halo'') and a piece
  dependent only on the underlying particle physics model. The
  number of gammas per unit area per unit time is given by:
  
  \begin{equation}
    \frac{dN_{\gamma}}{dAdt} =
    \int_{\Delta\Omega}\Big\{\int_{l(\Omega
      ')}\rho^{2}(\vec{r})ds\Big\}d\Omega '\cdot\frac{1}{4\pi}\frac{\tsigv}{2m_{\LSP}^{2}}\int^{E_{max}}_{E_{th}}\frac{dN^{\gamma,tot}}{dE_{\gamma}}dE_{\gamma}
    \label{gamrate}
    \end{equation}
  
  \noindent where, for a generic DM distribution $\rho(\vec{r})$,
  the first factor is the integral of $\rho^2(\vec{r})$ along the
  portion of each line-of-sight $l(\Omega ')$ for which
  $\rho(\vec{r}) \neq 0$, over the solid angle subtended
  by the distribution around our detector, $\Delta\Omega$. In the 
  second factor $m_{\LSP}$ is the LSP mass, $\sigv$ is the thermal-averaged
  total WIMP annihilation cross-section, $R$ is the quantity that
  was defined in Eqn. (\ref{relicrescale}) and the quantity: 
  
  \begin{equation}
    \frac{dN^{\gamma,tot}}{dE_{\gamma}} =
    \sum_{i}B_{i}\bigg\{\frac{dN_{i}^{\gamma,sec}}{dE_{\gamma}} +
    \frac{dN_{i}^{\gamma,FSR}}{dE_{\gamma}} + \frac{dN_{i}^{\gamma,VIB}}{dE_{\gamma}} + \frac{dN_{i}^{\gamma,line}}{dE_{\gamma}}\bigg\} 
    \label{totyield}
  \end{equation}
  
  \noindent\ is the total continuum ``yield curve'' with terms describing
  hadronization yield (``secondary'' gammas), final-state radiation (FSR), virtual internal
  bremsstrahlung (VIB) and possible monochromatic \gams for each
  annihilation channel\footnote{We use the language of \cite{Bringmann:2007nk} in
    discriminating FSR and VIB, although it has
    been pointed out that such a distinction is somewhat
    artificial (or not even gauge invariant)
    \cite{Barger:2009xe}.}. The sum runs over final state annihilation channels,
  indexed by $i$, and the weights $\sigv B_i$ are the annihilation
  rates into the $i^{th}$ final state ($\ie$, $\tau^{+}\tau^{-}$, $b\bar{b}$, $W^+W^-$,
  etc.). We integrate the total continuum
  yield curve from the lower threshold $E_{th}$, which is chosen 
  according to the performance of a given detector, up to
  $E_{max}=m_{\chi}$ for all experiments considered here.
  
  Although the calculation of yield curves is straightforward (given
  a particular particle physics model), estimations of the halo-dependent
  piece are typically fraught with great uncertainty. For instance, while
  much effort has been devoted to numerical simulations of DM halos
  similar to that of our own Milky Way Halo, the uncertainties
  associated with quantities such as the slope of the inner-halo cusp profile allow
  for the predicted flux of \gams from DM annihilating in a region
  near the Galactic Center (GC) to vary by more than an order of magnitude (even when
  specializing to fairly cuspy profiles \cite{Bertone:2008xr}). The
  estimation of halo integrals for dwarf galaxies is a
  simpler process: here one can directly observe the stellar
  kinematics of the dwarf system member stars. Of course this
  process enjoys its own difficulties, for example in estimating the 
  dynamical state of the dwarf DM distribution (i.e., the influence of
  tidal forces due to the Milky Way itself) or in determining the
  membership of individual stars (see e.g., \cite{Simon:2007dq}).    
  
  Still, the gamma-ray signals of DM annihilation incur much
  less uncertainty than their counterparts
  involving electrons and positrons. For the energies of
  importance here, the galaxy is essentially transparent to
  \gams so a non-trivial propagator is not necessary to
  encode their attenuation, in contrast to $e^{\pm}$'s. The halo
  integral above is carried out along a full line-of-sight 
  because, unlike $e^{\pm}$'s, gammas do not diffuse but rather
  point directly back to their source.
  
  Perhaps the best target for DM annihilation searches in dwarf
  galaxies is the ultra-faint dwarf 
  Segue 1. Discovered relatively recently in analyses of SDSS
  data \cite{Belokurov:2006ph}, Segue 1 is known to have a
  mass-to-light ratio $M/L\sim1000 (M_{\odot}/L_{\odot})$, and
  a preliminary analysis has been carried out to estimate its 
  DM content using the kinematics of 24 of its member stars 
  \cite{Geha:2008zr}. We will
  use the analysis presented in \cite{Essig:2009jx}, appropriate to the
  FERMI-LAT observations of Segue 1, in which the line-of-sight halo
  integral was estimated by marginalizing over halo distributions,
  including both cusped and cored types\footnote{Although, as emphasized in
    \cite{Essig:2009jx}, \cite{Strigari:2007at}, stellar kinematic data
    fix the normalization of $\rho$ and $\rho^{2}$ such that the 
    detailed nature of the inner-slope of the halo ends up having
    a relatively small effect on the total halo integral.}. The work 
  \cite{Essig:2009jx} has recently been updated and expanded in \cite{Essig:2010em}.

  While it may eventually be possible to obtain full spectral information,
  the first task at hand is to identify \emph{any} excess \gams
  coming from dwarfs, i.e., the integrated yield from
  Eqns. (\ref{gamrate}) and (\ref{totyield}). We calculate the yield
  curves for each of our SUSY models using the DarkSUSY package, version
  5.0.4 \cite{Gondolo:2004sc}, and for this analysis we do not
  include the contributions from $\gamma$ lines\footnote{A rough
    calculation of these contributions (without any attempt to include
    the LAT instrument response) showed that the line contribution to
    the total yield of gammas would be $>1\%$ in only $\approx 0.5\%$
    of the models in our pMSSM set. Further, none of these models have
    total yields that would be observable by the FERMI-LAT unless they
    had boost factors $B\gsim 10^4$.}. We estimate the $\gamma$ yield
  for FERMI-LAT observations of Segue 1 arising from each of our
  SUSY models by integrating these continuum yield curves from a
  low energy threshold of 100 MeV (corresponding to the FERMI-LAT
  threshold) up to $m_{\LSP}$ (this assumes that
  the LAT can dependably reconstruct \gams all the way up to our
  highest LSP mass, $\sim700 \gev$) and by taking the central value estimate
  for the Segue 1 halo integral provided in \cite{Essig:2009jx}. In
  the recent work \cite{Essig:2010em} this central value estimate has
  been lowered significantly from the value used here. We emphasize
  that the results presented here are approximate and an effort to
  more accurately calculate the dwarf signals that can be expected
  from our pMSSM model set is currently underway \cite{futuredwarf}. 

\begin{figure}[hbtp]
    \centering
    \includegraphics[width=0.65\textwidth]{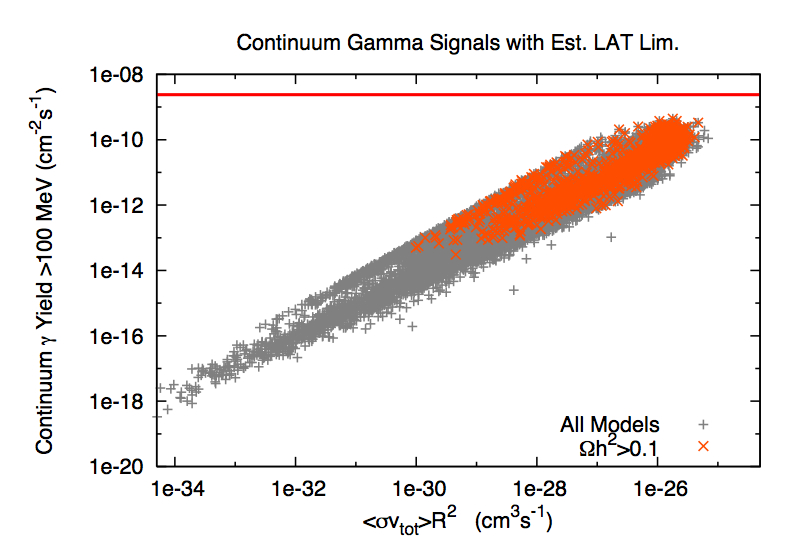}
    \caption{Continuum $\gamma$ yields integrated above
      $100\mev$ as a function of $\tsigv$. A point is shown for each model,
      orange points corresponding to pMSSM models for which 
      $\Omega h^2|_{LSP}\geq0.1$. The red line represents the estimated
      sensitivity for the LAT to detect a point source signal at high
      galactic latitudes with $3\sigma$ confidence \cite{Essig:2009jx}.}
    \label{figs:gyield-standard}
  \end{figure}

\begin{figure}[hbtp]
    \centering
    \includegraphics[width=0.65\textwidth]{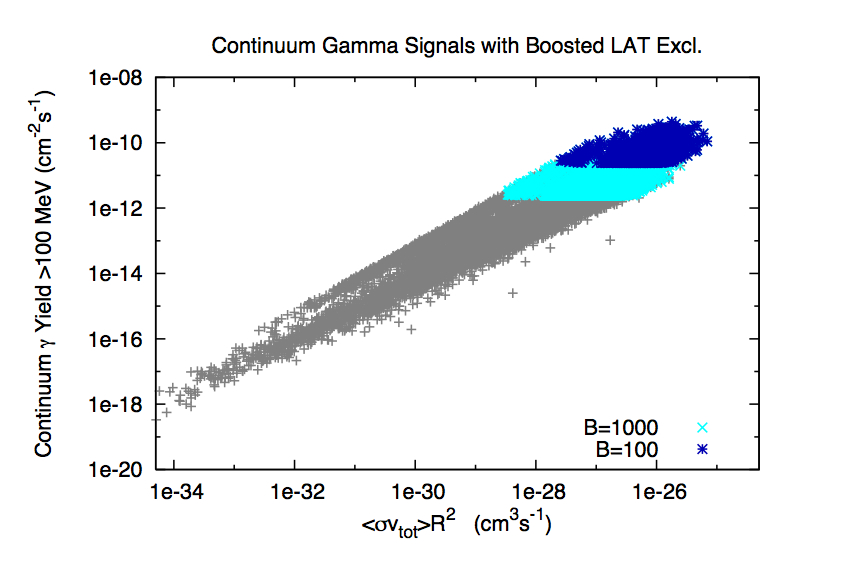}
    \caption{Same as in Fig. \ref{figs:gyield-standard} except models are
      colored according to whether they can be excluded assuming
      a boost factor of $B=100$ (dark blue) or $B=1000$ (light blue).}
    \label{figs:gyield-boosts}
  \end{figure}

  The result is shown in Figure \ref{figs:gyield-standard}, 
  where the horizontal line reflects the estimated sensitivity
  \cite{Essig:2009jx} for the LAT to detect a point source
  signal at high galactic latitudes with $3\sigma$ confidence 
  after 1 year of operation. One can see that
  this sensitivity is about an order of magnitude above what is
  necessary to detect any of the models in our set, assuming a
  relic abundance consistent with thermal cosmology. In Figure
  \ref{figs:gyield-boosts} we examine the effect of including a
  boost factor to see what fraction of our model set would be 
  excluded if effective boost factors of $B=100$ or
  $B=1000$ are incorporated. Here we are not trying to motivate such boost factors
  by, $\eg$, non-thermal cosmological history or novel halo
  distribution effects, rather we will use them in what follows to
  study the SUSY model dependence of these signals.
  
\begin{figure}[hbtp]
    \centering
    \includegraphics[width=0.65\textwidth]{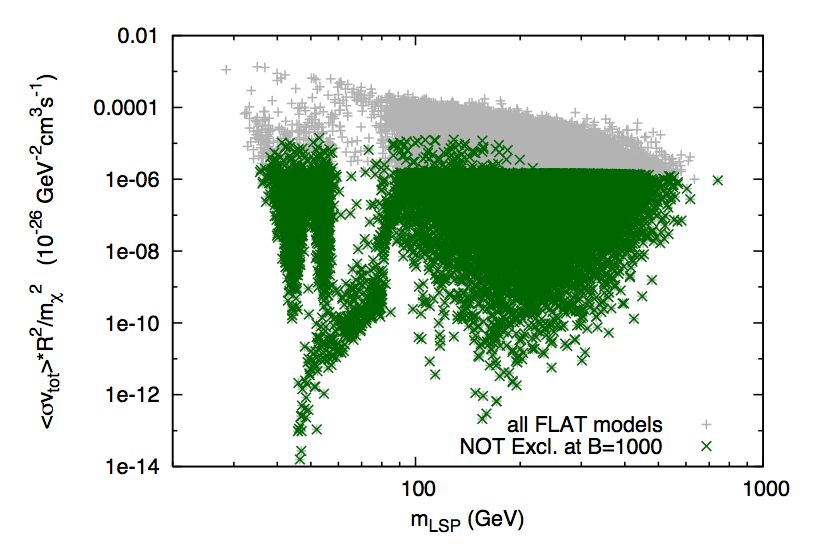}
    \caption{$\tsigv/m_{\LSP}^2$ versus $m_{\LSP}$. Models that are
      \emph{NOT} excluded by FERMI-LAT observations of Segue 1 with
      $B=1000$ are emphasized in dark green against
      the full pMSSM model set. Dark points that leak
      above the apparent ``exclusion line'' feature are pMSSM models
      whose continuum gamma yields are relatively difficult to detect.}
    \label{figs:DifficultScatter}
  \end{figure}

  The vertical width of the scatter in Figures
  \ref{figs:gyield-standard}-\ref{figs:gyield-boosts} represents
  the extent of the SUSY model dependence in relating $\tsigv$ for
  a given model to the total yield of gammas in our pMSSM model
  set. Of course there is the trivial $m_{\LSP}^{-2}$ dependence apparent in Eq.
  (\ref{gamrate}), but it is instructive to factor this out and look
  for further model dependence. In this aim, we show
  $\tsigv/m_{\LSP}^2$ as a function of $m_{\LSP}$, emphasizing
  models that are \emph{NOT} excluded with $B=1000$ in Figure \ref{figs:DifficultScatter}.
  In this figure the non-trivial SUSY model dependence is
  expressed as the extent to which models with similar $\tsigv$
  cluster along a horizontal line. Of course, for a given boost,
  we expect that many models cannot be excluded simply because 
  they have lower $\tsigv$ than the necessary minimum ($\eg$, 
  consider the improbable case of annihilations entirely into
  final state \gamsn). However, the interesting feature of this figure 
  is that there is a collection of $\sim100$ pMSSM models that are 
  difficult to detect relative to others with similar values of
  $\tsigv$ \nolinebreak ($\ie$, those scattered above the ``exclusion line''
  in the figure).

\begin{figure}[hbtp]
    \centering
    \includegraphics[width=0.65\textwidth]{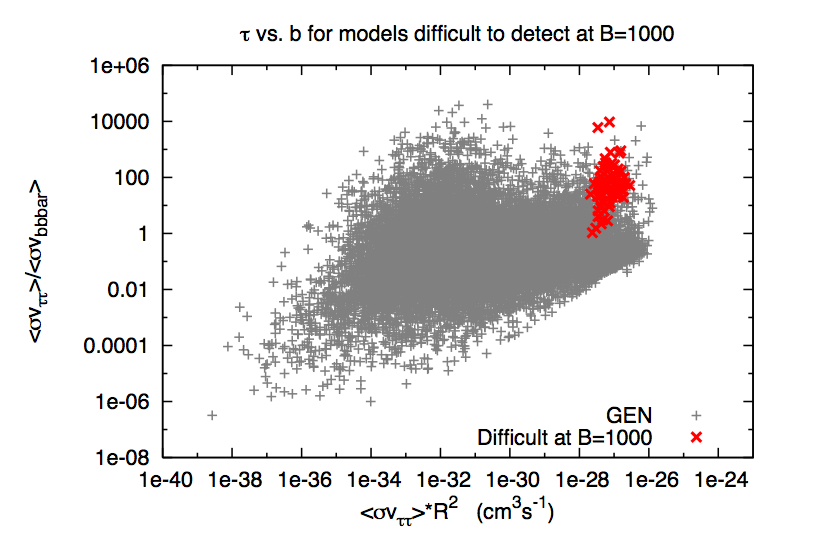}
    \caption{ The pMSSM models that are relatively
      difficult to observe by FERMI-LAT observations of Segue 1 with
      $B=1000$ (see Fig. \ref{figs:DifficultScatter}) are
	highlighted against the rest of our pMSSM model set in a
	figure similar to Fig. \ref{figs:allheavyonTauVB}. We note that
	these models lie in the	``leptophilic'' region 
	described in the text and that the models highlighted here are similar to
        those highlighted in Figure \ref{figs:goodmods} but with
        lower overall $\tsigv$.
    }
    \label{figs:DifficultOnTauVB}
  \end{figure}

  It is easy to see what these ``difficult'' models have in common
  by taking a look at the relative values of their annihilation
  rates into the various SM final states: one finds that they have
  especially large annihilation rates into the 
  $\tau^{+}\tau^{-}$ final state. For the set of models whose integrated 
  yields are more than a factor of 2 from being excluded by FERMI
  observations of Segue 1 with a boost factor $B=1000$, we find that in all
  cases the LSP annihilates into $\tau^{+}\tau^{-}$ $\gsim 80\%$ of
  the time, while the vast majority of models in this set have LSPs
  which annihilate into $\tau^{+}\tau^{-}$ $\gsim 94\%$ of the time.
  We highlight this set of models in Figure
  \ref{figs:DifficultOnTauVB}, where we show the ratio of
  annihilation rates into $\tau^{+}\tau^{-}$ and $b\bar{b}$ final
  states as a function of the annihilation rate into the 
  $\tau^{+}\tau^{-}$ final state. One can see that they lie in the 
  ``leptophilic'' region discussed in Section \ref{sec:pMSSM}. Of course the focus on
  exclusion with $B=1000$ is arbitrary and we note that
  an analysis of exclusion with $B=100$ tells a similar story; models that are more difficult to 
  detect have large annihilation rates to $\tau^{+}\tau^{-}$ and, if 
  highlighted on Figure (\ref{figs:DifficultOnTauVB}), would fall
  just to the right of those discussed here.

\begin{figure}[hbtp]
    \centering
    \includegraphics[width=0.65\textwidth]{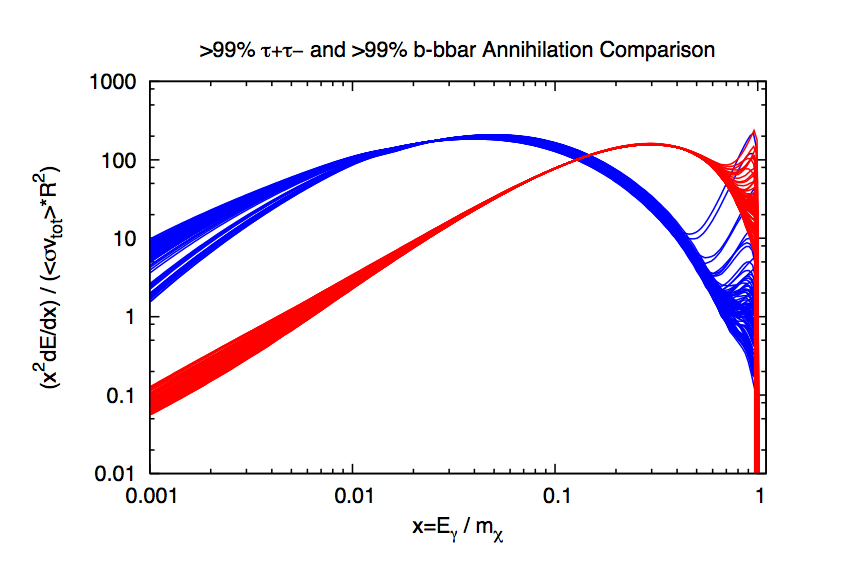}
    \caption{
      $\gamma$ continua (calculated with DarkSUSY 5.0.4) from models
      in our pMSSM set for which the LSP annihilates into the
      $\tau^{+}\tau^{-}$ final state greater than 99\% of the time
      (red), and those for which the LSP annihilates into the
      $b\bar{b}$ final state greater than 99\% of the time
      (blue). The curves have been rescaled according to $\tsigv$
      and are displayed as a function of $x=E_{\gamma}/m_{\LSP}$.
      Note that there is a \textit{universal} cross-over in the 
      differential flux spectra near $x\sim 0.15$.
    }
    \label{figs:puretauvspureb}
  \end{figure}

  This may seem contrary to the common intuition (see
  $\eg$, \cite{Cirelli:2009dv}\cite{Abdo:2010ex}) that large rates
  into the $\tau^{+}\tau^{-}$ final state are
  \emph{easier} to detect in \gams because of the relatively stiff
  $\pi^{0}$ decay spectrum. In Figure \ref{figs:puretauvspureb} we
  display the continuum $\gamma$ flux spectra from the 
  models in our set that annihilate into the $\tau^{+}\tau^{-}$
  final state $\gsim 99\%$ of the time (red) and from those that
  annihilate into the final state $b\bar{b}$ $\gsim 99\%$ of the time (blue).
  These models have a wide range of values for $m_{\LSP}$ and $\tsigv$ so we
  factor out this model dependence by dividing each curve by the
  appropriate value of $\tsigv$ and by using the variable
  $x=E_{\gamma}/m_{\LSP}$. What remains is highly model-dependent
  near the endpoint $x\rightarrow 1$ where VIB can become dominant
  \cite{Bringmann:2007nk}, but is relatively model-independent
  otherwise. One can see that generic $\tau^{+}\tau^{-}$
  spectra lead to higher continuum $\gamma$ fluxes than do generic
  $b\bar{b}$ spectra, but only above a \textit{universal} value 
  $x\sim 0.15$. Below this value the
  softer $b\bar{b}$ spectrum dominates and we see that the
  visibility becomes a function of the threshold energy
  appropriate to a given detector. The spectral crossover
  occurs at $E_{\gamma}\sim 15 \gev$ for $m_{\LSP}=100\gev$ and 
  at $E_{\gamma}\sim 150\gev$ for $m_{\LSP}=1\tev$. For Cerenkov
  telescopes, whose threshold energies are typically $E_{th}\sim 100
  \gev$, this means that the differential flux of continuum \gams
  from annihilation into $\tau^{+}\tau^{-}$ would be larger than
  that from annihilation into the $b\bar{b}$ final state over
  the entire energy range above threshold. Thus, for Cerenkov
  telescopes, we expect that DM which annihilates dominantly into $\tau^{+}\tau^{-}$ can
  be more easily detected. However, in the case of the FERMI-LAT, whose threshold energy can be
  taken as $E_{th}\sim 100 \mev$, the situation is quite different.
  The FERMI-LAT would be able to count \gams with energies well
  below the universal cross-over point (for $m_{\LSP}=100\gev$ this
  corresponds to integrating over the entire range shown in Figure 
  \ref{figs:puretauvspureb}), so that DM which annihilates mostly to
  $\tau^{+}\tau^{-}$ would actually be \emph{more difficult} for the
  FERMI-LAT to detect.
  
  Several things can be said by looking deeper into the detailed
  mass spectra of this leptophilic model set. Starting with the
  identity of the nLSP, according to
  frequency we find: $\tilde{\tau_1}$, $\tilde{\nu_{\tau}}$,
  $\tilde{e_{R}}$, $\chipm$ and
  $\tilde \chi_2^0$ nLSPs (the vast majority being of the first
  variety). As discussed above, it is clear that such large leptonic
  annihilation rates can be arranged for in models that have relatively light
  sleptons and relatively heavy squarks, thus suppressing or enhancing
  the associated t-channel annihilation graphs accordingly. Here, the
  $\tilde{\tau_1}$, $\tilde{\nu_{\tau}}$, $\tilde{e_{R}}$ nLSP cases
  attain leptophilia simply because the $\tilde{\tau_1}$ is always
  relatively light (either the nLSP or close to the nLSP). In the
  cases where the $\chipm$ or the $\tilde \chi_2^0$ are the nLSP, one
  generally expects a sizeable rate for annihilation to gauge bosons.
  For the model set discussed here, however, we have either that the LSPs are
  too light to annihilate into W or Z bosons, or that they are sufficiently purely
  bino-like that their couplings to gauge bosons are suppressed. These models
  are leptophilic simply because their colored superpartners are
  quite heavy (taking, $\eg$, the ratio of sbottom and stau masses, one finds that
  $m_{\tilde{b}_1}/m_{\tilde{\tau}_1}\sim2-7$).
  
  With a relatively light (often nLSP) $\tilde{\tau_1}$ one might
  expect that the relic densities of these models are significantly
  determined by stau co-annihilation (recall that models in our pMSSM set
  have much more generic spectra than the mSUGRA models that are
  usually referred to as stau co-annihilation models). One would then expect that the
  leptophilic models that have the largest relic density have some preferred
  range of mass splitting $(m_{\tilde{\tau}_1}-m_{\LSP})$, and indeed this is the case,
  typically $(m_{\tilde{\tau}_1}-m_{\LSP})\sim 10-20\gev$. Given
  that the flux of cosmic-ray positrons from DM annihilation
  scales as the relic density squared, we expect that this specific
  subset of leptophilic models may be the most likely
  candidates for producing a sizeable excess in the positron
  fraction. Indeed this is the case, as we have described in
  Section \ref{sec:results}.
  
  \section{Description of Best-Fit pMSSM Models}
  \label{sec:10models}
  
  In Section \ref{sec:results} we presented the fits to the CR data fits that
  result from our combined scan over pMSSM and astrophysical
  parameter spaces. We found that the pMSSM models that most
  frequently provide the best fits to the data have significant
  annihilation rates into $\tau$-pairs and are apparently
  leptophilic, as they were found to provide a significant
  supersymmetric DM contribution to the $\pamr$ spectrum without
  significant excesses over the measured $\pbarp$ data. We identified
  a set of ten ``best'' pMSSM models, which have $m_{\LSP}>100\gev$
  and most frequently (over the set of 524 astrophysical models) 
  satisfy $\rchisq<1.7$ with best fit boost factors $B<200$. We now
  discuss these pMSSM models and their CR spectra in more detail.

  We begin by presenting Table \ref{crphenotab} where, for each of
  these ten models, we display the quantities that are most important
  for the resulting DM annihilation phenomenology. We observe that all
  models have LSPs with $m_{\LSP}\sim 100-130\gev$ and account for a
  significant fraction of the total DM density measured by WMAP. We
  also see that, after a thermal relic rescaling and inclusion of the
  appropriate boost factor, $B\tsigv\approx 10^{-24} \: \mathrm{cm}^3
  \, \mathrm{s}^{-1}$ for all ten models. While all models in this set
  have significant annihilation rates into $\tau$ pairs, we note that
  seven of the ten annihilate into $\tau$ pairs $>80\%$ of the time.

  \begin{table}
    \renewcommand{\arraystretch}{1.1}
    \centering
    \begin{tabular}{|c|c|c|c|c|c|c|c|c|} \hline\hline
      Mod & $m_{\LSP}$ ($\gev$) & $R$ & $B$ &
      $B\tsigv$ &
      $\sigv_{\tau}/\sigv$ &
      $\sigv_{b}/\sigv$ &
      $\sigv_{Z}/\sigv$ &
      $\sigv_{W}/\sigv$ \\ \hline\hline      
      1 & 101 & 0.64 & 115 & 1.23 & 0.46 & 0.03 & 0.13 & 0.37 \\ \hline
      2 & 107 & 0.99 & 72 & 1.27 & 0.71 & 0.09 & 0.05 & 0.14 \\ \hline
      3 & 132 & 0.91 & 99 & 1.55 & 0.68 & 0.11 & 0.08 & 0.11 \\ \hline
      4 & 122 & 0.73 & 102 & 1.39 & 0.81 & 0.07 & 0.05 & 0.07 \\ \hline
      5 & 116 & 0.64 & 163 & 1.27 & 0.85 & 0.02 & 0.05 & 0.08 \\ \hline
      6 & 105 & 0.67 & 104 & 1.15 & 0.90 & 0.05 & 0.01 & 0.02 \\ \hline
      7 & 114 & 0.74 & 187 & 1.21 & 0.95 & 0.05 & $<$0.01 & $<$0.01 \\ \hline
      8 & 103 & 0.80 & 119 & 1.07 & 0.997 & $<$0.01 & $<$0.01 & $<$0.01 \\ \hline
      9 & 105 & 0.68 & 179 & 1.08 & 0.999 & $<$0.01 & $<$0.01 & $<$0.01 \\ \hline
      10 & 132 & 1.03 & 156 & 1.34 & 0.996 & $<$0.01 & $<$0.01 & $<$0.01 \\ \hline
      \hline
    \end{tabular}
    \caption{Parameters describing the DM annihilation phenomenology of the 10
      SUSY models which provide the best fit to CR data. For each
      model we display the LSP
      mass, $m_{\LSP}$, the fraction of DM attributed to the LSP, via
      the scaling factor $R$ (see Eqn. \ref{relicrescale}), the boost
      factor, $B$, that provides the best fit to the CR data, the
      boosted and scaled total annihilation cross-section $B\tsigv\:$ in units of $10^{-24} \:
      \mathrm{cm}^3 \: \mathrm{s}^{-1}$ and also the ratios of
      annihilation rates into various final states ($\tau$-pairs,
      $b\bar{b}$, $Z^0Z^0$ and $W^{+}W^{-}$) to the total annihilation
      cross-section $\sigv$.
    }
    \label{crphenotab}
  \end{table}
  
\begin{figure}[hbtp]
    \centering
    \includegraphics[width=0.75\textwidth]{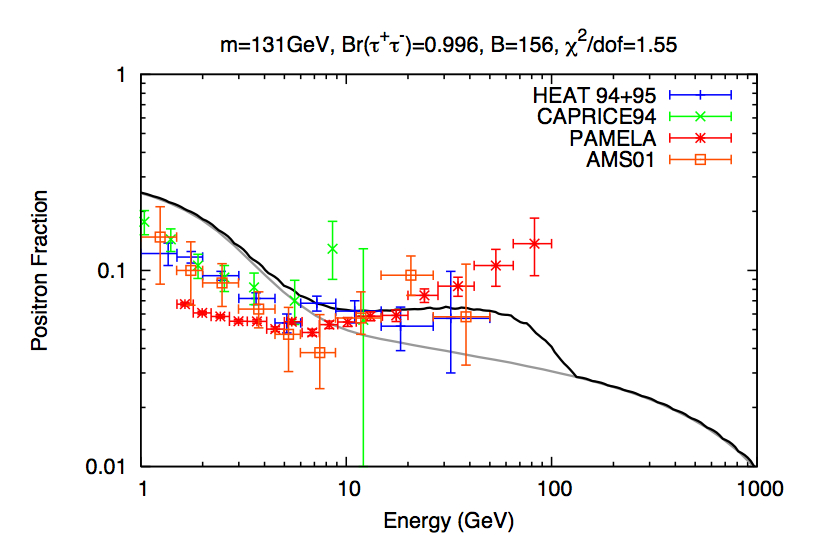}
    \caption{The positron fraction spectrum for a case with
      $\rchisq=1.55$ and $B=156$ (Model 10 of Table \ref{crphenotab}).
      We display curves with (black) and without (grey)
      the DM signal contribution. The DM contribution is seen
      to produce a sizeable excess.  Data shown are from 
      the PAMELA \cite{Adriani:2008zr}, HEAT \cite{DuVernois:2001bb},
      AMS01 \cite{Aguilar:2002ad}, and CAPRICE94 \cite{CAPRICE}
      collaborations. 
    }
    \label{figs:pep11417}
  \end{figure}

\begin{figure}[hbtp]
    \centering
    \includegraphics[width=0.75\textwidth]{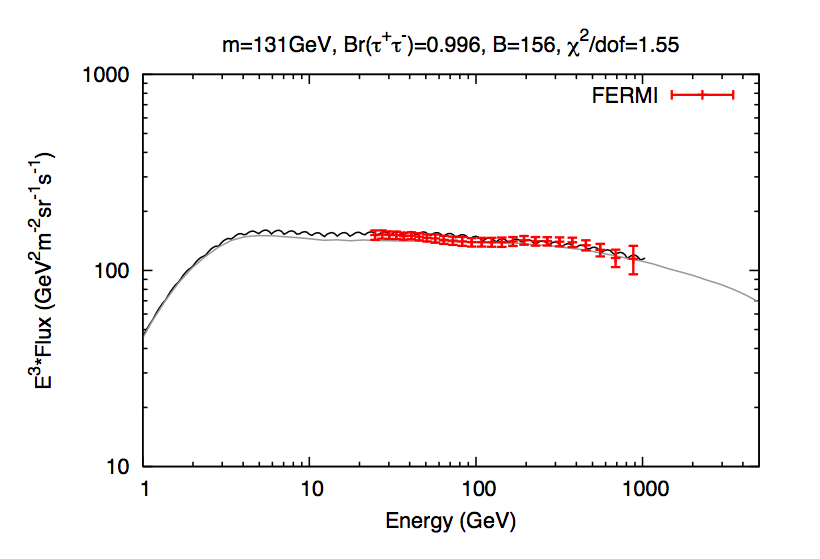}
    \caption{ The \alle spectrum for a case with
      $\rchisq=1.55$ and $B=156$ (Model 10 of Table \ref{crphenotab}).
      We display curves with (black) and without (grey) the DM signal 
      contribution. The FERMI \alle data \cite{Abdo:2009zk} is also shown.
    }
    \label{figs:alle11417}
  \end{figure}

\begin{figure}[hbtp]
    \centering
    \includegraphics[width=0.75\textwidth]{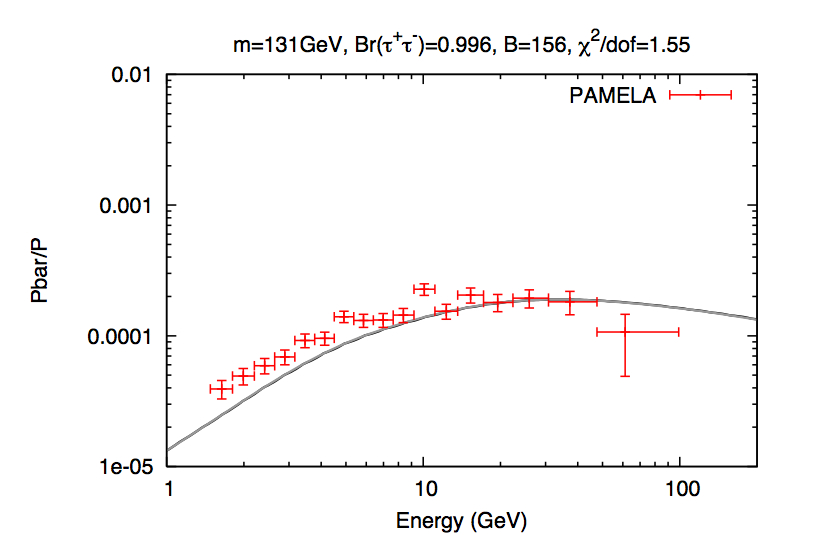}
    \caption{ The \pbarp spectrum for a case with
      $\rchisq=1.55$ and $B=156$ (Model 10 of Table \ref{crphenotab}).
      We display curves with (black) and without (grey)
      the DM signal contribution, though the two are
      indistinguishable in the figure. The PAMELA \pbarp data 
      \cite{Adriani:2008zq} is also shown.
    }
    \label{figs:ppb11417}
  \end{figure}

  Figures \ref{figs:pep11417}-\ref{figs:ppb11417} show the \pamr,
  \alle and \pbarp spectra, respectively, for a typical case (Model 10
  in Table \ref{crphenotab}) of these best-fit combinations of pMSSM 
  and astrophysical models. The DM contribution
  is seen to produce a sizeable excess in the positron flux fraction,
  though still lying significantly below the trend of the PAMELA data
  points and appearing to fit the smaller excesses measured by HEAT
  and AMS-01 quite well. In the \alle spectrum, one can see a slight
  bump due to the DM contribution near $100\gev$\footnote{The wiggle of the
  black curve in Figure \ref{figs:alle11417} is a numerical 
  effect incurred in interpolating and combining data in order to 
  display the results and does not enter into the calculation of
  the fit.}. In both the \alle and \pbarp spectra, we see that the 
  pMSSM DM contribution is barely visible above the astrophysical
  background fluxes and provides a good fit to the data.

  It is also interesting to consider the DM contributions to gamma-ray
  observations that arise from these pMSSM models. The relation between DM 
  contributions to CRs and DM contributions to gamma-rays is
  non-trivial, however, because it is not clear how to relate the boost factor
  appropriate for gamma-ray quantities to that which is
  required to best fit the CR data. We recall that 
  locally-measured CRE quantities give information about DM in the
  local $\sim\!\kpc$ neighborhood of the Milky Way DM Halo, while gamma-ray
  observables, such as the diffuse mid-latitude gamma-ray spectrum,
  are composed of line-of-sight integrals of $\rho^2(\vec{r})$ that
  extend throughout the halo. Even though the mid-latitude region is
  chosen in order to remove the large backgrounds along the galactic
  plane and to emphasize a more local diffuse gamma population, one
  still expects a different boost factor to apply to the gamma-ray signal
  (see $\eg$, \cite{Kamionkowski:2010mi}).
  Here, for simplicity, we will assume that the \emph{same} boost 
  factor that provides a best-fit the CR data should be applied to the
  DM contribution to the diffuse mid-latitude gamma-ray spectrum.

\begin{figure}[hbtp]
    \centering
    \includegraphics[width=0.75\textwidth]{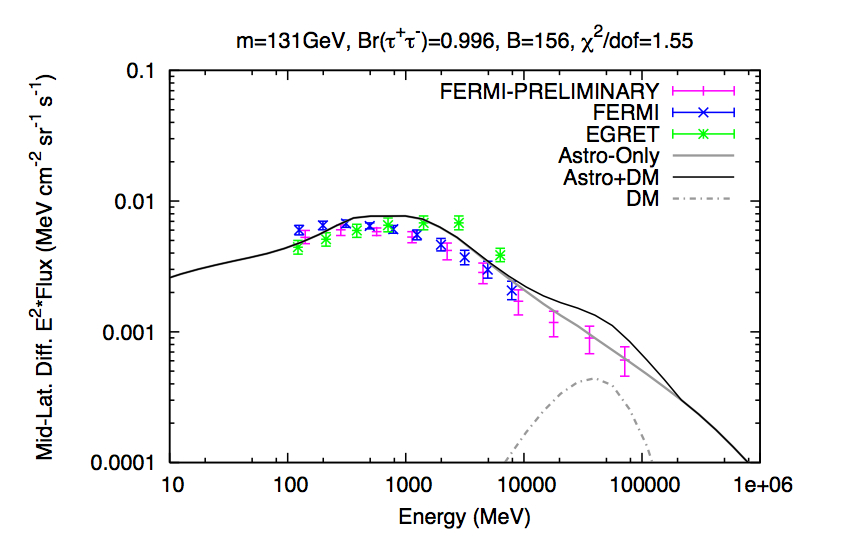}
    \caption{ The diffuse mid-latitude gamma-ray spectrum for a case with
      $\rchisq=1.55$ and $B=156$ (Model 10 of Table \ref{crphenotab}).
      We display curves with (black) and without (grey-solid) the DM 
      signal contribution, as well as the DM spectrum in isolation 
      (grey-dashed). The DM signal is calculated using the DarkSUSY
      default NFW halo profile. Data shown are published
      data from EGRET \cite{Hunger:1997we} and FERMI
      \cite{Abdo:2009mr}, as well as the preliminary FERMI data 
      presented in \cite{tporter:sympdiff}.
    }
    \label{figs:dml11417}
  \end{figure}

  Figure \ref{figs:dml11417} shows the diffuse mid-latitude gamma-ray
  spectrum that can be expected from the same example (Model 10) that 
  has been discussed in the previous figures
  \ref{figs:pep11417}-\ref{figs:ppb11417}. We calculate the DM
  contribution to the diffuse mid-latitude gamma-ray spectrum using
  the DarkSUSY default NFW halo profile. The influence of this choice
  on the resulting signal can be described by the dimensionless
  quantity (this is essentially just the first factor in Eq. \ref{gamrate}),

  \begin{equation}
    \bar{\mathcal{J}} =
    \frac{1}{\Delta\Omega} \int_{\Delta\Omega}d\Omega' \cdot\int_{l(\Omega
      ')}\frac{\rho^{2}(\vec{r})}{\rho^2_{\odot}}\frac{ds}{r_{\odot}} \:,
    \label{jbar}
  \end{equation}
  
  \noindent where, with $r_{\odot}=8.5\kpc$,
  $\rho_{\odot}=0.3\gev\,\mathrm{cm}^{-3}$ and
  $\Delta\Omega=2.12$ for the mid-latitude region
  ($0^{\circ}\leq l < 360^{\circ}$ and $10^{\circ}\leq |b| \leq 20^{\circ}$), we find
  $\bar{\mathcal{J}}_{\mathrm{NFW}}=3.21$. For comparison, we find that
  the smooth isothermal profile implemented in DarkSUSY
  gives a very similar value,
  $\bar{\mathcal{J}}_{\mathrm{ISO}}=3.16$, in this region.

  We observe that the DM annihilation contribution to the diffuse
  mid-latitude astrophysical background results in a bump
  in the spectrum, giving a slight excess above the preliminary FERMI data for
  energies $E\sim 20-50\gev$. We emphasize again that we use the same
  boost as that which best fits the CR data in calculating the DM 
  contribution to this quantity and that this is a non-trivial 
  assumption. These diffuse mid-latitude data sets were
  \emph{NOT} included in the simultaneous fit of the CR data in
  our combined scan of pMSSM and astrophysical parameters (due to the
  uncertainties inherent in this calculation as discussed at the end 
  of Section \ref{sec:crepars}). For the small number of pMSSM models
  considered here, we have repeated the global fit incorporating the
  diffuse mid-latitude data by including the highest-energy data bins
  in the preliminary FERMI data set. We find little change in the
  results, as the CR data sets dominate the determination of the
  best-fit boost factor and the resulting value of $\rchisq$. We 
  point out that if the appropriate boost for this quantity were 
  a factor of 2 lower than that for the CR data, then
  the resulting excess due to DM would be consistent with the 
  preliminary data at the $1\sigma$ level for all high-energy
  bins.

  It is also interesting to include the boost factors that best fit
  the CR data sets in computing the gamma-ray yields from DM 
  annihilation in dwarf galaxies. We do this in the
  context of the FERMI-LAT observations of Segue 1 by simply multiplying
  the yields for each of the models described in Table \ref{crphenotab} 
  by the corresponding boost factors. This is again a
  non-trivial assumption, and in principle a very different one that
  has been made in the diffuse mid-latitude gamma case; the
  systematics involved in modeling the DM distribution of a dwarf are
  different than those involved in modeling the DM distribution of the
  Milky Way halo on much larger scales. The results are displayed in 
  Table \ref{dwfboosttab}. We observe that the expected DM
  annihilation signals range from a factor of $\sim\!5$ \emph{above}
  the estimated sensitivity to a factor of $\sim\!2$ \emph{below}, 
  and that, in keeping with the results of 
  Section \ref{sec:gammas}, the models which annihilate most purely
  to $\tau$-pairs are the most difficult to detect\footnote{A similar
    analysis was carried out to
    compare boosted monochromatic gamma fluxes to the FERMI-LAT limits
    presented in \cite{Abdo:2010nc}. Here we have found that expected
    signals are below the limits in all cases by factors ranging from 
    $\sim\! 2$ to several orders of magnitude.}.

  \begin{table}
    \renewcommand{\arraystretch}{1.1}
    \centering
    \begin{tabular}{|c|c|c|c|c|c|c|} \hline\hline
      Mod & $B\times$(Yield) $\mathrm{cm}^{-2}\,\mathrm{s}^{-1}$ & $B\times$(Yield)/(Est. Sens.) & 
      $\sigv_{\tau}/\sigv$ \\ \hline\hline
      1 & $1.12 \times 10^{-8}$ & 4.69 & 0.46  \\ \hline
      2 & $6.76 \times 10^{-9}$ & 2.82 & 0.71  \\ \hline
      3 & $6.26 \times 10^{-9}$ & 2.61 & 0.68  \\ \hline
      4 & $4.40 \times 10^{-9}$ & 1.83 & 0.81  \\ \hline
      5 & $3.65 \times 10^{-9}$ & 1.52 & 0.85  \\ \hline
      6 & $3.32 \times 10^{-9}$ & 1.38 & 0.90  \\ \hline
      7 & $2.27 \times 10^{-9}$ & 0.95 & 0.95  \\ \hline
      8 & $1.53 \times 10^{-9}$ & 0.64 & 0.997  \\ \hline
      9 & $1.47 \times 10^{-9}$ & 0.61 & 0.999  \\ \hline
      10 & $1.21 \times 10^{-9}$ & 0.51 & 0.996  \\ \hline
      \hline
    \end{tabular}
    \caption{ Continuum gamma-ray yields ($> \! 100\mev$) estimated for
    FERMI-LAT observations of the dwarf Segue 1 for each of the best
    ten models in our pMSSM model set. For each model we display the
    boosted yield, $B\times$(Yield) $\mathrm{cm}^{-2}\,\mathrm{s}^{-1}$,
    and the ratio of this quantity with the estimated 1 yr. $3\sigma$
    sensitivity $2.4 \times 10^{-9}\mathrm{cm}^{-2}\,\mathrm{s}^{-1}$ \cite{Essig:2009jx}.
    We also display the ratio of the annihilation rate into
    $\tau$-pairs over the total annihilation rate.
    }
    \label{dwfboosttab}
  \end{table}

  The result that the DM models which provide the best fits to the CR
  data sets should give gamma-ray contributions that are comparable to the
  current experimental sensitivities, to within a factor of a few, provides an
  exciting opportunity to confirm or refute these contributions to the
  anomalous CR data with measurements and analyses that will 
  be carried out in the very near term.

  We also investigate the direct detection signatures
  that can be expected from the pMSSM models in Table
  \ref{crphenotab}. Here, again, there is some ambiguity in determining
  the boost factor that should be applied to the DM direct detection
  signals, although the fact that these signals arise from DM
  \emph{scattering} instead of DM \emph{annihilation} necessitates
  further discussion.

  While we have remained agnostic as to the origin of the boost factor
  throughout our analysis, simply defining it as the factor by which
  $\tsigv$ must deviate from the canonical value that would result
  from thermal cosmological evolution in order to best fit the CR
  data, the origin of the boost factor has a significant impact on the
  resulting direct detection rates. For example, in scenarios where
  the annihilation signal is enhanced due to the presence of a
  resonance \cite{resDM}, there
  will be no corresponding boost to the direct detection signal, which
  arises via t-channel scattering diagrams. In scenarios which employ
  non-thermal cosmological evolution \cite{gordy},
  there will be an excess relic density of DM over that which would be
  expected from the usual thermal relation between $\sigv$ and $\Omega
  h^2|_{\rm LSP}$, and the direct detection signal should be boosted
  accordingly. If one imagines the boost factor to originate entirely from
  non-trivial substructure in the local DM distribution, then
  scattering in direct detection experiments may or may not incur a
  boost, as the overdensities responsible for enhanced annihilation
  may or may not coincide with terrestrial detectors.

\begin{figure}[hbtp]
    \centering
    \includegraphics[width=0.75\textwidth]{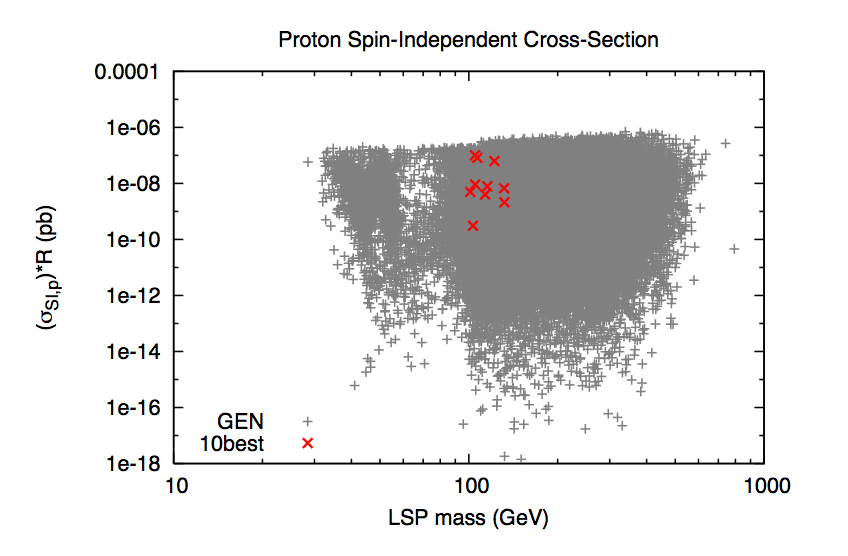}
    \caption{Proton spin-independent elastic scattering cross
    sections. \emph{Un-boosted} cross sections for the 10 models
    described in Table \ref{crphenotab} are highlighted (red points)
    against the cross sections from the entire $\sim 69$k pMSSM model
    set. Neutron spin-independent scattering cross sections are
    similar and are not presented here.
    }
    \label{figs:best10sip}
  \end{figure}

\begin{figure}[hbtp]
    \centering
    \includegraphics[width=0.75\textwidth]{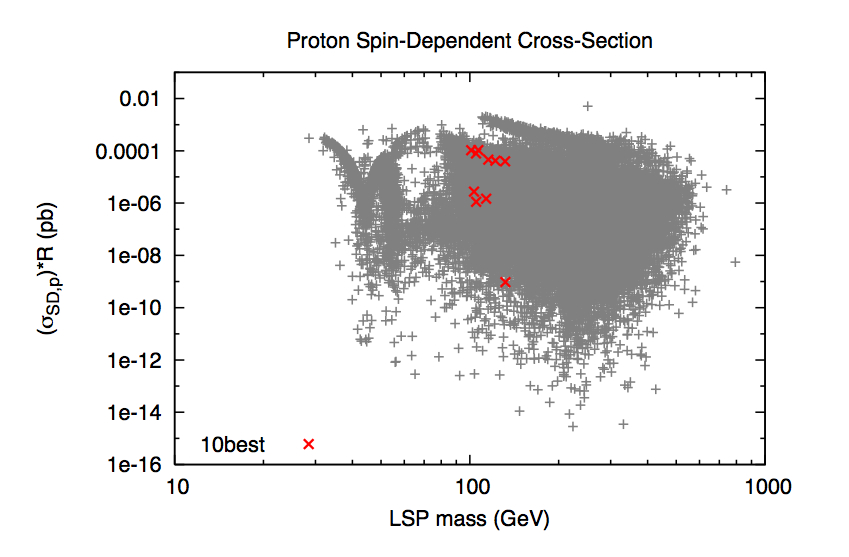}
    \caption{Proton spin-dependent elastic scattering cross
    sections. \emph{Un-boosted} cross sections for the 10 models
    described in Table \ref{crphenotab} are highlighted (red points)
    against the cross sections from the entire $\sim 69$k pMSSM model
    set. Neutron spin-dependent scattering cross sections are
    similar and are not presented here.
    }
    \label{figs:best10sdp}
  \end{figure}

  In light of this, we present Figures
  \ref{figs:best10sip}-\ref{figs:best10sdp}, where we highlight the
  direct detection cross sections for the ten models in Table
  \ref{crphenotab} against the rest of our pMSSM model set,
  \emph{without any boost factors included}. We note that the cross
  sections predicted for these models span several orders of
  magnitude, with some being on the edge of discovery.

\begin{figure}
    \centering
    \subfloat [Model 1]
    {\label{fig:48785spec}\includegraphics[width=0.48\textwidth]{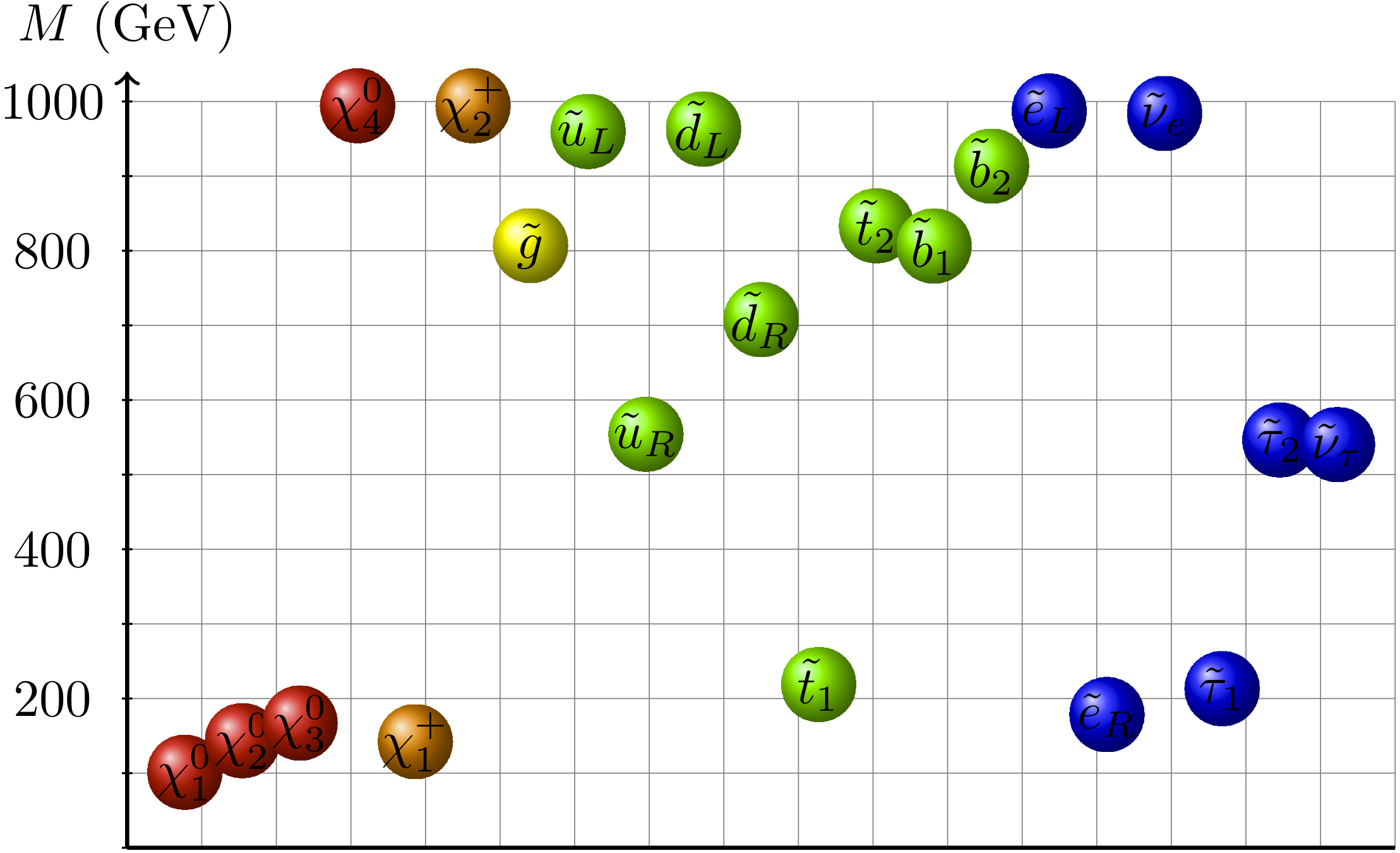}}
    \subfloat [Model 2]
    {\label{fig:39820spec}\includegraphics[width=0.48\textwidth]{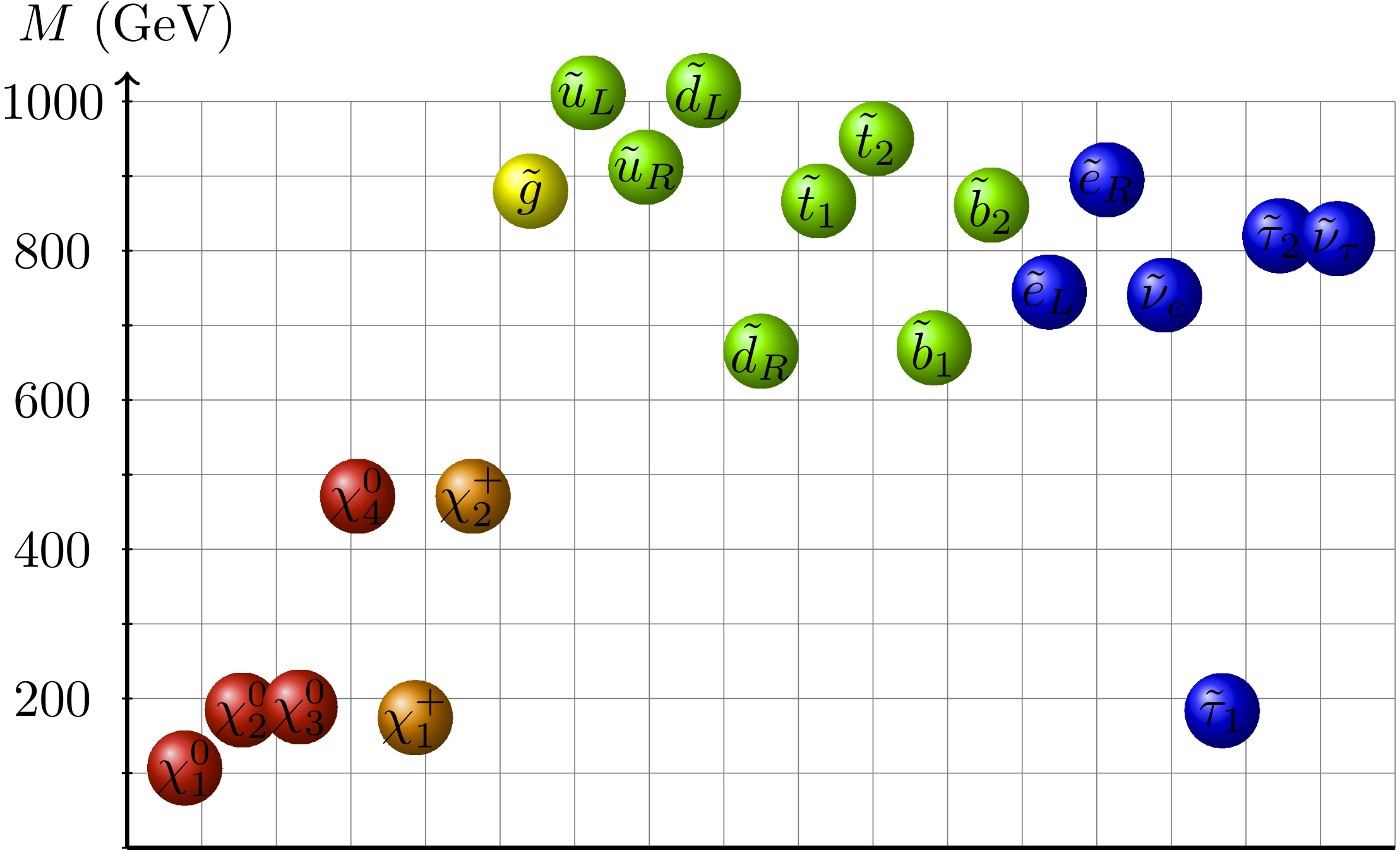}} \\
    \subfloat [Model 3]
    {\label{fig:11963spec}\includegraphics[width=0.48\textwidth]{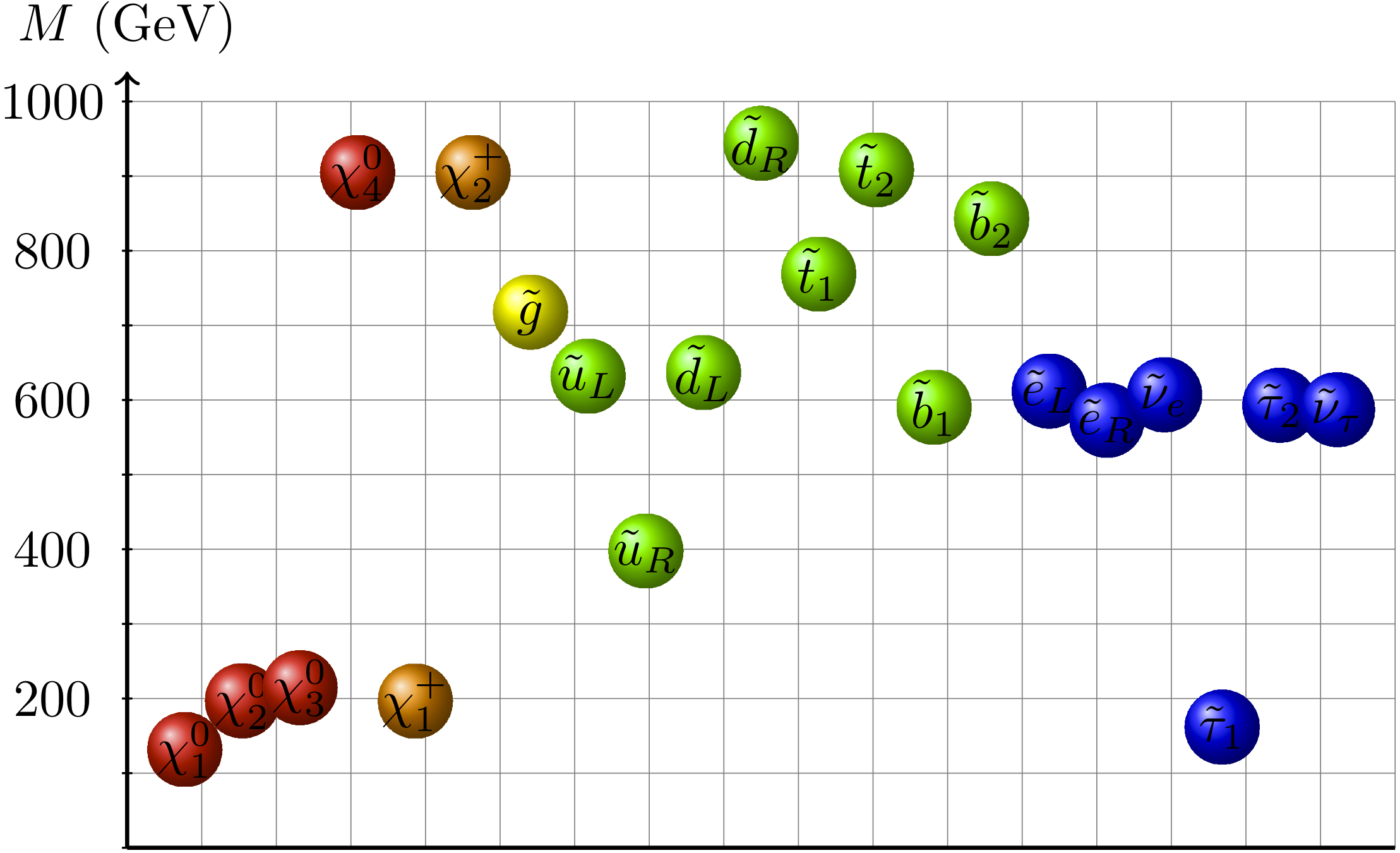}}
    \subfloat [Model 4]
    {\label{fig:23999spec}\includegraphics[width=0.48\textwidth]{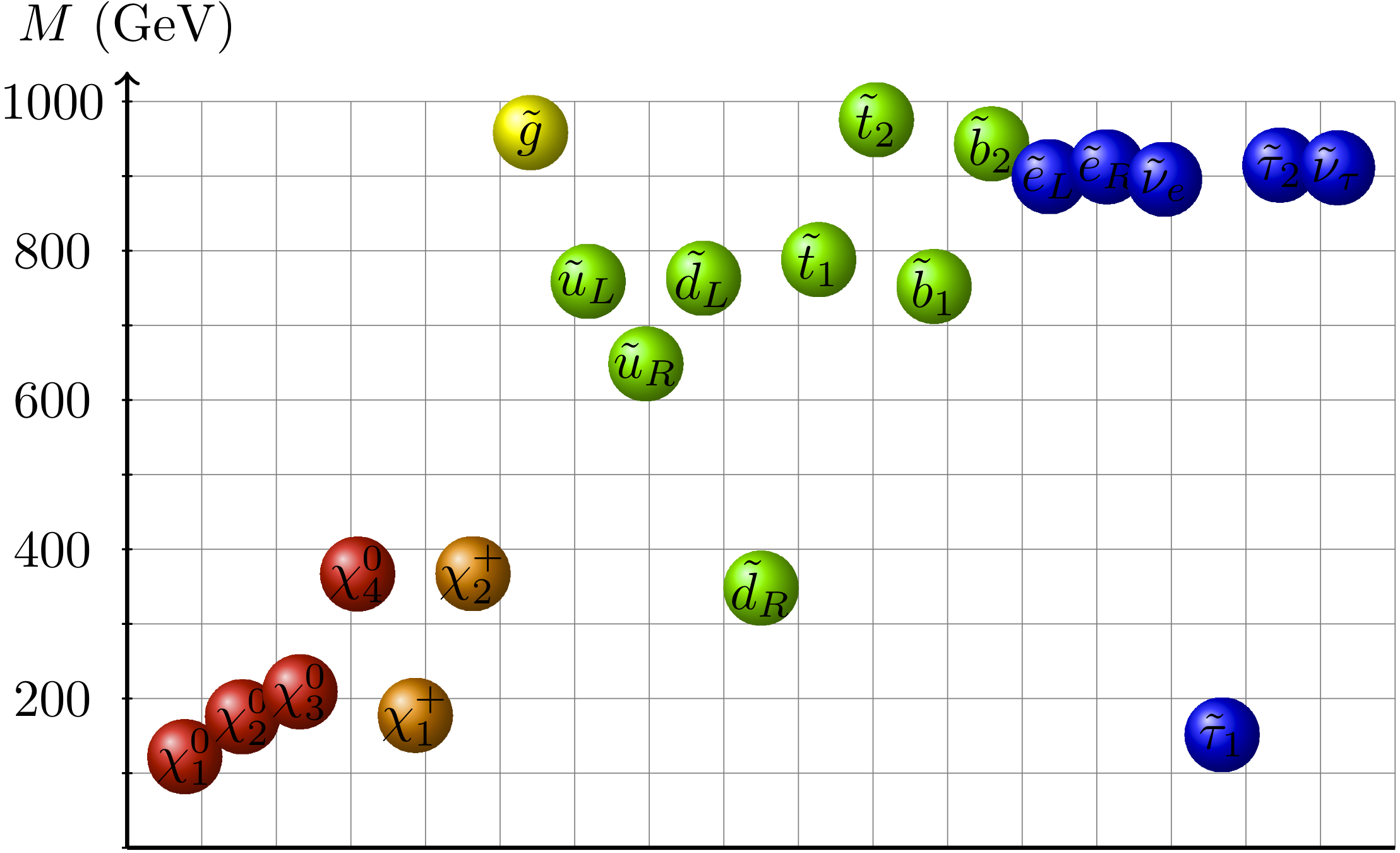}} \\
    \subfloat [Model 5]
    {\label{fig:9720spec}\includegraphics[width=0.48\textwidth]{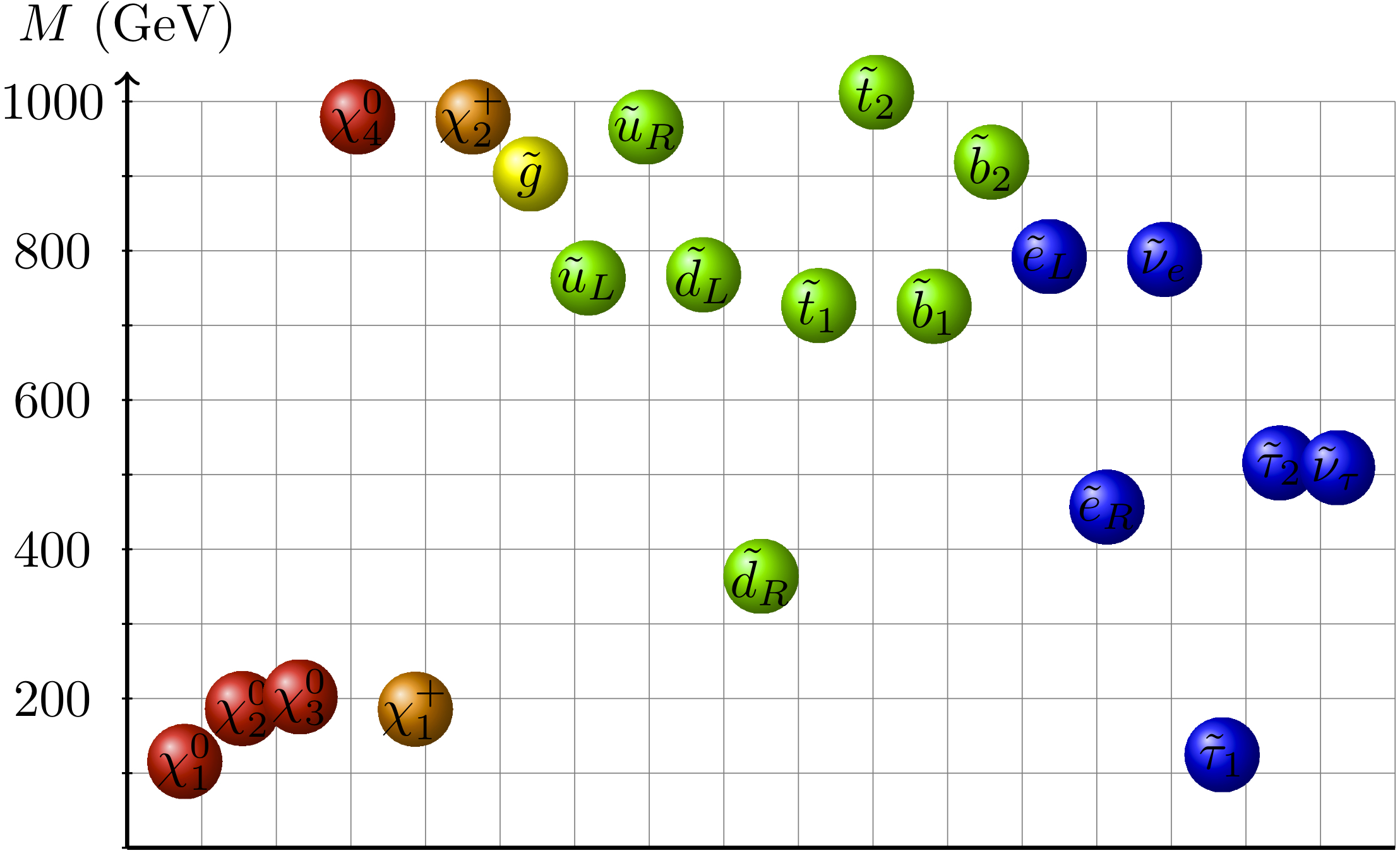}}
    \subfloat [Model 6]
    {\label{fig:61642spec}\includegraphics[width=0.48\textwidth]{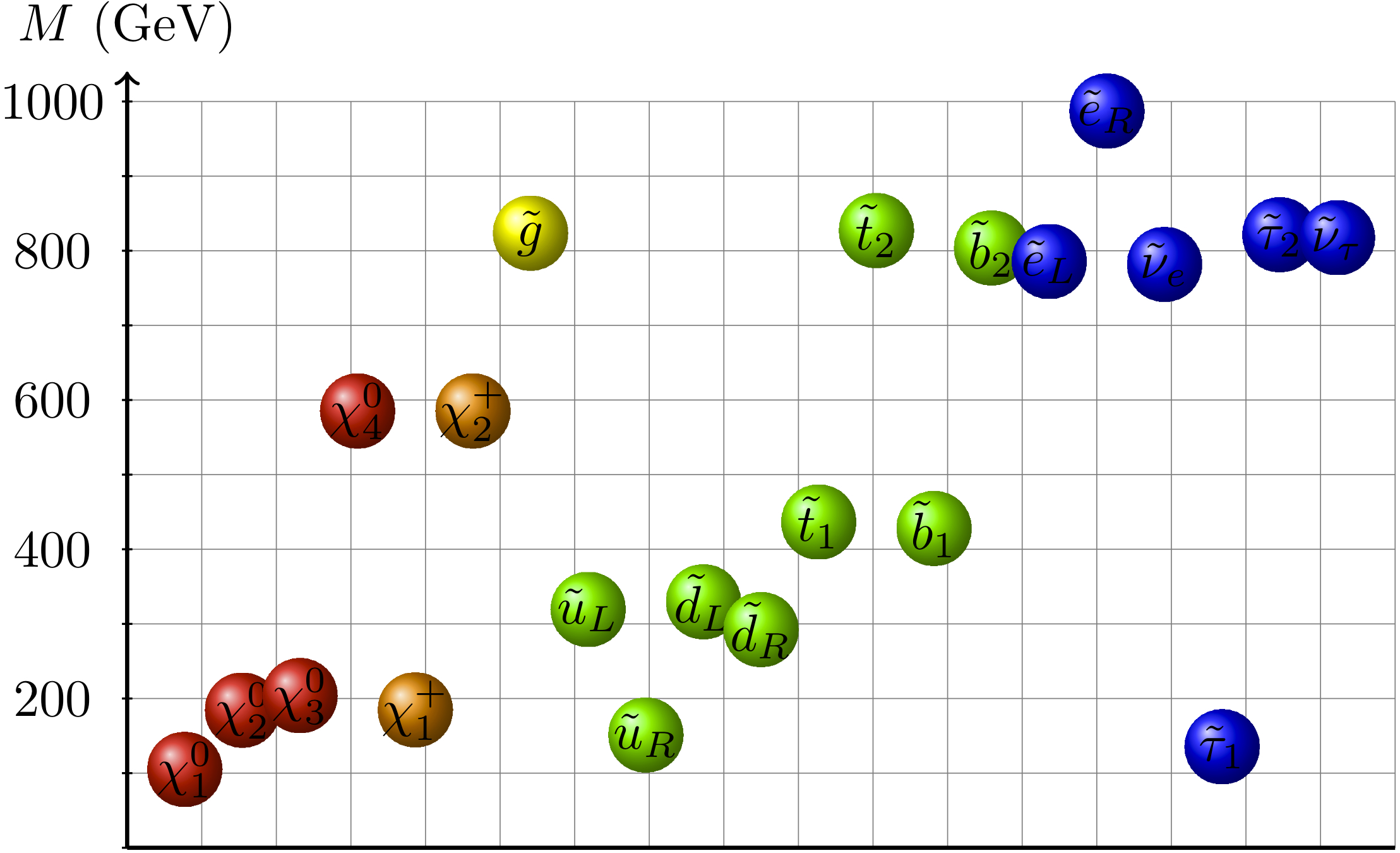}}
    \caption{Superparticle mass spectra for the pMSSM models discussed in
    Table \ref{crphenotab}. Colored balls represent masses for (left
    to right) $\tilde{\chi}^0_1$, $\tilde{\chi}^0_2$,
    $\tilde{\chi}^0_3$, $\tilde{\chi}^0_4$, $\tilde{\chi}^{+}_1$,
    $\tilde{\chi}^{+}_2$, $\tilde{g}$, $\tilde{u}_L$, $\tilde{u}_R$,
    $\tilde{d}_L$, $\tilde{d}_R$, $\tilde{t}_1$, $\tilde{t}_2$,
    $\tilde{b}_1$, $\tilde{b}_2$, $\tilde{e}_L$, $\tilde{e}_R$,
    $\tilde{\nu}_e$, $\tilde{\tau}_1$, $\tilde{\tau}_2$ and
    $\tilde{\nu}_{\tau}$, respectively.
    }
    \label{fig:ballplots1}
  \end{figure}

\begin{figure}
    \centering
    \subfloat [Model 7]
    {\label{fig:33623spec}\includegraphics[width=0.48\textwidth]{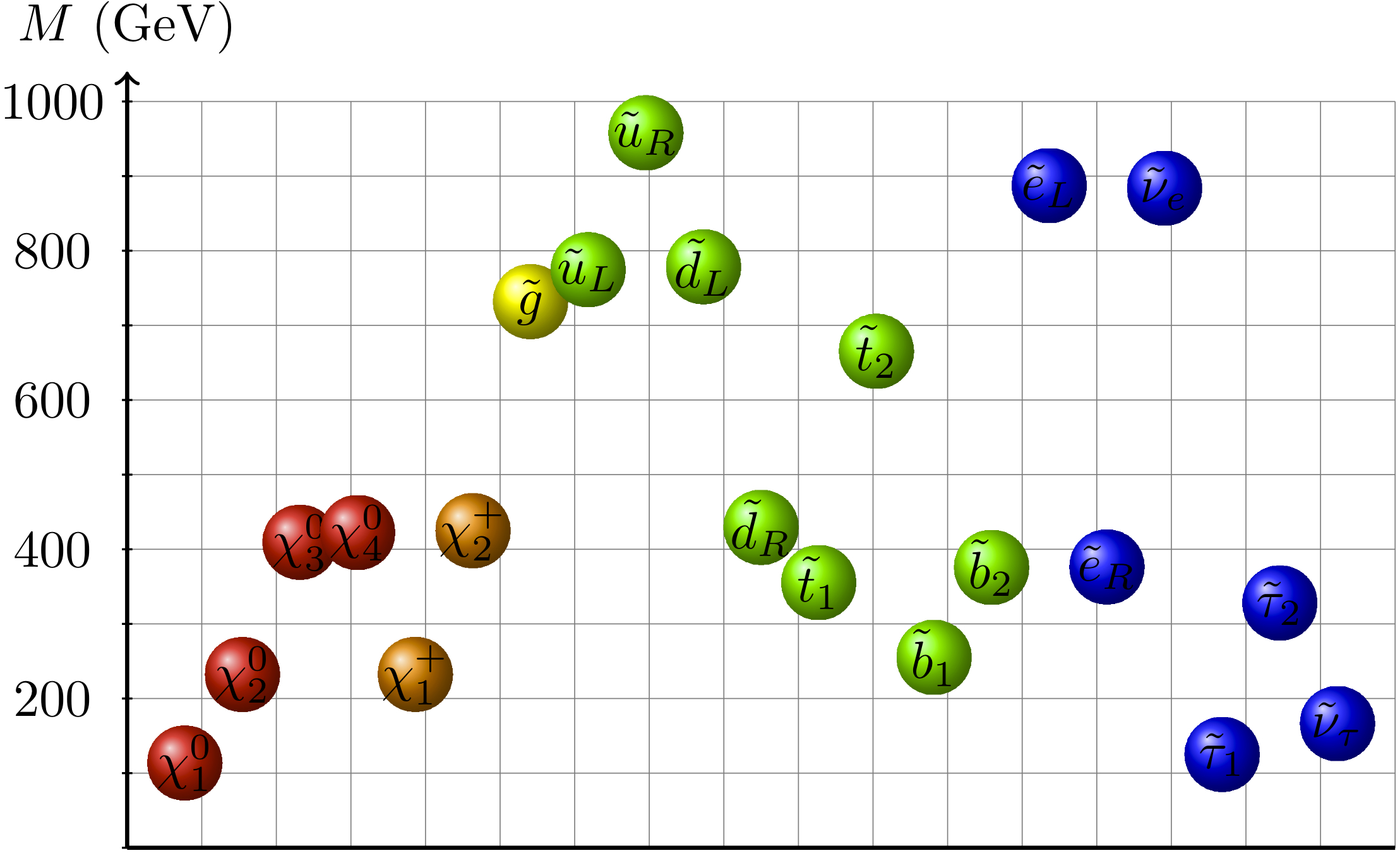}}
    \subfloat [Model 8]
    {\label{fig:58768spec}\includegraphics[width=0.48\textwidth]{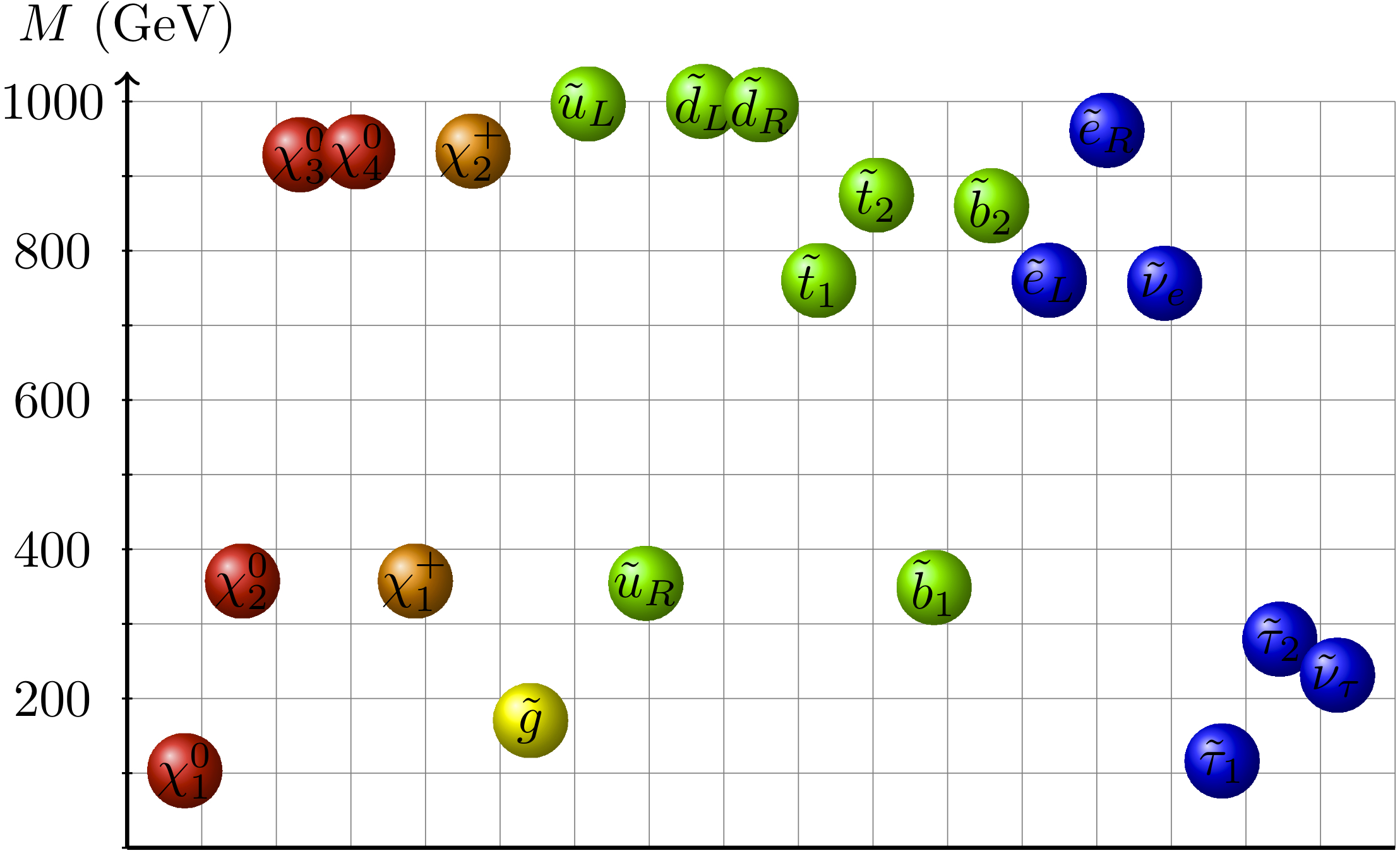}} \\
    \subfloat [Model 9]
    {\label{fig:12485spec}\includegraphics[width=0.48\textwidth]{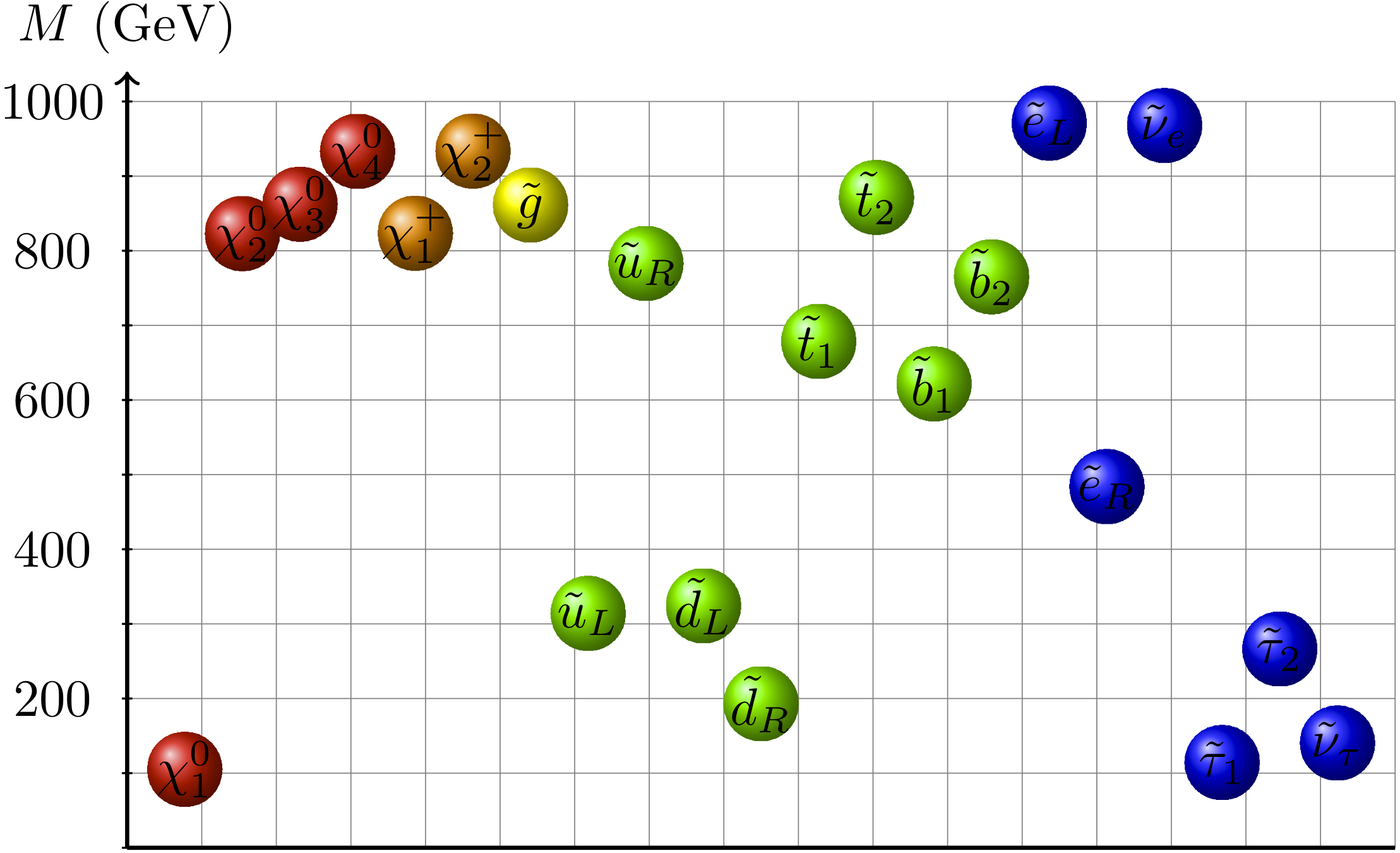}}
    \subfloat [Model 10]
    {\label{fig:11417spec}\includegraphics[width=0.48\textwidth]{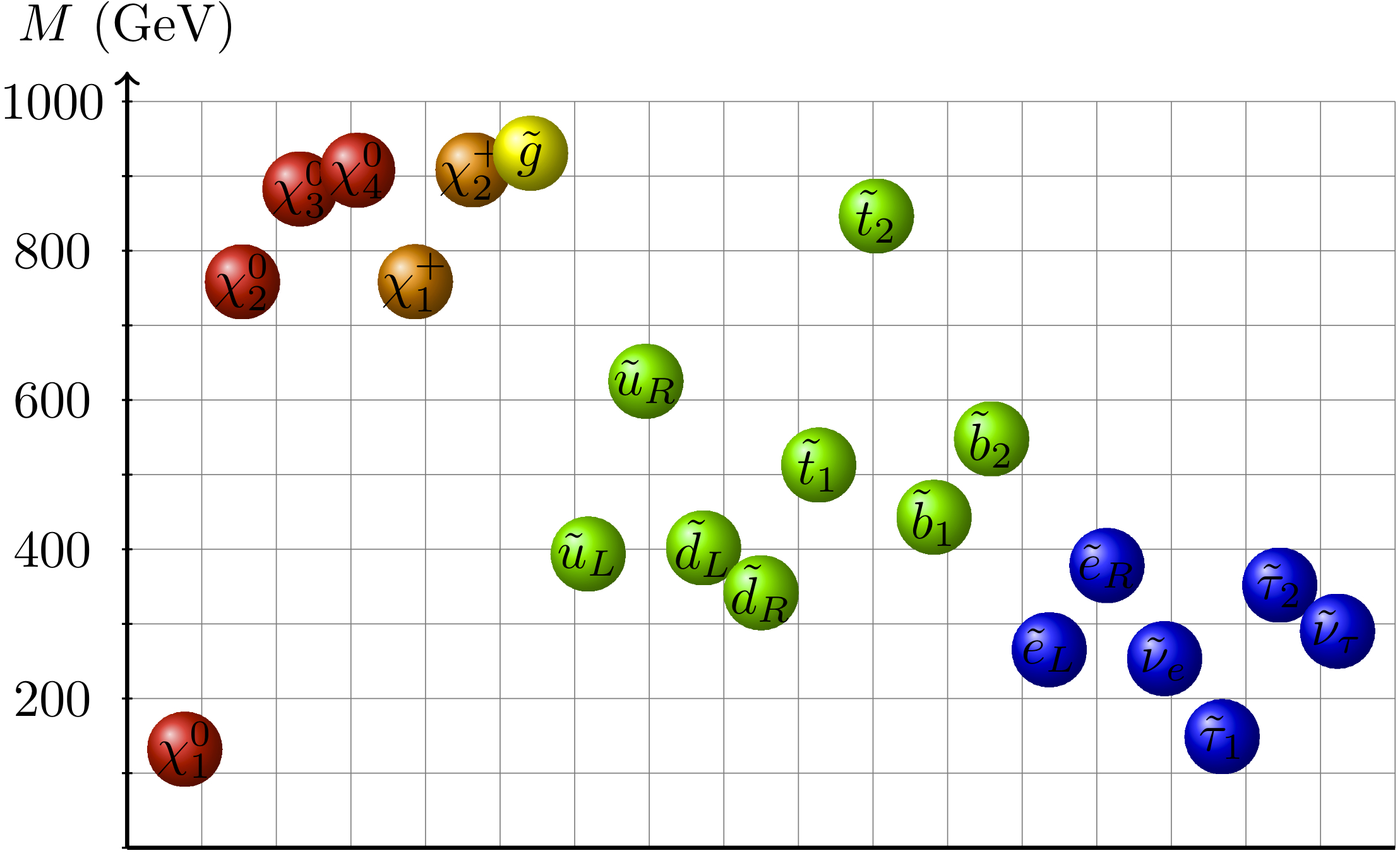}}
    \caption{Superparticle mass spectra for the pMSSM models discussed in
    Table \ref{crphenotab}. Colored balls represent masses for (left
    to right) $\tilde{\chi}^0_1$, $\tilde{\chi}^0_2$,
    $\tilde{\chi}^0_3$, $\tilde{\chi}^0_4$, $\tilde{\chi}^{+}_1$,
    $\tilde{\chi}^{+}_2$, $\tilde{g}$, $\tilde{u}_L$, $\tilde{u}_R$,
    $\tilde{d}_L$, $\tilde{d}_R$, $\tilde{t}_1$, $\tilde{t}_2$,
    $\tilde{b}_1$, $\tilde{b}_2$, $\tilde{e}_L$, $\tilde{e}_R$,
    $\tilde{\nu}_e$, $\tilde{\tau}_1$, $\tilde{\tau}_2$ and
    $\tilde{\nu}_{\tau}$, respectively.
    }
    \label{fig:ballplots2}
  \end{figure}

  Lastly, in Figures \ref{fig:48785spec}-f and 
  \ref{fig:33623spec}-d we display the full
  superparticle mass spectrum for each of the ten pMSSM
  models, described in Table \ref{crphenotab}, that provide the
  best-fit to the CR data with $B\leq 200$. The most important feature common to all models in
  this set is the light $\tilde{\tau}_1$; this being necessary 
  for annihilations dominantly to $\tau$-pairs. All LSPs in this set
  are (mostly-bino) mixtures of bino and higgsino gauge eigenstates
  but, we observe that the LSPs which are most purely bino-like
  undergo annihilations most purely into the $\tau^{+}\tau^{-}$ final state.

  \newpage
  
  \section{Discussion and Conclusions}
  \label{sec:conclusion}
  
  In this paper we have investigated the ability of recently measured
  astrophysical data to either constrain or shed light on the
  nature of particle dark matter in supersymmetric models.
  To this end we have employed the large set of phenomenologically 
  viable MSSM models generated in \cite{Berger:2008cq}. We have used
  the software package DarkSUSY \cite{Gondolo:2004sc} to
  calculate the products of DM annihilations using the full details
  (masses, mixing angles, etc.) that describe each pMSSM model. We have stressed
  the need to consider the uncertainty inherent in
  determining astrophysical backgrounds to DM signals, and have
  presented an analysis in which we used the software package GALPROP
  \cite{Strong:1998pw}\cite{Moskalenko:1997gh} to account for these
  uncertainties by scanning over phenomenologically viable CR 
  astrophysics models. We performed a simultaneous fit to the
  PAMELA \pamr and \pbarp data sets as well as to the FERMI \alle
  data for all possible combinations of pMSSM models and 
  astrophysical background models, using both DarkSUSY
  and GALPROP to consistently combine signal and background CR fluxes.
  
  As a result of our analyses we were able to identify an interesting
  class of pMSSM models. These models have CR signals that are unique
  in the sense that they come the closest to producing the steep rise
  in the positron fraction observed by the PAMELA collaboration,
  while requiring boost factors in the neighborhood of only
  $B\sim100$. This set of pMSSM models also provide good agreement
  with the FERMI \alle and PAMELA \pbarp data. These models are also
  unique in the sense that they are particularly difficult to
  discover or constrain from measurements of the total flux of
  gamma-rays coming from quiet astrophysical objects, such as the
  ultra-faint dwarf Segue 1 used as an example here. The contributions
  of these models to the diffuse mid-latitude gamma-ray spectrum
  yields a small excess over the preliminary FERMI data in the high
  energy bins, but is most likely within the uncertainties inherent in
  this calculation. In addition, we computed the DM direct detection
  rates for these pMSSM models and found that they span several orders
  of magnitude, with some models being on the edge of discovery. The
  CR fluxes generated by this class of pMSSM models
  stand out because they are
  produced by DM annihilations which are dominantly into the $\tau^{+}\tau^{-}$
  final state, a property which is due to having very light (often
  nLSP) $\tilde{\tau_1}$ and relatively heavy colored superpartner
  mass spectra. Models in this class which give rise to the
  largest annihilation fluxes, $\ie$ the largest \tsigv, have mass
  splittings $(m_{\tilde{\tau}_1}-m_{\LSP})\sim 10-20\gev$, which are
  not so small that they excessively dilute the relic density, $R$, and not so
  large that they excessively decrease the total cross-section, $\sigv$.
  
  Finally we remark on a few caveats concerning our results and on
  prospects for future observations or constraints on DM.
  
  We first note that while we produced simultaneous
  fits to the $e^+/(e^++e^-)$, \pbarp and \alle data with sizeable excesses 
  over background in \pamr and no
  corresponding excess in $\bar p/p$, the quality of the best fits is
  not as good as is perhaps desired. It is clear that the shape
  of the DM contribution to the \pamr fraction from this class 
  of ``best'' pMSSM models is not a perfect fit to the sharp rise 
  observed by PAMELA, although it would easily explain the more 
  modest excesses observed in the AMS-01 and HEAT data. There is also
  the question of how to obtain a boost factor in the range $B\sim
  70-190$, which is a number that is probably still too large to be
  explained by the current uncertainty in the normalization
  or substructure of the DM halo distribution of the Milky Way 
\cite{Brun:2009aj}\cite{Lavalle:2006vb}.
  Other works have attempted the explanation of such factors in terms of
  non-standard cosmology, or non-trivial modifications to
  the BSM particle model, while here we have remained 
  strictly agnostic as to the origin of the boost, seeking only
  to quantify its size.
  
  We also note that more prosaic astrophysical explanations ($\eg$ a
  local population of pulsars
  \cite{Profumo:2008ms}) could account for the
  rise in the positron fraction spectrum observed by PAMELA. In order
  to confirm a SUSY DM explanation from scenarios such as those
  discussed in this paper, one would likely need to see that subsequent 
  measurements of the positron fraction suggest an electroweak scale
  ($\sim100-200\gev$) edge feature, and that the upward slope of the 
  positron fraction is not as steep as that observed by PAMELA. Such
  assertions would require an experiment with exceptional 
  proton/positron ID, even for particles of energy in the $\sim100-200\gev$
  range, and large enough effective volume to beat down statistical errors in 
  such higher-energy bins. There is at least one near-future
  experiment, AMS-02, that may be capable of providing such a 
  measurement on the $\sim 5$ year time scale\footnote{We note the
    recent decision of the AMS-02 collaboration
    to replace the planned superconducting spectrometer magnet with
    a non-superconducting permanent magnet. It has been estimated that the
    silicon tracking system can be augmented such that the expected
    overall performance will remain unchanged \cite{Ting}.}.

  In the same time frame the LHC is expected to provide its first
  exploration of the $\tev$ scale. If the BSM particle physics 
  describing Nature is similar to the class of models discussed
  above, with $m_{\LSP}\sim100-200\gev$, a (typically nLSP) $\tilde{\tau_1}$
  of mass $m_{\tilde{\tau}_1}\approx m_{\LSP}+(10-20)\gev$, and
  relatively heavy colored superpartners, one may expect
  to observe many events with energetic $\tau$ leptons and
  $\slashed{E}_T$. Conversely, observations of such events at the LHC
  may be used to inform indirect detection strategies or, with 
  non-observation of DM signals in $\gamma$ and CR fluxes, to
  provide an indirect bound on the relic density of the observed
  $\LSP$. Such a limit may be an important complement to the constraints
  that may be attained via direct detection, especially in the case
  of a leptophilic LSP.

  One may also expect that the astrophysical uncertainties discussed here 
  may gradually decrease as data from experiments such as FERMI \cite{url-fermi},
  AMS-02 \cite{url-ams2}, CREAM \cite{url-cream} and TRACER
  \cite{url-tracer} are expected to enable a host of analyses
  that may improve and refine CR propagation models, though one
  certainly expects the case of CR $e^{\pm}$, whose transport depends
  strongly on the detailed features of the local galactic
  neighborhood, to remain quite a challenge.
  
  \section{Acknowledgments}
  \label{sec:Acknowledgments}

  The authors would like to thank M.P. Le, T. Porter and G. Tarl\'{e}
  for discussions related to this work. They would also like to thank
  M.P. Le for computational aid.  The work of R.C.C. is supported in part by an NSF Graduate
  Fellowship. The work of J.A.C. is supported by the BMBF 
  ``Verbundprojekt HEP-Theorie'' under contract 05H09PDE.  The work of J.S.G. is
  supported in part by the U.S. Department of Energy under contracts 
  No. DE-AC02-06CH11357 and No. DE-FG02-91ER40684.

  \newpage

  %
  %


\begin{thebibliography}{99}

\bibitem{silk} See talk given by J. Silk at {ICHEP2010}, 22-28 July,
  2010,
Paris, France.

\bibitem{Bergstrom:2010zz}
For a recent review, see,
  L.~Bergstrom,
  AIP Conf.\ Proc.\  {\bf 1241}, 49 (2010).

\bibitem{Komatsu:2008hk}
  E.~Komatsu {\it et al.}  [WMAP Collaboration],
  Astrophys.\ J.\ Suppl.\  {\bf 180}, 330 (2009)
  [arXiv:0803.0547 [astro-ph]].

\bibitem{Adriani:2008zr}
  O.~Adriani {\it et al.}  [PAMELA Collaboration],
  Nature {\bf 458}, 607 (2009)
  [arXiv:0810.4995 [astro-ph]].

\bibitem{Abdo:2009zk}
  A.~A.~Abdo {\it et al.}  [The Fermi LAT Collaboration],
  Phys.\ Rev.\ Lett.\  {\bf 102}, 181101 (2009)
  [arXiv:0905.0025 [astro-ph.HE]].

\bibitem{Adriani:2008zq}
  O.~Adriani {\it et al.},
  Phys.\ Rev.\ Lett.\  {\bf 102}, 051101 (2009)
  [arXiv:0810.4994 [astro-ph]].

\bibitem{Abdo:2010ex}
  A.~A.~Abdo {\it et al.},
  Astrophys.\ J.\  {\bf 712}, 147 (2010)
  [arXiv:1001.4531 [astro-ph.CO]].

\bibitem{Abdo:2010nc}
  A.~A.~Abdo {\it et al.},
  Phys.\ Rev.\ Lett.\  {\bf 104}, 091302 (2010)
  [arXiv:1001.4836 [astro-ph.HE]].

\bibitem{Abdo:2009mr}
  A.~A.~Abdo {\it et al.}  [Fermi LAT Collaboration],
  Phys.\ Rev.\ Lett.\  {\bf 103}, 251101 (2009)
  [arXiv:0912.0973 [astro-ph.HE]].

\bibitem{tporter:sympdiff}
  Preliminary results for a larger range of energies were presented in
  the talk of T. Porter at the \emph{2009 FERMI Symposium}, 
  Washington, DC, 2-5 Nov, 2009. http://www.slac.stanford.edu/econf/C0911022/.

\bibitem{SUSYrev}
H.~E.~Haber and G.~L.~Kane,
  Phys.\ Rept.\  {\bf 117}, 75 (1985);
S.~P.~Martin,
  arXiv:hep-ph/9709356;
D.~J.~H.~Chung, L.~L.~Everett, G.~L.~Kane, S.~F.~King, J.~D.~Lykken
and L.~T.~Wang,
  Phys.\ Rept.\  {\bf 407}, 1 (2005)
  [arXiv:hep-ph/0312378];
 L.~Pape and D.~Treille,
  Rept.\ Prog.\ Phys.\  {\bf 69}, 2843 (2006);
H.K. Dreiner, H.E. Haber, S.P. Martin,
``Two-component spinor techniques and Feynman rules for quantum field
theory
and supersymmetry,''
SCIPP-08/08, submitted to {\bf {Phys. Rep.}};
 M.~Drees, R.~Godbole and P.~Roy,
{\it  Hackensack, USA: World Scientific (2004) 555 p};
H.~Baer and X.~Tata,
{\it  Cambridge, UK: Univ. Pr. (2006) 537 p}.

\bibitem{toobig}
V.~Barger, W.~Y.~Keung, D.~Marfatia and G.~Shaughnessy,
  Phys.\ Lett.\  B {\bf 672}, 141 (2009)
  [arXiv:0809.0162 [hep-ph]];
I.~Cholis, L.~Goodenough, D.~Hooper, M.~Simet and N.~Weiner,
  Phys.\ Rev.\  D {\bf 80}, 123511 (2009)
  [arXiv:0809.1683 [hep-ph]];
M.~Cirelli, M.~Kadastik, M.~Raidal and A.~Strumia,
  Nucl.\ Phys.\  B {\bf 813}, 1 (2009)
  [arXiv:0809.2409 [hep-ph]].

\bibitem{Lavalle:2006vb}
  J.~Lavalle, J.~Pochon, P.~Salati and R.~Taillet,
  arXiv:astro-ph/0603796;
  J.~Lavalle, Q.~Yuan, D.~Maurin and X.~J.~Bi,
  arXiv:0709.3634 [astro-ph].

\bibitem{resDM}
D.~Feldman, Z.~Liu and P.~Nath,
  Phys.\ Rev.\  D {\bf 79}, 063509 (2009)
  [arXiv:0810.5762 [hep-ph]].
M.~Ibe, H.~Murayama and T.~T.~Yanagida,
  Phys.\ Rev.\  D {\bf 79}, 095009 (2009)
  [arXiv:0812.0072 [hep-ph]];
W.~L.~Guo and Y.~L.~Wu,
  Phys.\ Rev.\  D {\bf 79}, 055012 (2009)
  [arXiv:0901.1450 [hep-ph]];
K.~Kadota, K.~Freese and P.~Gondolo,
  Phys.\ Rev.\  D {\bf 81}, 115006 (2010)
  [arXiv:1003.4442 [hep-ph]].

\bibitem{summf}
A. Sommerfeld, Annalen der Physik {\bf 403}, 257 (1931);
N.~Arkani-Hamed, D.~P.~Finkbeiner, T.~R.~Slatyer and N.~Weiner,
  Phys.\ Rev.\  D {\bf 79}, 015014 (2009)
  [arXiv:0810.0713 [hep-ph]];
M.~Pospelov and A.~Ritz,
  Phys.\ Lett.\  B {\bf 671}, 391 (2009)
  [arXiv:0810.1502 [hep-ph]];
P.~J.~Fox and E.~Poppitz,
  Phys.\ Rev.\  D {\bf 79}, 083528 (2009)
  [arXiv:0811.0399 [hep-ph]];
J.~L.~Feng, M.~Kaplinghat, H.~-B.~Yu,
  Phys.\ Rev.\ Lett.\  {\bf 104}, 151301 (2010).
  [arXiv:0911.0422 [hep-ph]];
 M.~Backovic, J.~P.~Ralston,
  Phys.\ Rev.\  {\bf D81}, 056002 (2010).
  [arXiv:0910.1113 [hep-ph]];
  J.~L.~Feng, M.~Kaplinghat, H.~-B.~Yu,
    [arXiv:1005.4678 [hep-ph]].

\bibitem{Bai:2009ka}
  Y.~Bai, M.~Carena, J.~Lykken,
  Phys.\ Rev.\  {\bf D80}, 055004 (2009).
  [arXiv:0905.2964 [hep-ph]].

\bibitem{Bajc:2010qj}
  B.~Bajc, T.~Enkhbat, D.~K.~Ghosh {\it et al.},
  JHEP {\bf 1005}, 048 (2010).
  [arXiv:1002.3631 [hep-ph]]

\bibitem{gordy}
P.~Grajek, G.~Kane, D.~Phalen, A.~Pierce and S.~Watson,
  Phys.\ Rev.\  D {\bf 79}, 043506 (2009)
  [arXiv:0812.4555 [hep-ph]];
G.~Kane, R.~Lu and S.~Watson,
  Phys.\ Lett.\  B {\bf 681}, 151 (2009)
  [arXiv:0906.4765 [astro-ph.HE]];
B.~S.~Acharya, G.~Kane, S.~Watson and P.~Kumar,
  Phys.\ Rev.\  D {\bf 80}, 083529 (2009)
  [arXiv:0908.2430 [astro-ph.CO]].

\bibitem{Berger:2008cq}
  C.~F.~Berger, J.~S.~Gainer, J.~L.~Hewett and T.~G.~Rizzo,
  JHEP {\bf 0902}, 023 (2009)
  [arXiv:0812.0980 [hep-ph]];
R.~C.~Cotta, J.~S.~Gainer, J.~L.~Hewett and T.~G.~Rizzo,
  New J.\ Phys.\  {\bf 11}, 105026 (2009)
  [arXiv:0903.4409 [hep-ph]].

\bibitem{D'Ambrosio:2002ex}
  G.~D'Ambrosio, G.~F.~Giudice, G.~Isidori {\it et al.},
  Nucl.\ Phys.\  {\bf B645}, 155-187 (2002).
  [hep-ph/0207036].

\bibitem{uslhc}J.A. Conley, J.S. Gainer, J.L. Hewett, M.P. Le, and T.G.
Rizzo, in preparation.

\bibitem{Jungman:1995df}
  G.~Jungman, M.~Kamionkowski, K.~Griest,
  Phys.\ Rept.\  {\bf 267}, 195-373 (1996).
  [hep-ph/9506380].

\bibitem{Gates:1995dw}
  E.~I.~Gates, G.~Gyuk and M.~S.~Turner,
  Astrophys.\ J.\  {\bf 449}, L123 (1995)
  [arXiv:astro-ph/9505039].

\bibitem{Diemand:2008in}
  J.~Diemand, M.~Kuhlen, P.~Madau, M.~Zemp, B.~Moore, D.~Potter and
  J.~Stadel,
  Nature {\bf 454}, 735 (2008)
  [arXiv:0805.1244 [astro-ph]].

\bibitem{Strigari:2007at}
  L.~E.~Strigari, S.~M.~Koushiappas, J.~S.~Bullock {\it et al.},
  [arXiv:0709.1510 [astro-ph]].

\bibitem{Delahaye:2007fr}
  T.~Delahaye, R.~Leroy's, F.~Donato, N.~Fornengo and P.~Salati,
  Phys.\ Rev.\  D {\bf 77}, 063527 (2008)
  [arXiv:0712.2312 [astro-ph]];
  T.~Delahaye, F.~Donato, N.~Fornengo, J.~Lavalle, R.~Lineros,
  P.~Salati and R.~Taillet,
  Astron.\ Astrophys.\  {\bf 501}, 821 (2009)
  [arXiv:0809.5268 [astro-ph]].

\bibitem{Delahaye:2010ji}
  T.~Delahaye, J.~Lavalle, R.~Lineros, F.~Donato and N.~Fornengo,
  arXiv:1002.1910 [astro-ph.HE].

\bibitem{Grasso:2009ma}
  D.~Grasso {\it et al.}  [FERMI-LAT Collaboration],
  Astropart.\ Phys.\  {\bf 32}, 140 (2009)
  [arXiv:0905.0636 [astro-ph.HE]].

\bibitem{Strong:2007nh}
  A.~W.~Strong, I.~V.~Moskalenko and V.~S.~Ptuskin,
  Ann.\ Rev.\ Nucl.\ Part.\ Sci.\  {\bf 57}, 285 (2007)
  [arXiv:astro-ph/0701517].

\bibitem{Ginzburg:1990sk}
  V.~L.~Ginzburg, V.~A.~Dogiel, V.~S.~Berezinsky, S.~V.~Bulanov and
  V.~S.~Ptuskin,
{\it  Amsterdam, Netherlands: North-Holland (1990) 534 p};
  R.~Schlickeiser,
{\it  Berlin, Germany: Springer (2002) 519 p}.

\bibitem{Strong:1998pw}
  A.~W.~Strong and I.~V.~Moskalenko,
  Astrophys.\ J.\  {\bf 509}, 212 (1998)
  [arXiv:astro-ph/9807150].

\bibitem{Moskalenko:1997gh}
  I.~V.~Moskalenko and A.~W.~Strong,
  Astrophys.\ J.\  {\bf 493}, 694 (1998)
  [arXiv:astro-ph/9710124].

\bibitem{Strong:2004td}
  A.~W.~Strong, I.~V.~Moskalenko, O.~Reimer, S.~Digel and R.~Diehl,
  Astron.\ Astrophys.\  {\bf 422}, L47 (2004)
  [arXiv:astro-ph/0405275].

\bibitem{Porter:2005qx}
  T.~A.~Porter and A.~W.~Strong,
  arXiv:astro-ph/0507119.

\bibitem{Gondolo:2004sc}
  P.~Gondolo, J.~Edsjo, P.~Ullio, L.~Bergstrom, M.~Schelke and
  E.~A.~Baltz,
  JCAP {\bf 0407}, 008 (2004)
  [arXiv:astro-ph/0406204].

\bibitem{GALPROP}
    http://galprop.stanford.edu/web\_galprop/galprop\_home.html.

\bibitem{Shikaze:2006je}
  Y.~Shikaze {\it et al.},
  Astropart.\ Phys.\  {\bf 28}, 154 (2007)
  [arXiv:astro-ph/0611388].

\bibitem{Gleeson:1968zz}
  L.~J.~Gleeson and W.~I.~Axford,
  Astrophys.\ J.\  {\bf 154}, 1011 (1968).

\bibitem{clemetal}
  Clem, J.~M. {\it et al.},
  Astrophys.\ J.\  {\bf 464}, 507 (1996).

\bibitem{Tarle}
  See G. Tarl\'{e} talk at \emph{the Ninth UCLA Symposium on Sources and
  Detection of Dark Matter and Dark Energy in the Universe}, Marina del
  Rey, CA, 24-26 Feb, 2010. http://www.physics.ucla.edu/hep/dm10/index.html.

\bibitem{Haino:2004nq}
  S.~Haino {\it et al.},
  Phys.\ Lett.\  B {\bf 594}, 35 (2004)
  [arXiv:astro-ph/0403704].

\bibitem{Alcaraz:2000vp}
  J.~Alcaraz {\it et al.}  [AMS Collaboration],
  Phys.\ Lett.\  B {\bf 490}, 27 (2000).

\bibitem{Boezio:2002ha}
  M.~Boezio {\it et al.},
  Astropart.\ Phys.\  {\bf 19}, 583 (2003)
  [arXiv:astro-ph/0212253].

\bibitem{Panov:2006kf}
  A.~D.~Panov {\it et al.},
  arXiv:astro-ph/0612377.

\bibitem{deltalow}
  A.~W.~Strong and I.~V.~Moskalenko,
  Astrophys.\ J.\  {\bf 509}, 212 (1998)
  [arXiv:astro-ph/9807150].
  G.~Di Bernardo, C.~Evoli, D.~Gaggero, D.~Grasso and L.~Maccione,
  arXiv:0909.4548 [astro-ph.HE].
  M.~Simet and D.~Hooper,
  JCAP {\bf 0908}, 003 (2009)
  [arXiv:0904.2398 [astro-ph.HE]].

\bibitem{deltahi}
  D.~Maurin, F.~Donato, R.~Taillet and P.~Salati,
  Astrophys.\ J.\  {\bf 555}, 585 (2001)
  [arXiv:astro-ph/0101231].
  A.~Putze, L.~Derome and D.~Maurin,
  arXiv:1001.0551 [astro-ph.HE].
  D.~Maurin, A.~Putze and L.~Derome,
  arXiv:1001.0553 [astro-ph.HE].

\bibitem{Cox:1987ww}
  D.~P.~Cox and R.~J.~Reynolds,
  Ann.\ Rev.\ Astron.\ Astrophys.\  {\bf 25}, 303 (1987);
  F.~Donato, D.~Maurin and R.~Taillet,
  Astron.\ Astrophys.\  {\bf 381}, 539 (2002)
  [arXiv:astro-ph/0108079].

\bibitem{Simet:2009ne}
  M.~Simet and D.~Hooper,
  JCAP {\bf 0908}, 003 (2009)
  [arXiv:0904.2398 [astro-ph.HE]].

\bibitem{Ahn:2008my}
  H.~S.~Ahn {\it et al.},
  Astropart.\ Phys.\  {\bf 30}, 133 (2008)
  [arXiv:0808.1718 [astro-ph]].

\bibitem{Engelmann:1990zz}
  J.~J.~Engelmann, P.~Ferrando, A.~Soutoul, P.~Goret and E.~Juliusson,
  Astron.\ Astrophys.\  {\bf 233}, 96 (1990).

\bibitem{Panov:2007fe}
  A.~D.~Panov {\it et al.},
  arXiv:0707.4415 [astro-ph].

\bibitem{picozza:tevpa}
  See P. Picozza talk at \emph{TeV Particle Astrophysics 2009}, SLAC
  National Accelerator Laboratory, Menlo Park, CA, 13-17 July, 2009.
  http://www-conf.slac.stanford.edu/tevpa09/Talks.asp.



\bibitem{Regis:2009md}
  M.~Regis and P.~Ullio,
  Phys.\ Rev.\  D {\bf 80}, 043525 (2009)
  [arXiv:0904.4645 [astro-ph.GA]].

\bibitem{Moderksi:2005jw}
  R.~Moderksi, M.~Sikora, P.~S.~Coppi and F.~A.~Aharonian,
  [arXiv:astro-ph/0504388];
  G.~R.~Blumenthal and R.~J.~Gould,
  Rev.\ Mod.\ Phys.\  {\bf 42}, 237 (1970).

\bibitem{Jones:1968zza}
  F.~C.~Jones,
  Phys.\ Rev.\  {\bf 167}, 1159 (1968).

\bibitem{Schlickeiser:2009qq}
  R.~Schlickeiser and J.~Ruppel,
  New J.\ Phys.\  {\bf 12}, 033044 (2010)
  [arXiv:0908.2183 [astro-ph.HE]].

\bibitem{Stawarz:2009ig}
  L.~Stawarz, V.~Petrosian and R.~D.~Blandford,
  Astrophys.\ J.\  {\bf 710}, 236 (2010)
  [arXiv:0908.1094 [astro-ph.GA]].

\bibitem{Hunger:1997we}
  S.~D.~Hunter {\it et al.},
  Astrophys.\ J.\  {\bf 481}, 205 (1997).

\bibitem{Ptuskin:2005ax}
  V.~S.~Ptuskin, I.~V.~Moskalenko, F.~C.~Jones {\it et al.},
  Astrophys.\ J.\  {\bf 642}, 902-916 (2006).
  [astro-ph/0510335].

\bibitem{Kamae:2006bf}
  T.~Kamae, N.~Karlsson, T.~Mizuno {\it et al.},
  Astrophys.\ J.\  {\bf 647}, 692-708 (2006).
  [astro-ph/0605581].

\bibitem{Cumberbatch:2010ii}
  D.~T.~Cumberbatch, Y.~-L.~S.~Tsai, L.~Roszkowski,
  [arXiv:1003.2808 [astro-ph.HE]].

\bibitem{NFW}
  J.~F.~Navarro, C.~S.~Frenk and S.~D.~M.~White,
  Astrophys.\ J.\  {\bf 490}, 493 (1997)
  [arXiv:astro-ph/9611107].
  J.~F.~Navarro, C.~S.~Frenk and S.~D.~M.~White,
  Astrophys.\ J.\  {\bf 462}, 563 (1996)
  [arXiv:astro-ph/9508025].

\bibitem{DuVernois:2001bb}
  M.~A.~DuVernois {\it et al.} [HEAT Collaboration],
  Measurements with
  Astrophys.\ J.\  {\bf 559}, 296 (2001);
  S.~W.~Barwick {\it et al.}  [HEAT Collaboration],
  Astrophys.\ J.\  {\bf 482}, L191 (1997)
  [arXiv:astro-ph/9703192].

\bibitem{Aguilar:2002ad}
  M.~Aguilar {\it et al.}  [AMS Collaboration],
  Station.
  Phys.\ Rept.\  {\bf 366}, 331 (2002)
  [Erratum-ibid.\  {\bf 380}, 97 (2003)].

\bibitem{CAPRICE}
  Boezio, ~M. {\it et al.}  [CAPRICE Collaboration],

\bibitem{Bringmann:2007nk}
  T.~Bringmann, L.~Bergstrom and J.~Edsjo,
  JHEP {\bf 0801}, 049 (2008)
  [arXiv:0710.3169 [hep-ph]].

\bibitem{Barger:2009xe}
  V.~Barger, Y.~Gao, W.~Y.~Keung and D.~Marfatia,
  Phys.\ Rev.\  D {\bf 80}, 063537 (2009)
  [arXiv:0906.3009 [hep-ph]].

\bibitem{Bertone:2008xr}
  G.~Bertone, M.~Cirelli, A.~Strumia and M.~Taoso,
  JCAP {\bf 0903}, 009 (2009)
  [arXiv:0811.3744 [astro-ph]].

\bibitem{Simon:2007dq}
  J.~D.~Simon, M.~Geha,
  Astrophys.\ J.\  {\bf 670}, 313-331 (2007).
  [arXiv:0706.0516 [astro-ph]].

\bibitem{Belokurov:2006ph}
  V.~Belokurov {\it et al.}  [SDSS Collaboration],
  Astrophys.\ J.\  {\bf 654}, 897 (2007)
  [arXiv:astro-ph/0608448].

\bibitem{Geha:2008zr}
  M.~Geha, B.~Willman, J.~D.~Simon, L.~E.~Strigari, E.~N.~Kirby,
  D.~R.~Law and J.~Strader,
  Astrophys.\ J.\  {\bf 692}, 1464 (2009)
  [arXiv:0809.2781 [astro-ph]].

\bibitem{Essig:2009jx}
  R.~Essig, N.~Sehgal and L.~E.~Strigari,
  Phys.\ Rev.\  D {\bf 80}, 023506 (2009)
  [arXiv:0902.4750 [hep-ph]].

\bibitem{Essig:2010em}
  R.~Essig, N.~Sehgal, L.~E.~Strigari, M.~Geha and J.~D.~Simon,
  arXiv:1007.4199 [astro-ph.CO].

\bibitem{futuredwarf}
  Work in progress.

\bibitem{Cirelli:2009dv}
  M.~Cirelli, P.~Panci and P.~D.~Serpico,
  arXiv:0912.0663 [astro-ph.CO].

\bibitem{Kamionkowski:2010mi}
  M.~Kamionkowski, S.~M.~Koushiappas, M.~Kuhlen,
  Phys.\ Rev.\  {\bf D81}, 043532 (2010).
  [arXiv:1001.3144 [astro-ph.GA]].

\bibitem{Brun:2009aj}
  P.~Brun, T.~Delahaye, J.~Diemand, S.~Profumo and P.~Salati,
  Phys.\ Rev.\  D {\bf 80}, 035023 (2009)
  [arXiv:0904.0812 [astro-ph.HE]].

\bibitem{Profumo:2008ms}
  S.~Profumo,
  [arXiv:0812.4457 [astro-ph]];
  L.~Gendelev, S.~Profumo, M.~Dormody,
  JCAP {\bf 1002}, 016 (2010).
  [arXiv:1001.4540 [astro-ph.HE]].

\bibitem{Ting}
  See presentation by S. Ting at http://www.er.doe.gov/hep/files/pdfs/HEPAP\_AMS.pdf.

\bibitem{url-fermi}
  http://fermi.gsfc.nasa.gov/

\bibitem{url-ams2}
  http://ams.cern.ch/

\bibitem{url-cream}
  http://cosmicray.umd.edu/cream/

\bibitem{url-tracer}
  http://tracer.uchicago.edu/


\end{thebibliography}
\end{document}